\documentclass[reqno,12pt]{article}
\usepackage[english]{babel}

\usepackage{ae} % or {zefonts}0
\usepackage[T1]{fontenc}
\usepackage{amsmath}
\numberwithin{equation}{section}

\usepackage{amssymb}
\usepackage{mathrsfs}
\usepackage{graphicx}
\usepackage{mdframed}
\usepackage{amsfonts,mathtools}
\usepackage{color}
\definecolor{darkblue}{cmyk}{0.9,0.9,0,0}
\usepackage[colorlinks=true,linkcolor=darkblue,citecolor=darkblue,urlcolor=darkblue]{hyperref}
\usepackage{epsfig}

\usepackage{graphicx}
\usepackage{tikz}

\newcommand{\beq}{\begin{equation}}
\newcommand{\eeq}{\end{equation}}
\newcommand\beqa{\begin{eqnarray}}
\newcommand\eeqa{\end{eqnarray}}
\newcommand\bea{\begin{array}}
\newcommand\eea{\end{array}}

\def\XXint#1#2#3{{\setbox0=\hbox{$#1{#2#3}{\int}$}
\vcenter{\hbox{$#2#3$}}\kern-.5\wd0}}

\newcommand{\nn}{\nonumber}

\newcommand{\neqa}{\nonumber\end{eqnarray}}
\newcommand{\la}[1]{\label{#1}}

\newcommand{\ie}{{\it i.e.,\ }}
\newcommand{\eg}{{\it e.g.,\ }}
\newcommand{\mt}[1]{\textrm{\tiny #1}}
\newcommand{\GN}{G_\mt{N}}
\newcommand{\tz}{\tilde{z}}
\newcommand{\tx}{\tilde{x}}
\newcommand{\bx}{{\bf x}}
\newcommand{\by}{{\bf y}}
\newcommand{\bz}{{\bf z}}
\newcommand{\hrho}{{\pmb \rho}} %{\hat\rho}

\renewcommand{\d}{\partial}

\newcommand{\<}{{\langle}}
\renewcommand{\>}{{\rangle}}

\newcommand{\re}{\relax{\rm I\kern-.18em R}}

\renewcommand{\sp}{p\hspace{-.40em}/}

\def\su2{{SU(2)}}

\def\[{\left[}
\def\]{\right]}

\def\({\left(}
\def\){\right)}
\def\[{\left[}
\def\]{\right]}

\def\<{\langle}
\def\>{\rangle}

\def\i2{\frac{i}{2}}

\def\spi{\relax{\rm \pi\kern-0.5em /}}
\def\sA{\relax{\rm A\kern-0.5em /}}
\def\sp{\relax{\rm p\kern-0.5em /}}
\def\sd{\relax{\rm \d\kern-0.5em /}}
\def\sk{\relax{\rm k\kern-0.5em /}}
\def\sn{\relax{\rm n\kern-0.5em /}}
\def\sl{\relax{\rm l\kern-0.5em /}}
\def\sP{\relax{\rm P\kern-0.7em /}}
\def\sBethe{\relax{\rm \Bethe\kern-0.5em /}}

\def\bx{{\bf x}}

	\def\i{\textrm{i}}

\usepackage{varioref}
\usepackage{makeidx}
\makeindex

\newcommand\blfootnote[1]{%
  \begingroup
  \renewcommand\thefootnote{}\footnote{\hspace{-6mm}#1}%
  \addtocounter{footnote}{-1}%
  \endgroup
}

\def\OOmega{\Theta}

\begin{document}

\thispagestyle{empty}

\renewcommand{\thefootnote}{\fnsymbol{footnote}}
\setcounter{page}{1}
\setcounter{footnote}{0}
\setcounter{figure}{0}

%\begin{flushright}
%CERN-TH-2017-162
%\end{flushright}
\vspace{-0.4in}

\begin{center}
$$~$$
{\Large
Holography and Correlation Functions of Huge Operators:\\ 
Spacetime Bananas
\par}
\vspace{1.0cm}

{Jacob Abajian$^\text{\tiny 1,\tiny 2}$, Francesco Aprile$^\text{\tiny 3,\tiny 4}$, Robert C. Myers$^\text{\tiny 1}$, Pedro Vieira$^\text{\tiny 1,\tiny 3}$}
\blfootnote{\tt{faprile@ucm.es, jacobmabajian@gmail.com, pedrogvieira@gmail.com, rmyers@pitp.ca}}
\\ \vspace{1.2cm}
\footnotesize{\textit{
$^\text{\tiny 1}$Perimeter Institute for Theoretical Physics,
Waterloo, Ontario N2L 2Y5, Canada \\
$^\text{\tiny 2}$University of Waterloo,
Waterloo, Ontario N2L 3G1, Canada  \\
$^\text{\tiny 3}$ICTP South American Institute for Fundamental Research, 
IFT-UNESP, S\~ao Paulo, SP Brazil 01440-070\\ 
$^\text{\tiny 4}$Dept.~de F\'isica Te\'orica, Facultad de Ciencias F\'isicas,
Universidad Complutense, 28040 Madrid, Spain
}
\vspace{4mm}
}
\end{center}

\vspace{2mm}
\begin{abstract}
We initiate the study of holographic correlators for 
operators whose dimension scales with the central charge of the CFT.   
Differently from  light correlators or probes, the insertion 
of any such maximally heavy operator 
changes the AdS metric, so that the correlator itself is 
dual to a backreacted geometry with marked points at the Poincar\'e boundary.
We illustrate this new physics for two-point functions.
Whereas the bulk description of light or probe operators 
involves Witten diagrams or extremal surfaces in an AdS background, 
the maximally heavy two-point functions are described by nontrivial 
new geometries which we refer to as ``spacetime bananas".
As a universal example, we discuss the two-point function of maximally 
heavy scalar operators described by the Schwarzschild black hole in 
the bulk and we show that its onshell action reproduces the expected CFT result.
This computation is nonstandard, and adding boundary terms to the 
action on the stretched horizon is crucial. 
Then, we verify the conformal Ward Identity from the holographic stress tensor
and discuss important aspects of the Fefferman-Graham patch. 
Finally we  study a Heavy-Heavy-Light-Light correlator 
by using geodesics propagating in the banana background.
Our main motivation here is to set up the formalism to 
explore possible universal results for three- and 
higher-point functions of maximally heavy operators.

\end{abstract}

\newpage

\addtocontents{toc}{\protect\setcounter{tocdepth}{2}}

\setcounter{page}{1}
\renewcommand{\thefootnote}{\arabic{footnote}}
\setcounter{footnote}{0}

{
\tableofcontents
}

\newpage

%=====================================================

\section{Introduction}

%=====================================================

%\cite{Maldacena:1997re}Witten:1998qjAharony:1999ti

\textit{As the players on the court fiercely rally, exchanging shots with precision and grace, so too do correlation functions in AdS/CFT, a match of wits and calculation where the stakes are nothing less than unlocking the mysteries of quantum gravity,} says \texttt{chatgpt}.\footnote{The rest of this article was written by humans.}

%Correlation functions play an important role in understanding the emergence of gravity in the AdS/CFT correspondence.
%The nature of the holographic calculation of CFT correlation functions depends qualitatively on the dimensions of the operators involved.

The holographic dual of a CFT correlation functions depends 
qualitatively on the dimensions of the operators involved.
Holographic correlators might feature the insertion of
light operators dual to particles, \eg Kaluza-Klein modes 
on AdS, and/or heavier operators.    
In principle, all light correlators are computed  by Witten diagrams.
For heavier operators, the correct picture depends on some of the details of the operator. 
Strings, branes or bricks of branes can all be considered emerging 
from the insertion points, and each category might come with decorations thereof. 
Holographic correlators for strings and branes receive the leading order contribution  
from the action associated to an extended surface in the bulk anchored at the insertion points,  
as is well established. But when it comes to huge operators, such as large bricks of branes,
the bulk geometry itself is deformed. This is a scenario in which much less 
has been explored. Hence the question we would like to investigate here is:  
``How do we compute correlators of huge operators from the bulk?''

As a first  step, in this paper we discuss the  
simplest case of two-point functions, and the construction of 
the corresponding two-point function geometries.
In order to explain what these are we will use the 
AdS-Schwarzschild black hole in $D$ dimensions as a guiding 
example.\footnote{The generalisation to include electric charge and matter will be presented elsewhere \cite{USthree}.}
Of course, our main motivation here is to establish a general 
formalism, which can be applied to more general solutions such as 
those that are dual to higher point correlation functions \cite{USthree}.

CFT two-point functions are very simple, i.e.
\beq
\langle O_{\Delta_i}(\vec{x}_1)\, O_{\Delta_j}(\vec{x}_2)\rangle
\simeq \frac{\delta_{ij}}{|\vec{x}_{1}-\vec{x}_{2}|^{2\Delta_i}}\,.
\label{start}
\eeq
Our goal, however, is to recover this result from a bulk calculation 
that involves huge operators. %in a holographic CFT from a bulk calculation.
The information that we need to find is the dimension $\Delta_i$ of the scalar operators, and 
the spacetime dependence with respect to their separation $\vec{x}_{12}$.
Recall that the spectrum of dimensions of operators in a holographic CFT 
coincides (up to a shift by the Casimir energy) with the spectrum of energies 
with respect to global time $\tau$ in AdS. 
Therefore a way to read off $\Delta$ is to compute the energy $E$ of the 
dual operator in a frame where the geometry is asymptotically $\mathbb{R}\times S^{d-1}$. 
This is true for any operator: light, heavy or huge.
Thus in global coordinates, a two-point function geometry is an 
asymptotically AdS geometry with a backreacted interior which 
depends on the operator. Using the natural cutoff in these coordinates 
corresponds to placing the operators at $\tau=\pm\infty$, 
and one might be satisfied with this.

However, the global AdS perspective is not the end of the story. 
In this paper, we will present another computation that is intrinsically Euclidean  
and such that the operators are inserted at the boundary of AdS in Poincar\'e coordinates.
A similar idea was explored already in the context of classical spinning strings 
in global AdS in \cite{Janik:2010gc}. The strategy there was 
to embed the surface of the string into Euclidean AdS with Poincar\'e slices. 
In these coordinates, the string looks like a fattened geodesic that connects the 
points at the boundary, where the dual operators are inserted, and 
its onshell action was shown to compute the corresponding two-point function. 

\begin{figure}[t]
\begin{center}
{\includegraphics[scale=0.52]{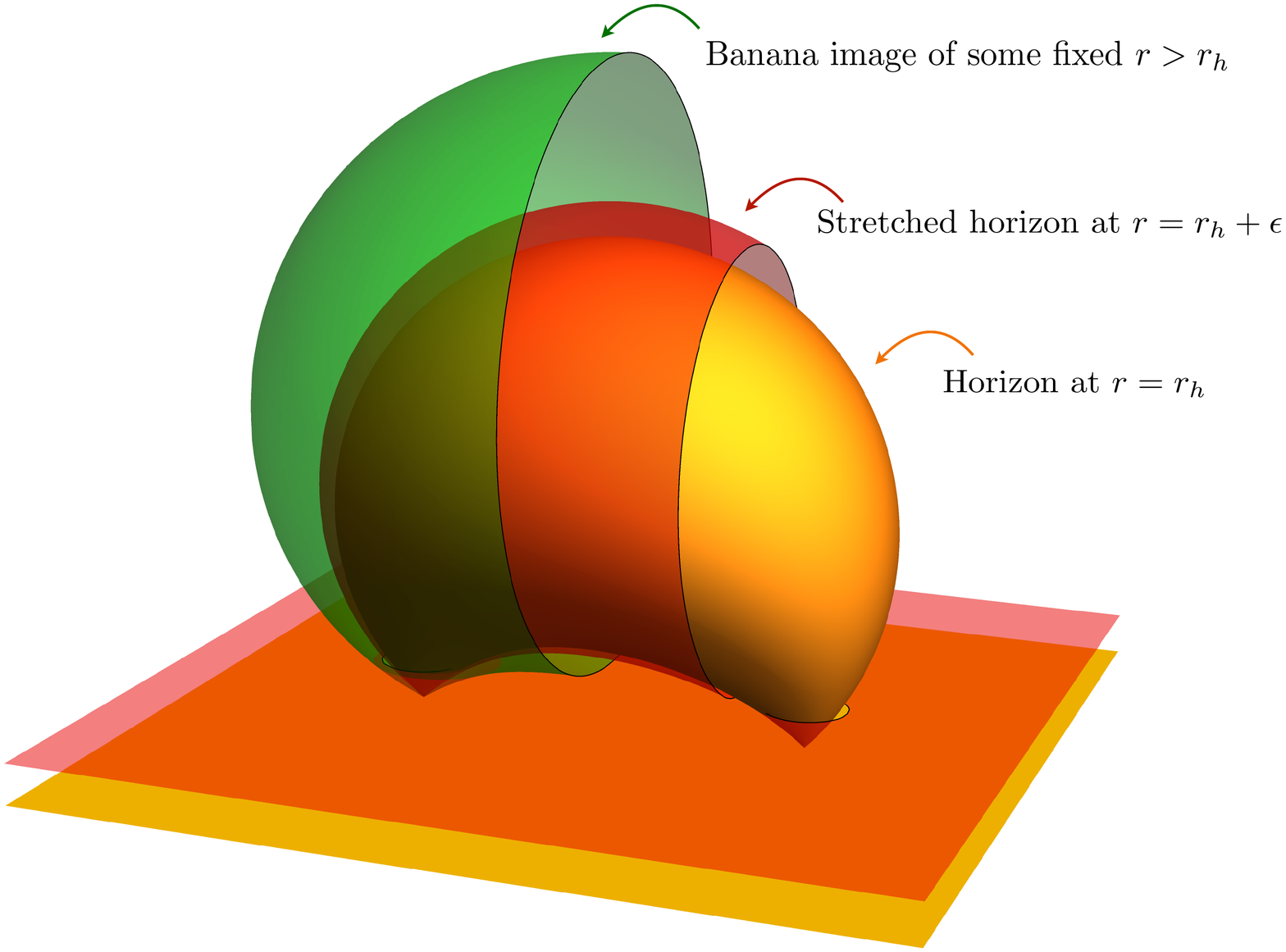}}
\end{center}
\vspace{-0.5cm}
\caption{The \textit{banana} foliation. (See also figure \ref{globalConeBanana}.)
\label{foliationBananas}}
\end{figure}

For two-point function geometries,
the relationship between global and Poincar\'e coordinates is a bit more subtle. 
The reason is that, unlike the case of embedded objects, the metric itself 
transforms in a nontrivial way. Nevertheless, we will construct 
a change of coordinates that maps a geometry with 
global asymptotics to a geometry with Poincar\'e asymptotics. 
We will call this the Global-to-Poincar\'e (GtP) map.
The spacetime we end up with has novel features which we will describe in detail.
To start with, we can picture how it looks by visualising the foliation 
defined by mapping surfaces of constant $r$ in global coordinates 
into the Poincar\'e picture. This foliation is composed of ``spacetime bananas'' 
that originate from the marked points, \eg see figure \ref{foliationBananas}.

The induced metric on the spacetime bananas is what characterises 
the backreaction of the operators inserted. For gravitational solutions 
which  have a smooth interior, we can follow the foliation 
to the point where it shrinks to the geodesic connecting the 
two boundary points. In a neighbourhood of that geodesic, the metric 
depends on the specifics of the operators inserted,  
and it will deviate strongly from empty AdS.

When the operator insertions have created a black hole connecting the two insertion points, 
we can follow the foliation in figure \ref{foliationBananas}  only up to 
the innermost banana which is the GtP image of the black hole horizon at $r=r_h$. 
%However, as we will see in the following, a
As is characteristic of a black hole horizon, the induced geometry 
on this innermost banana maintains a finite size in the transverse 
directions (i.e.~the horizon area) but has zero length in the 
longitudinal direction (i.e.~the analog of the $\tau$ direction). 
The latter results in a conical singularity on this surface.  
This zero-length direction goes between the insertion points, passing through the bulk.
Therefore, we have to be careful when picturing the horizon as 
a banana (as in figure \ref{foliationBananas}), since the proper 
length along this banana actually vanishes. For the purpose of 
visualization, it is useful to think instead of a ``stretched horizon", 
a banana that is some small distance outside of the horizon.
This perspective is also closely related to the membrane paradigm \cite{Thorne:1986iy}, %,Parikh:1997ma}. 
which constructs a simplified model to describe the black hole 
by replacing it with a physical surface (or membrane) at a vanishingly 
close distance from the event horizon.\footnote{While this approach 
is traditionally employed to model the dynamical behaviour of black 
holes in the context of Lorentzian signature, we may also employ 
the membrane paradigm in our Euclidean calculations here -- see footnote \ref{footy97}.}

At this point, it may beneficial for our reader if we step back 
to compare our approach to more traditional calculations with Euclidean 
black holes. First, a comment on nomenclature is that we continue 
to use ``horizon", an intrinsically Lorentzian concept, to refer to 
the innermost surface in our Euclidean geometry.  This is a codimension-two 
surface since the length along the (Euclidean) ``time" direction vanishes, as noted above. 
Traditionally, one makes the Euclidean time direction periodic and fixes 
the periodicity to maintain a smooth geometry across this surface. 
Of course this is the starting point in using Euclidean black hole 
geometries to study black hole thermodynamics, \eg \cite{Gibbons:1976ue}. 
In our approach, we are not enforcing any periodicity for the time coordinate. 
In this regard, our geometries are analogous to ``fixed area" states 
\cite{Akers:2018fow,Dong:2018seb}, which have recently appeared in 
discussions of quantum information aspects of holography.\footnote{We note, 
however, that generally one expects fixed-area states to have a finite 
conical deficit at the horizon, while in our geometries, the horizon 
develops an infinite angular excess. That is, in keeping with our 
discussion of the conformal dimensions and global coordinates, we 
allow the black holes to ``propagate" for an infinite amount to Euclidean time.}  
In either of these contexts, one is considering an ensemble of high 
energy states in the boundary CFT while we wish to consider a single 
state, \ie the state created by the insertion of our huge operator. 
The latter requires that our Euclidean action includes a boundary 
contribution, \ie the Gibbons-Hawking-York (GHY) term \cite{Gibbons:1976ue,York:1972sj}, 
at a stretched horizon to fix the boundary conditions there -- again, 
in contrast to the ensemble calculations.

The importance of the latter is also emphasized by the 
following holographic considerations: 
What we want for our geometry is that the onshell action computes 
a CFT two-point function, rather than say the Gibbs free energy 
at fixed temperature $\beta^{-1}$. Thus, our computation is more 
closely related to a Euclidean time translation amplitude  
rather than a trace over an ensemble of states with a thermal circle.
Schematically,
\beq\label{goalE}
\<\text{BH}|e^{-H \delta \tau} |\text{BH}\>
= e^{-E \delta \tau} 
\qquad \text{rather than}\qquad Z={\rm tr}( e^{-\beta \hat{H}})=e^{-\beta E+S}\,.
\eeq
In both cases, the right-hand side is computed by 
the onshell action of a gravitational background, and  roughly 
$\delta \tau \sim \log|x_{12}|^2$. 
But for a %black hole 
two-point function, 
there should be no entropy contribution!
It will turn out that the GHY boundary term on the stretched 
horizon is crucial to remove the entropy appearing in the calculation 
of the thermal ensemble. Let us also note that one can interpret 
this new boundary term as the ``membrane action"~\cite{Parikh:1997ma} 
within the framework of the membrane paradigm.

The remainder of our paper is organised as follows: In 
section \ref{geo_bh_sect}, we describe the salient aspects of our construction.  
Specifically, we introduce the GtP change of coordinates and the bananas, 
then we discuss a version of the onshell action computation, 
whose immediate aim is to highlight 
the emergence of nontrivial spacetime dependence,
and the crucial role played by the GHY boundary term at the stretched horizon.
In section \ref{FG_expansion}, we describe further aspects 
of the two-point function geometry. In particular, we verify 
the conformal Ward identity from the holographic stress tensor, 
and we revisit the onshell action computation by discussing the 
implications of the fact that the Fefferman-Graham coordinates 
only extend to a finite surface in the bulk, which we call ``the wall".
In section \ref{geodesic_sec}, we build on the idea of the horizon 
as a membrane by showing that -- absent fine tuning -- geodesics 
anchored at the boundary always remain outside the horizon banana. 
Then, we compute the action of such geodesics and show in examples 
that at leading order in the black hole mass,
the result is simply the stress tensor conformal block.  
Finally, we conclude with a discussion of our results and 
future directions in section \ref{discuss_sect}.
There, we sketch an outline of our plan to extend our 
calculations to three- and higher-point geometries, which 
we will investigate in the future \cite{USthree}.

%=======================================================================

\section{The Banana Geometry}\label{geo_bh_sect}

%=======================================================================

In 1916, Schwarzschild discovered the first black hole solution of Einstein gravity.  
Here we will give it a new outfit, and show how it looks when 
we think of it as a two-point function geometry. As described in the introduction,
from this new point of view, the black hole will look like a spacetime banana.

%======================================================================
\subsection{AdS-Schwarzschild}\label{ads_sch_sect}
%======================================================================

In global coordinates, the Euclidean metric of the $(d+1)$-dimensional AdS-Schwarzschild black hole reads 
\beq
ds^2_\texttt{global} = f(r) d\tau^2 + \frac{dr^2}{f(r)}  + r^2 d\Omega_{d-1}^2 \la{globalE}
\eeq
where the blackening factor is
\beq\label{blackening_fact}
%f(r)=1 + \frac{r^2}{L_{AdS}^2} - \frac{\alpha M L_{AdS}^{d-1}}{\,\,r^{d-2}}.
f(r)=1 + r^2 - \frac{\alpha M }{\,\,r^{d-2}}.
\eeq
Here $M$ is the black hole \textit{mass} (= \textit{energy}) 
%-- measured in units of the $AdS$ radius %-- or \textit{dimension} of the dual CFT operator, 
if the parameter $\alpha$ takes the canonical value, 
\beq\label{value_alpha}
\alpha = \frac{16\pi \GN}{(d-1) \Omega_{d-1} } \,, \qquad{\rm where}
\quad \Omega_{d-1}=\texttt{Vol(unit (d-1)-sphere)} = \frac{2\pi^{d/2}}{\Gamma\!\left({d}/2\right)}
\eeq
and $\GN$ is Newton's constant in the $(d+1)$-dimensional bulk.
In practice, we will set $\alpha=1$ to avoid cluttering our 
computations, and spell it out only in the final formulae.
Further, we have implicitly set the AdS curvature scale to $L_{AdS}=1$.

The signature of the black hole can be changed from Euclidean 
to Lorentzian by taking~$\tau=-i t$. In either signature, the vector 
along the time direction is Killing and has norm square proportional 
to~$f(r)$. This norm vanishes at the real value of $r=r_h$ where~$f(r_h)=0$, \ie the Killing vector becomes null in Lorentzian 
signature and vanishes in Euclidean signature. This value, $r=r_h$, 
sets the location of the horizon, and it depends on $M$. Very 
light black holes with $M\ll 1$ have a very small horizon radius $r_h$, i.e.~they are 
effectively small excitations in the middle of AdS. Very heavy black holes with 
$M\gg 1$ have a horizon radius scaling with $M^{1/d}$ and thus occupy most of the AdS spacetime. 

Of course, in Lorentzian signature, the horizon $r=r_h$ is the locus where light 
accumulates from the point of view of an observer at infinity. 
Free falling observers nevertheless cross the horizon in a finite proper time.
On the other hand, as discussed in the introduction, in Euclidean signature, the spacetime ends at the horizon $r=r_h$. 
There will be a conical singularity at the horizon unless Euclidean time 
is compactified to a circle, $\tau \sim \tau + \beta$, where the 
periodicity is given by the inverse Hawking temperature
\beq
T_\mt{H}\equiv\beta^{-1}=\frac{f'(r_h)}{4\pi}\,. \label{Hawking}
\eeq
Then, the $(\tau,r)$ geometry is that of a ``cigar'' ending at $r=r_h$ (see 
\cite{Cassani_lectures} for a nice review).
%
%\footnote{
%With $r=r_h+x^2$, the near-horizon metric reads $ds^2\simeq
%\frac{4}{f'(r_h)}(dx^2+ x^2 d\varphi^2)  +r_h^2 d\Omega^2_{d-1}$ where 
%$\varphi=\tfrac{1}{2}f'(r_h) \tau$. If $\varphi$ has $2\pi$ periodicity, 
%the metric describes a smooth cigar $\times$ a sphere. Translating back to 
%$\tau$ leads to eq.~(\ref{Hawking}). \rcm{Let's find a good ref and drop the footnote} }  
%
The horizon area is $A= r_h^{d-1}\Omega_{d-1}$, 
and it determines the black hole entropy $S={A}/{4 \GN}$ 
in the thermodynamic interpretation of the black hole \cite{Gibbons:1976ue}.

What we want for a Euclidean two-point function geometry 
is a black hole connecting the insertion points 
at the boundary of AdS in Poincar\'e coordinates. 
This picture would match with what we expect for very light 
black holes behaving as structureless point particles travelling through the AdS vacuum.
In that case, the two-point function would be computed by a 
geodesics anchored at the insertion points.

%--------------------------------

\begin{figure}[t]
\begin{center}
{\includegraphics[scale=0.47]{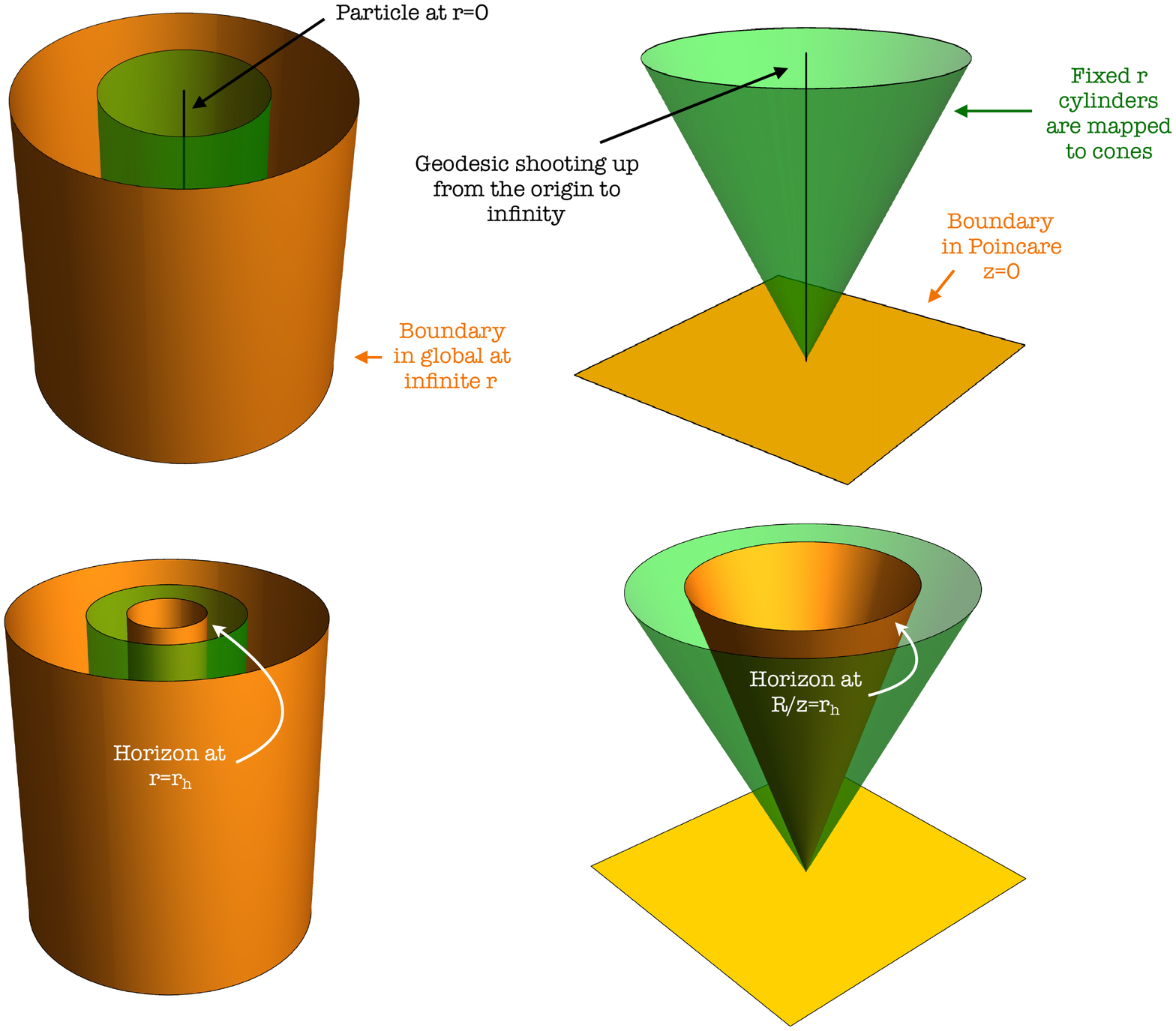}}
\end{center}
\caption{Global ({\bf Left}) and Poincar\'e ({\bf Right}) are related 
by the map in eq.~(\ref{poincareCoords}). The trajectory of a particle 
at rest in the middle of global AdS is mapped to a geodesic shooting 
from the origin to infinity in Poincar\'e AdS, and the $\mathbb{R}\times S^{d-1}$ 
boundary at $r=\infty$ is mapped to the Poincar\'e boundary at $\tz=0$. 
Cylinders of varying $r$ that interpolate between the center and the 
boundary of global AdS correspond to cones in Poincar\'e coordinates. 
For black holes, there is an horizon at $r=r_h$ which corresponds 
to a minimal cone at $R/\tz=r_h$.
\label{guessing}}
\end{figure}

%--------------------------------

To match with our expectation, we look for a change of coordinates 
such that the center of global AdS, namely $r=0$ and $-\infty \leq \tau\leq \infty$, 
is mapped precisely to that geodesic in Poincar\'e AdS anchored 
at the insertion points of the two-point function.
Further, we require that the $\mathbb{R}\times S^{d-1}$ boundary 
of global AdS is conformally mapped to the $\mathbb{R}^{d}$ boundary of Poincar\'e AdS.
With the insertion points at $0$ and $\infty$, we can use $SO(d-1)$ invariance
to restrict ourselves to $\tau=\tau(z,R)$ and $r=r(z,R)$ where $R$ is the radial coordinate in the  
Poincar\'e boundary. Then, the two conditions above are enough to suggest the 
change of variables\footnote{It 
is also possible to deduce this transformation   
by generalising the argument of \cite{Janik:2010gc}, 
in particular, by passing from $(\tau,r)$ coordinates to embedding 
coordinates, and from embedding coordinates to $(\tz,R)$ in the Poincar\'e patch. 
We give the details in appendix \ref{AdSunitary_app}.} 
\beq\label{poincareCoords}
\tau = \frac{1}{2} \log(\tz^2+R^2)\qquad;\qquad
r = \frac{R}{\tz}\,.
\eeq
Under this map, cylinders of constant radius in global coordinates are mapped to cones in Poincar\'e coordinates,
and translations in global time are mapped to dilations preserving
these cones -- see Figure \ref{guessing}. We call this map the Global to Poincar\'e (GtP) map.

After the GtP mapping \eqref{poincareCoords}, the black hole metric takes the form
\beq\label{cone_metric}
ds^2_{cone} = \frac{1}{\tz^2}\left[ 
\frac{d\tz^2}{h(\tfrac{R}{\tz})} + h(\tfrac{R}{\tz}) \left(dR + \tfrac{R}{\tz}\,v(\tfrac{R}{\tz}) d\tz\right)^2 + R^2 d\Omega_{d-1}^2 \right]\,,
\eeq
where
\beq
h(r)= \frac{1}{f(r) }+ \frac{r^2f(r) }{ (1+r^2)^2} \qquad{\rm and}\qquad
 v(r)= \frac{1}{f(r) h(r)}\left[\frac{f(r)^2}{ (1+r^2)^2 }-1\right]\,. 
\eeq
With $M=0$, we have $f=1+r^2$. As a result, one 
finds $h=1$ and $v=0$, and the above metric \eqref{cone_metric} 
reduces to exactly Poincar\'e AdS. 

%---------------------------------------------------------------

\begin{figure}[t]
\begin{center}
{\includegraphics[scale=0.57]{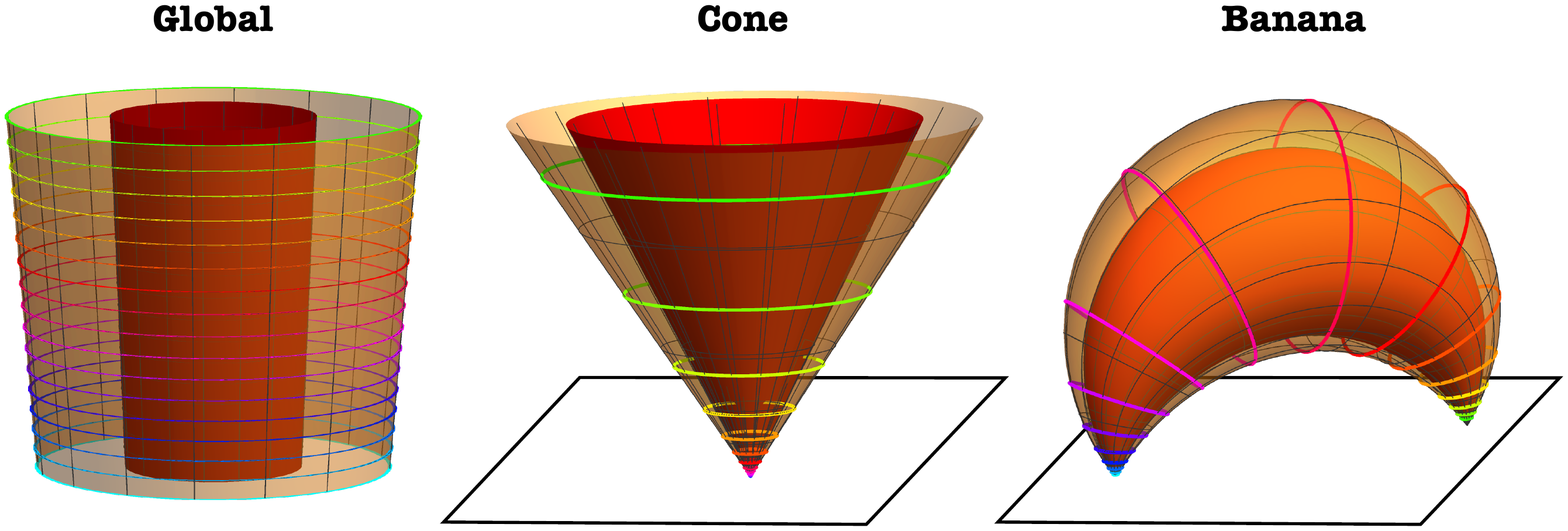}}
\end{center}
\vspace{-6.5cm}
\caption{Black hole in global AdS (left), as a cone (middle) and a 
banana (right). The red corresponds to the horizon at $r=r_h$, while 
the semi-transparent surface corresponds to some constant $r$ 
surface with $r>r_h$. There is a conical singularity at the 
horizon, since we did not make $\tau$ periodic. %remove "2" to get picture without the lines
\label{globalConeBanana}}
\end{figure}

%---------------------------------------------------------------

%{\color{blue} FA: I have a suggestion regarding $\tilde z$. Can we use $z$ all over? even above, 
%and then in the paragraph below say in words that: Finally, we can bring 
%the insertion point at $\infty$ to a finite distance by a change 
%of coordinates that acts  as a special conformal transformation (SCT) 
%on the boundary. For clarity, let us rename coordinates in \eqref{cone_metric} 
%as $z\rightarrow \tilde z$ and replace $R,\Omega_i$ on the boundary 
%with Cartesian coordinates $\tx^i$. The desired change of coordinates is
%\beq\label{SCT}
%\begin{array}{ll} 
%\displaystyle{x}^i= \frac{ {\tx}^{i} - b^{i}\, (  %\tx^2 +  z^2 ) }{  \widetilde\OOmega^2 }\\
%\displaystyle z =\frac{ \tz}{\, \widetilde\OOmega^2}
%\end{array}
%\qquad{\rm with}\quad  \widetilde\OOmega^2=1- 2\, b %\cdot {\tx}  + b^2(\tx^2 +  \tz^2 )\,.
%\eeq} \rcm{Personally I like the clarity of the current presentation. FA. OK}

Finally, we can bring the insertion point at $\infty$ to a finite distance by a change of 
coordinates that acts  as a special conformal 
transformation (SCT) on the boundary. Upon introducing Cartesian coordinates $\tx^i$ to replace the polar coordinates
$R,\Omega_i$ on the boundary, desired change of coordinates in the bulk is
\beq\label{SCT}
\begin{array}{ll} 
\tx^{i}&\rightarrow\displaystyle{x}^i= \frac{ {\tx}^{i} - b^{i}\, (  \tx^2 +  z^2 ) }{  \widetilde\OOmega^2 }\\
\tz&\rightarrow \displaystyle z =\frac{ \tz}{\, \widetilde\OOmega^2}
\end{array}
\qquad{\rm with}\quad  \widetilde\OOmega^2=1- 2\, b \cdot {\tx}  + b^2(\tx^2 +  \tz^2 )\,.
\eeq
We call the above transformation the SCT mapping. It is useful to 
recall that the inverse of the SCT mapping with shift parameter 
$b^i$ is simply another SCT map with shift $-b^{i}$.

We denote the points at which the operators are inserted as $\vec{x}_{\bullet1}=0$ and 
$\vec{x}_{\bullet2}=-\frac{b^i}{b^2}$. Without loss of generality, 
we can choose $\vec{x}_{\bullet1}-\vec{x}_{\bullet2}$ to lie along the 
${x}^1$-axis, \ie $b^i = b\, \delta^{i1}$. Then the geometry maintains 
a rotational $SO(d-2)$ symmetry in the remaining Cartesian directions. 
If we denote the radius in these transverse directions as 
$\rho=(\sum_{i=2}^d ({x}^i)^2)^{1/2}$, then the metric takes the form
\beq\label{3var_banana}
ds^2_{banana}= N_{z}^2\, dz^2 + \sum_{a,b={x}^1,\,\rho} 
h_{ab}(dy^a +N^a_{z}\, dz)(dy^b+N^b_{z}\, dz) + \frac{\rho^2\, d\Omega_{d-2}^2}{z^2}
\eeq
The explicit form of the components is not very illuminating, and hence we omit them here. 
Of course, the symmetry is enhanced to $SO(d-1)$ with $b\to0$, 
since the cone metric \eqref{cone_metric} is recovered in this limit. %

\begin{figure}[t]
\begin{center}
{\includegraphics[scale=0.5]{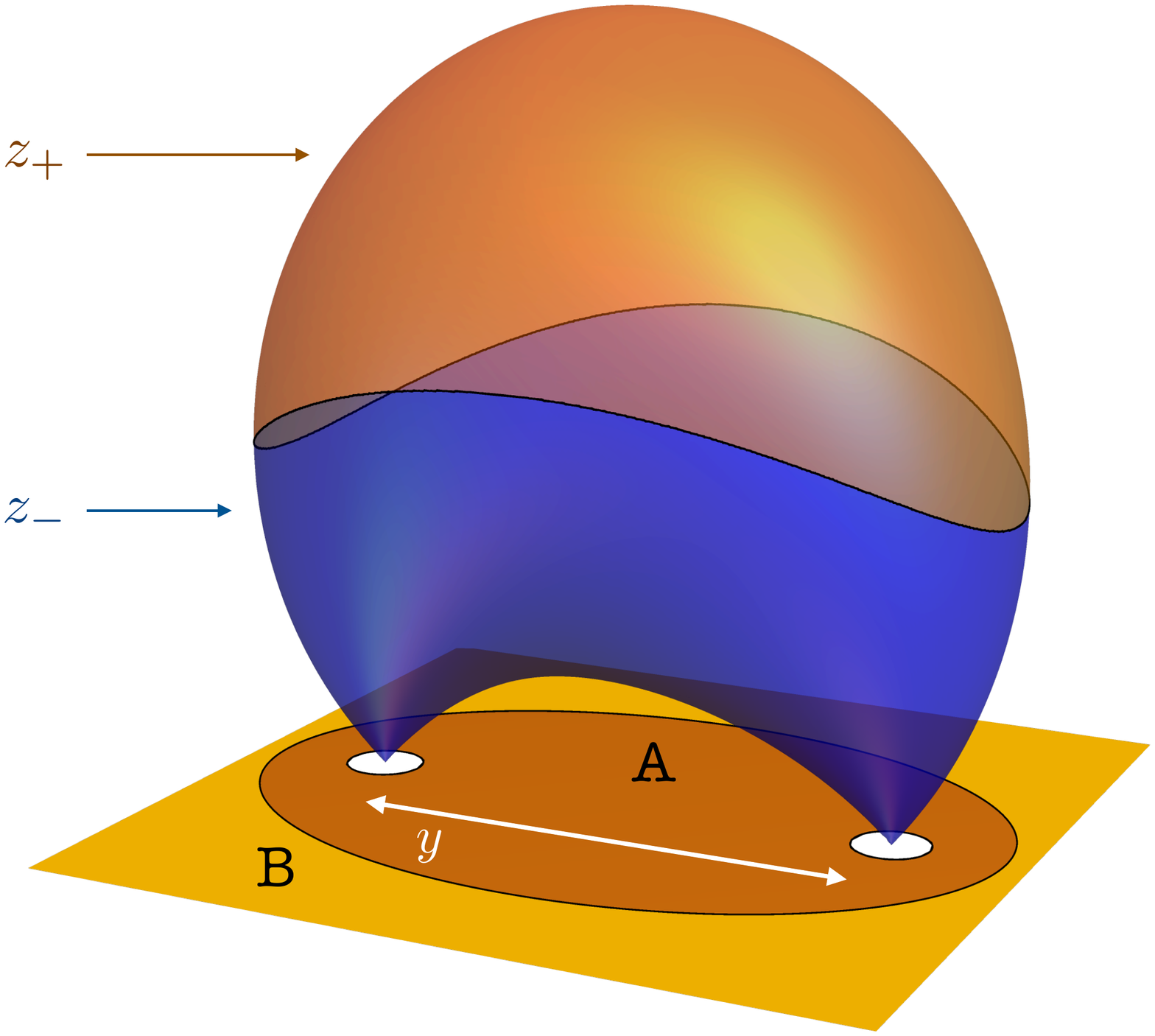}}
\end{center}
\vspace{-0.5cm}
\caption{
The two solutions $z_\pm$ of the banana equation (\ref{banana_eq}). 
We denote the (boundary) distance between the insertion points 
as $|\vec{x}_{\bullet1}-\vec{x}_{\bullet2}|=y=\frac{1}{b}$. 
The $z_-$ component has the shape of a pair of pants, while $z_+$ is a cap. For $r=r_h$, 
we denote the inside and outside of the ellipse below the stretched horizon as $A$ and $B$, respectively.
\label{BananaGeometry}}
\end{figure}

In figure \ref{globalConeBanana}, we illustrate the various coordinate transformations. 
Upon performing the SCT map, the foliation by cones (with $b=0$) becomes 
a foliation by bananas (with $b\ne0$).  More precisely, each 
constant $r$ cylinder in global coordinates is mapped to the banana
\beq\label{banana_eq}
%
%r^2= \frac{ \tilde \Theta^2 (\tilde R^2+\tilde z^2) -\tilde z^2  }{\tilde z^2}
%\qquad;\qquad \tilde \OOmega^2=1+ 2\, b. \tilde{x}  + b^2( \tilde R^2 + \tilde z^2 )
%
r^2= \Theta^2 \left( \frac{x^2}{z^2}+ 1\right) - 1 \qquad{\rm with}\quad  \OOmega^2=1+ 2\, b \cdot {x}  + b^2(  x^2 +  z^2 ) 
\eeq
Solving this equation for $z$ gives two positive components $z_{\pm}(r)\ge 0$, 
distinguished by the sign of a square root -- see figure \ref{BananaGeometry} 
and also eq.~\eqref{zmp_def}. The $z_-$ component has the shape of a pair of pants, while
the $z_+$ component is a cap. In the limit $b\to0$, the former 
transitions to the cone surfaces and $z_+$ goes off to infinity.

Of course, the induced metric on each banana is the same as 
that of a surface of constant $r$ in global coordinates,
even if the metric looks initially more complicated. For example, 
restricting to a cone $\tz=\frac{R}{r}$ with a fixed value 
$r\in\mathbb{R}$ in eq.~\eqref{cone_metric}, a direct computation yields
\beq\label{example_ind_cone}
 ds^2_{\tt cone}\big|_{\tz=\frac{R}{r}}= f(r)\,\frac{dR^2}{R^2} + r^2 d\Omega_{d-1}^2\,. 
 \eeq
Further, eq.~\eqref{poincareCoords} yields $\frac{dR}{R}=d\tau$ 
with fixed $r$, and hence eq.~\eqref{example_ind_cone} reduces 
to precisely the same induced metric as in global coordinates \eqref{globalE} with fixed $r$.

Although the induced metric on the bananas is the same as in global coordinates, 
the main purpose of the GtP map is to give us  
a Poincar\'e boundary,  which allows us to study the 
backreaction of the operators along slices of constant $z$, \ie the 
bulk evolution picture. The %is crucial since the 
surface $z=\epsilon$ is also a simple cut-off surface\footnote{This is a slightly 
unconventional choice here since the metric \eqref{3var_banana} 
is not in the standard Fefferman-Graham gauge -- see eq.~\eqref{choco2} below.} 
for computing the onshell action, which is ultimately %this is 
what we want to compute to match with a CFT two-point function.
The main novelty here is that the $z=\epsilon$ surface contains 
points which are both far and close to the operator insertions, 
\ie far and close to the black hole horizon.  
Thus the pre-image of $z=\epsilon$ in global coordinates 
is a surface that  explores the space from near the asymptotic boundary 
to deep into the bulk, and therefore the corresponding boundary 
conditions probes %disparate 
properties of the geometry that are usually 
viewed as distinct, \ie near infinity and in the interior. 
While this distinction will be formalised in section \ref{FG_expansion} 
where we discuss the Fefferman-Graham patch, 
we first need to understand how to deal with the horizon. 

As mentioned in the introduction, traditionally Euclidean 
black holes appear as the bulk saddle point describing a
n ensemble of high energy states in the boundary CFT. In contrast, 
we wish to consider a single unique state, \ie the state created 
by the insertion of our huge operator. For this purpose, we 
introduce a stretched horizon at $r=r_h(1+\epsilon')$.
away from the horizon. Then to fix the boundary conditions at 
the horizon, we introduce Gibbons-Hawking-York (GHY) 
term \cite{Gibbons:1976ue,York:1972sj} on the stretched 
horizon and then take the limit $\epsilon'\to 0$.\footnote{As noted 
above, we might alternatively consider our calculation 
within the framework of the membrane paradigm \cite{Thorne:1986iy}, %,Parikh:1997ma}. 
where the stretched horizon corresponds to a physical membrane. 
In order to describe this membrane, one needs introduce an action 
as described in \cite{Parikh:1997ma}. The interested reader is 
referred there for its derivation but here, we simply note 
that onshell, this membrane action reduces to a GHY term. \label{footy97}}

In sum, the total onshell action for our two-point function black hole is
\beq\label{total_action_FA}
I_{ \begin{tikzpicture} \draw (0,0) rectangle (.5,.3); \draw[fill=black] (.14,.15) circle (1pt); \draw[fill=black] (.35,.15) circle (1pt); \end{tikzpicture} }=I_{bulk}\ +\ I_{boundary}\ +\ I_{ct}
\eeq
where the first term is the Einstein-Hilbert bulk action (with a 
negative cosmological constant). The boundary action is comprised 
of two GHY terms, one on the asymptotic boundary and the other on the stretched horizon, \ie
\beq
 I_{boundary}= I_{GHY}(\partial_{AdS})+ I_{GHY}({\rm stretch})\,.
\eeq
Finally, there are the boundary counterterms $I_{ct}$ which are evaluated on the asymptotic boundary, \eg see \cite{Emparan:1999pm,Skenderis:1999nb,Skenderis:2002wp,Gibbons:2004ai,Liu:2004it,Batrachenko:2004fd}.
In the next section, we discuss the precise definition of all of these terms $I_{bulk},\ I_{boundary}$ and $I_{ct}$, 
and further we will compute their values.

%=======================================================================
\subsection{CFT two-point function from gravity: the onshell action}\label{oneshell_sec}
%=======================================================================

%----------------------------------

The goal of this section is to evaluate the gravitational action 
$I_{ \begin{tikzpicture} \draw (0,0) rectangle (.5,.3); 
\draw[fill=black] (.14,.15) circle (1pt); \draw[fill=black] (.35,.15) circle (1pt); \end{tikzpicture}}$ 
for the banana geometry and show that it reproduces a scalar two-point CFT correlator. 
Of course, given eq.~\eqref{start}, we know what to expect: 
\beq\la{goalS}
I_{\begin{tikzpicture} \draw (0,0) rectangle (.5,.3); 
\draw[fill=black] (.14,.15) circle (1pt); \draw[fill=black] (.35,.15) circle (1pt); 
\end{tikzpicture}}=
\Delta\, \log\! |\vec{x}_{\bullet1}-\vec{x}_{\bullet2}|^2  
\,+\, \texttt{distance-independent\,constant}
\eeq
so that 
\beq\label{rewriting_goals}
\langle O(\vec{x}_{\bullet1})O(\vec{x}_{\bullet2})\rangle= 
e^{-I_{\begin{tikzpicture} \draw (0,0) rectangle (.5,.3); 
\draw[fill=black] (.14,.15) circle (1pt); 
\draw[fill=black] (.35,.15) circle (1pt); 
\end{tikzpicture}}} \simeq \frac{1}{|\vec{x}_{\bullet1}-\vec{x}_{\bullet2}|^{2\Delta}} \,.
\eeq 
We will focus first on reproducing the spacetime dependence as function of $\Delta$.
We will also comment on the overall normalisation, which was omitted in eq.~\eqref{rewriting_goals}.

As pointed out already in \eqref{total_action_FA}, there are three 
contributions to the total action, which come from: 
\textit{bulk}, \textit{boundaries} and \textit{counterterms},
\beq
I_{ \begin{tikzpicture} \draw (0,0) rectangle (.5,.3); 
\draw[fill=black] (.14,.15) circle (1pt); 
\draw[fill=black] (.35,.15) circle (1pt); 
\end{tikzpicture} }= 
 I_{ bulk }\ + I_{GHY}(\partial_{AdS})+ I_{GHY}({\rm stretch}) +\ I_{ct}
 \eeq
We will analyse each of them in turn.

\begin{figure}[t]
\begin{center}
{\includegraphics[scale=0.5]{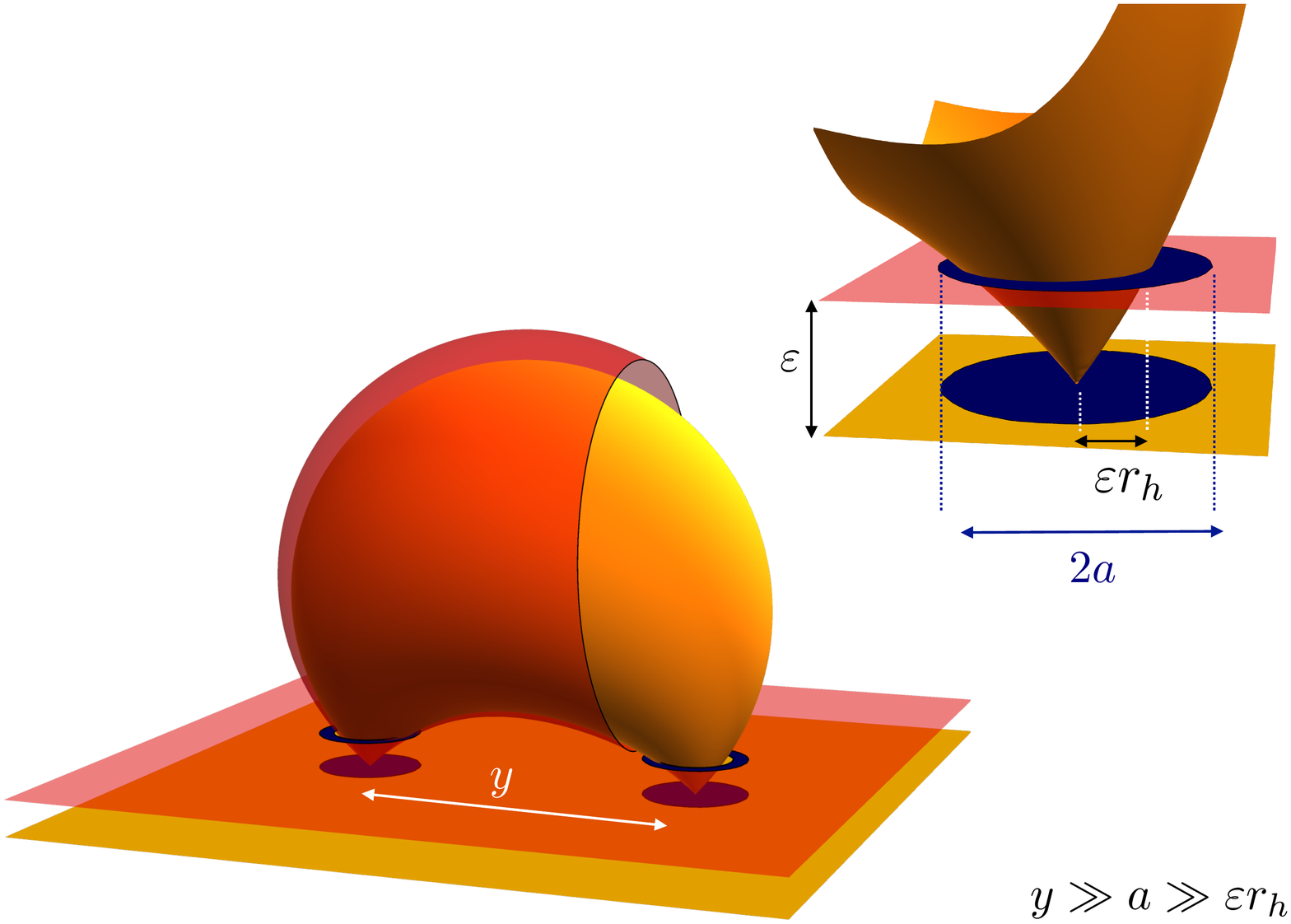}}
\end{center}
\vspace{-0.5cm}
\caption{To regulate the onshell action, we introduce 
a cut-off surface at $z=\epsilon$ near the asymptotic boundary 
and a stretched horizon at $r=r_h(1+\epsilon')$. These two surfaces 
intersect and effectively excise a disk of radius $r_h\epsilon$ to 
leading order in $\epsilon$, close to the insertion points. To compute 
the boundary integrals, we will excise a larger disk of radius $a$, 
much smaller than the separation between the insertions $y=\frac{1}{b}$, 
but much larger than $\epsilon r_h$. We will show that this scheme 
reproduces the expected spacetime dependence $y^{-2\Delta}$ of a 
two-point function, up to a distance-independent constant. . 
\label{BananaRegulators}}
\end{figure}

%==========================================================================
\subsubsection{The bulk action} 
%==========================================================================

The Einstein-Hilbert action is
\beq
I_{bulk}=-\frac{1}{16 \pi \GN}
\int\!d^{d}x\,dz\, \sqrt{g}\left(R+\frac{d(d-1)}{L_{AdS}^2} \right)\,.\label{action99}
\eeq 
Here and below, we are implicitly working with the metric 
\eqref{3var_banana} using the coordinates introduced by the 
SCT mapping~\eqref{SCT}. Then upon using the Einstein equations\footnote{We 
use $R_{\mu \nu}=-d\, g_{\mu \nu}$ after setting $L_{AdS}=1$. Further, 
with the coordinates introduced by the SCT map~\eqref{SCT}, we have 
$\sqrt{g} \, dz \, d^{d}x = \frac{1}{z^{d+1}} \, dz \, d^{d}x$, just as 
in empty AdS. \label{barn3}} and carrying out the $dz$ integration 
for $z\ge\epsilon$, we find
\beq
I_{bulk}=\frac{1}{8\pi \GN}
\int_A\!d^{d}x \left[\frac{1}{z_+(r_h)^d}-\frac{1}{z_-(r_h)^d}\right] + 
\frac{1}{8\pi \GN}\int_{A\cup B}\!\!d^{d}x\ \frac{ 1 }{ \epsilon^d } %\right).
\label{bulkFirst}
\eeq 
where the  domain of integration $A$ is the ellipse given 
by projecting the banana describing the horizon into the 
boundary coordinates -- see figure \ref{BananaGeometry} and 
eq.~\eqref{ellipse}. $B$ is the domain outside this ellipse 
and hence, the second term, which is the usual divergence 
of the $AdS$ volume, is integrated over the entire AdS boundary.

The first term consists of the boundary contributions coming 
from where the $z$ integration reaches the banana at the 
(stretched) horizon,\footnote{Here and for some quantities below, 
the distinction between horizon and stretched horizon does not matter. 
That is, we do not encounter any divergences if we evaluate these 
quantities on the stretched horizon and take then the limit 
$\epsilon'\to0$. Hence we can evaluate them directly at $r=r_h$.} 
and $z_\pm(r_h)$ are the two components of this banana described 
below eq.~\eqref{banana_eq}. Substituting $r=r_h$ into eq.~\eqref{banana_eq}, we find
\beq\label{zmp_def}
z_{\pm}(r_h)=\Bigg[ \tfrac{r_h^2}{2b^2} - x^1 \(x^1+\tfrac{1}{b}\)-
\rho^2  \pm \frac{r_h}{b}\, \sqrt{ \tfrac{r_h^2}{4b^2}-  
x^1\(x^1+\tfrac{1}{b}\)-\rho^2\(1+\tfrac{1}{r_h^2}\) }\Bigg]^{\frac{1}{2}}\,,
\eeq
as depicted in figure \ref{BananaGeometry}. 
The ellipse dividing the domains $A$ and $B$ corresponds 
to the vanishing locus of the square root in the above expression. 
Hence A is given by the inequality,\footnote{
Recall that $\rho^2=\sum_{i=2}^d(x^i)^2$.}  
\beq\label{ellipse}
x^1\(x^1+\tfrac{1}{b}\)+\rho^2\left(1+\tfrac{1}{r_h^2} \right)\le \tfrac{r_h^2}{4b^2}
\eeq
where as discussed below eq.~\eqref{SCT}, 
we have placed the two insertion points on the $x^1$-axis at~$x^1=0$ and $-1/b$.

The bulk construction with the stretched 
horizon implies that we should also excise a small 
disk around each insertion point from the inner domain 
$A$ for the term involving $z_{-}(r_h)$ -- see figure \ref{BananaRegulators}.
As we shall see, this feature of the computation is precisely the reason why the first 
contribution in eq.~\eqref{bulkFirst} is nonstandard and interesting to evaluate.

Let us examine how the $z_-$ component reaches the 
insertion points, by zooming there with a small $\delta$ expansion,
$(x^1,\rho)\rightarrow \delta \,(x^1,\rho)$ 
and $(x^1,\rho)\rightarrow (-\frac{1}{b},0)+\delta\,(x^1,\rho)$. We find
\beq\label{CFT_factor_2pt}
\frac{1}{z_{-}(r_h)^{d}}=r_h^d \,
\underbrace{ \frac{ 1 }{ ((x^1)^2+\rho^2)^{d/2}\,( (1+b x^1)^2+ b^2 \rho^2 )^{{d}/{2} }}}_{f_{2pt}}\,+\,
{\cal D }(x_1,x_2,r_h;b)
\eeq
where the difference function $\cal{D}$ vanishes when $b=0$. 
The first term in its most general form is given by
\beq
f_{2pt}=\frac{|\vec{x}_{\bullet1}-\vec{x}_{\bullet 2}|^d}{|\vec{x}-\vec{x}_{\bullet 1}|^d\,|\vec{x}-\vec{x}_{\bullet 2}|^d}.
\eeq
\ie a simple function of distances. 
The integral of $f_{2pt}$ in $A$ diverges logarithmically,\footnote{For 
example, in the cone coordinates (with $b=0$), we have 
$\int d^d x\,f_{2pt}=\int\frac{dR}{R}\,d\Omega_{d-1}$ and the radial 
integral over $R$ would yield a logarithmic divergence.} while instead 
the reminder ${\cal D}$ is integrable in $A$. On the other hand, 
$f_{2pt}$ is integrable at infinity, and thus in $B$.
Therefore, by adding and subtracting $f_{2pt}$ in $B$, we can write
the bulk action as
\begin{align}
I_{bulk}=&\frac{1}{8\pi \GN}
\int_{A\cup B}\!\!d^{d}x\ \bigg[ \frac{ 1 }{ \epsilon^d } 
- r_h^d\, f_{2pt}(x)\bigg] +{\cal N}_{bulk}
\label{bulkSecond}
\end{align}
where we defined the constant
\beq
{\cal N}_{bulk}= \frac{1}{8\pi \GN}\int_B d^dx 
 \ r_h^d\, f_{2pt}+\frac{1}{8\pi \GN}\int_A d^dx \bigg[\frac{1}{z_+(r_h)^d}-{\cal D}\bigg]\,.
 \label{barn2}
\eeq

We emphasize that ${\cal N}_{bulk}$ is a constant independent 
of the seperation $|\vec{x}_{\bullet 1}-\vec{x}_{\bullet 2}|=1/b$. 
Indeed, if we rescale the coordinates, say $(x^1,\rho)\rightarrow (x^1/b,\rho/b)$, 
the dependence on the separation factors out completely from $z_{\pm}$ 
in eq.~\eqref{zmp_def} and from $f_{2pt}$ and ${\cal D}$ in eq.~\eqref{CFT_factor_2pt}. 
Further, this overall factor cancels precisely that coming from the 
measure $d^dx$ in the two integrals comprising ${\cal N}_{bulk}$ 
in eq.~\eqref{barn2}. Similarly, the $b$ dependence also scales 
out of the ellipse separating $A$ and $B$, \eg see eq.~\eqref{ellipse}.
Finally, since there is no divergence in ${\cal D}$ near the insertion 
points, we can close the disks opened in $A$. Hence we conclude 
that ${\cal N}_{bulk}$ is completely independent of the separation 
between insertion points, and in fact eq.~\eqref{barn2} yields a 
finite constant depending only on $r_h$ in the $\epsilon\rightarrow 0$ limit.

The contribution coming from the first term in eq.~\eqref{bulkSecond} 
is different because $f_{2pt}$ yields a logarithmic divergence, 
and so it is really crucial to excise a disk of radius $a$ around 
each of the insertion points. Thus, although the dependence 
on $|\vec{x}_{\bullet 1}-\vec{x}_{\bullet 2}|$ can be scaled 
away between the integrand and the measure, as above, 
this rescaling changes the domain of integration. 
Hence the integral \emph{does} depend on the distance.

%================================================================================
\subsubsection{The asymptotic boundary action} 
%================================================================================

%
The action has two contributions on the asymptotic AdS boundary. 
The first is the GHY term\footnote{\label{foot_useful}Recall that for 
any vector $V^{\mu}$, the divergence $\nabla_{\mu} V^{\mu}=
\frac{1}{\sqrt{g}}\,\partial_{\mu}(\sqrt{g}\,V^{\mu})$. Further, 
$\sqrt{g}$ coincides with that in empty AdS space, as noted in footnote \ref{barn3}.}
\beq\label{GHgeneral}
I_{GHY}(\partial_{AdS}) = %\frac{1}{8 \pi G} 
\frac{1}{8\pi \GN}\int_{\d} d^dx \sqrt{h}\, {K}_{\partial_{AdS}} 
\qquad{\rm with}\quad {K}_{\partial_{AdS}}= -\nabla_{\mu}\frac{N_z^{\mu}}{|N_z|}
\eeq
and the second, the counterterm action\footnote{The ``$\cdots$" contain 
terms whose number and form depend on the number of bulk dimensions. 
Note that we are following the approach of \cite{Emparan:1999pm} where 
the counterterms are expressed in terms of the induced metric on the asymptotic boundary. 
}
\beq\label{counter_term_action}
I_{ct}= \frac{1}{8\pi \GN} 
\int_{\d}\!\!d^d{x} \,\sqrt{h} \, \Bigg((d-1) + \frac{1}{2(d-2)}\,{\cal R}[h] +\cdots \Bigg)\,.
\eeq
Here we are using the boundary metric $h_{ij}$ at $z=\epsilon$ 
and the outward-pointing unit normal vector $N_z^{\mu}=(-1,N_z^i)$ 
in the banana metric. The quantities $N_z^i$ here and $N_z$ in eq.~\eqref{GHgeneral}
correspond to those appearing in the metric \eqref{3var_banana}.

As already mentioned, we find points that are both far 
and close to the horizon on the $z=\epsilon$ cut-off. For 
example, in the cone coordinates \eqref{cone_metric}, on the 
cut-off surface $z=\epsilon$, we can approach $R=r_h\epsilon$ 
where the blackening factor vanishes, or we can consider large 
$R$ where the metric is Poincar\'e AdS at leading order in $\epsilon$. 
The behavior of the metric $h_{ij}$ as $\epsilon\rightarrow 0$, 
as well as that of $N_z^i$, depends on this distinction.
To properly disentangle these two regions, we  
introduce Fefferman-Graham coordinates in section \ref{FG_expansion}, 
and here we adopt a simpler strategy instead. Our approach is to 
choose the radius $a$ of the disks that we excise from $A$ so 
that the expansion of $h_{ij}$ and $N^{i}_z$, in $z=\epsilon$ 
with $\vec{x}$ fixed, is valid.\footnote{Our 
strategy will work here because we are studying a solution of the vacuum Einstein equations, 
and the powers of the $z$ expansion have the same gap as the FG expansion.}
This is illustrated in figure \ref{BananaRegulators}.

Given the approximation $|\vec{x}_{\bullet 1}-\vec{x}_{\bullet 2}|\gg a\gg \epsilon r_h$, 
the various quantities entering $I_{GHY}$ and $I_{ct}$ automatically 
have an expansion in $z\big/|\vec{x}-\vec{x}_{\bullet 1}||\vec{x}-\vec{x}_{\bullet 2}|$  
because this is the expansion of the SCT map, when we go from 
the expansion in $z/R$ of the cone to the banana coordinates.
For example, we find
\beq
d^dx \sqrt{h}=\frac{1}{z^d}\bigg[1-\frac{\alpha M}{2}\frac{z^d}{((x^1)^2+\rho^2)^{\frac{d}2}((1+b x^1)^2+ b^2 \rho^2)^{\frac{d}{2} } }+\ldots\bigg]dx^1\rho^{d-2}d\rho\,d\Omega_{d-2}
\eeq
Then, at leading order $K_{\partial_{AdS}}=-d+ O(z^{d+1})$ 
and  the counterterms involving the boundary curvature are 
suppressed because of the flat metric on the asymptotic boundary.
Combining the asymptotic boundary contributions with the bulk 
action \eqref{bulkSecond}, we find\footnote{At an intermediate step, 
we have
$$I_{bulk}+I_{GHY}(\partial_{AdS}) + I_{ct}\simeq \frac{1}{8\pi\GN}\int\!\!\!d^d{x} \bigg[   
\underbrace{ \frac{1}{\epsilon^d}  - r_h^d\, f_{2pt}(x) }_{I_{bulk}} + 
\underbrace{ \left( - \frac{d }{\epsilon^d} +\frac{dM}{2}f_{2pt}\right) }_{I_{GHY}} 
+ \underbrace{  \left(\frac{(d-1) }{\epsilon^d}-  \frac{(d-1)M}{2} f_{2pt}\right) }_{I_{ct}} \bigg]$$}
\beq\label{intermezzo_act_onshell}
I_{bulk}+I_{GHY}(\partial_{AdS}) + I_{ct} \simeq  
\frac{(\alpha M-2 r_h^d)}{16\pi \GN} \int_{\begin{tikzpicture} 
\draw (0,0) rectangle (.5,.3); 
\draw[fill=black] (.14,.15) circle (1pt); 
\draw[fill=black] (.35,.15) circle (1pt); \end{tikzpicture} }\!\!\!d^d{x}\, f_{2pt} \,,
\eeq
up to the constant term ${\cal N}_{bulk}$.
The domain of integration is the entire boundary minus the 
small disks around the insertion points, which we denote $\begin{tikzpicture} 
\draw (0,0) rectangle (.5,.3); 
\draw[fill=black] (.14,.15) circle (1pt); 
\draw[fill=black] (.35,.15) circle (1pt); 
\end{tikzpicture}=A\cup B-{\rm disks}$.

The integral of $f_{2pt}$ can be performed by introducing 
bipolar coordinates.\footnote{Use
$$x^1=\frac{1+T\cos\sigma}{b(1+T^2+2T\cos\sigma)}\qquad{\rm and}\qquad 
\rho=\frac{T\sin\sigma}{b(1+T^2+2T\cos\sigma)}$$ 
where $\sigma\in[0,\pi]$ for $d>2$. For $T\rightarrow 0$, 
we encircle $\vec{x}_{\bullet 1}=0$, while for $T\rightarrow\infty$, 
we encircle $\vec{x}_{\bullet 2}=-\frac{1}{b}\,\hat{x}^1$.}
Let us simply quote the final result,
\beq
\int_{\begin{tikzpicture} \draw (0,0) rectangle (.5,.3); 
\draw[fill=black] (.14,.15) circle (1pt); 
\draw[fill=black] (.35,.15) circle (1pt); \end{tikzpicture} }\!\!\!d^d{x}\, f_{2pt} 
= \Omega_{d-1}\log\frac{|\vec{x}_{\bullet 1}-\vec{x}_{\bullet 2}|^2}{a^2}
\label{barn10}
\eeq

The overall coefficient in eq.~\eqref{intermezzo_act_onshell} then becomes  
\beq
\frac{\Omega_{d-1}}{16\pi \GN }\,\(\alpha M-2 r_h^d\)= M - S\,T_H= F_\mt{Gibbs} 
\eeq
where  
\beq
E=\frac{\Omega_{d-1}(d-1)\alpha M}{16\pi \GN}=M\,,
\qquad\quad S=\frac{\Omega_{d-1}r_h^{d-1}}{4\GN}\,,\qquad\quad T_H=\frac{f'(r_h)}{4\pi}
\eeq
with the blackening factor $f(r)$ and the coefficient $\alpha$ 
given in eqs.~\eqref{blackening_fact} and \eqref{value_alpha}, 
respectively.\footnote{We will also find $E=M$ by evaluating  
the holographic stress tensor in section \ref{holo_stress_t_sec}.} Therefore, 
\beq
I_{bulk}+I_{GHY}(\partial_{AdS}) + I_{ct} = (M-S\, T_H) 
\,\log\frac{|\vec{x}_{\bullet 1}-\vec{x}_{\bullet 2}|^2}{a^2} +{\cal N}_{bulk}\,.
\label{barn4}
\eeq

We have nearly reproduced the desired CFT spacetime dependence, 
since we have a logarithm coming from $f_{2pt}$.
However, the prefactor for this logarithm is the Gibbs free energy, 
similarly to what we would have expected had we done the 
computation in global AdS -- but for a possible Casimir energy.  
However, as discussed in eq.~\eqref{goalS}, we expect the prefactor 
to the CFT dimension $\Delta=M$. That is, we would like to subtract 
the entropy contribution in eq.~\eqref{barn4}.
In the following subsection, we will see that the GHY contribution 
on the stretched horizon does precisely this. This confirms that 
this surface term is precisely what is needed to fix the calculation 
to that of a single operator inserted at the asymptotic AdS boundary.

%==========================================================================
\subsubsection{The stretched horizon action}  %{\bf The membrane action onshell} 
%==========================================================================

Recall that we introduced a final boundary at the stretched 
horizon $r=r_h(1+\epsilon')$ and supplemented the action with 
a GHY term there, namely
\beq\label{GH_hor}
I_{GHY}({\rm stretch}) = \frac{1}{8 \pi \GN} 
\int_{\d} d^dx \lim_{\epsilon'\rightarrow 0} 
\sqrt{\sigma}\, K_\mt{stretch}\qquad{\rm with}\quad K_\mt{stretch}=-\nabla_{\mu}\frac{V^{\mu}}{|V|\,}
\eeq
where  the extrinsic curvature $K$ is now determined by 
the divergence of a vector $V^\mu$ orthogonal to the stretched horizon,
and $\sigma$ is the determinant of the induced metric on 
the stretched horizon.

We can find $V^\mu$ by considering the image of $\partial_r$
through the GtP and SCT transformations. 
This gives 
\beq
\frac{V_{\mu}}{|V|\,}= %- \sqrt{f(r)}\, \frac{\partial_{\mu} r(\vec{x},z)}{(\frac{x^2}{z^2}+1)\frac{\Theta^2}{z^2}}
- \frac{1}{\sqrt{f(r)}}\, \partial_{\mu} r(\vec{x},z)
\eeq
where $r(\vec{x},z)$ in eq.~\eqref{banana_eq}. 
Simplifying the divergence as in footnote \ref{foot_useful}, 
and further using~$f(r_h)=0$ with $f'(r_h)=4\pi T_H$,
we arrive at
\beq\label{Kstrethor}
K_\mt{stretch}=\frac{2\pi T_H}{\sqrt{f(r)}}\,.
\eeq
%
%since $\partial_{\mu} r\partial_{\mu} r=(\frac{R^2}{z^2}+1)\frac{\Theta^2}{z^2}$. 
%
On the other hand,  if we only had the banana metric 
\eqref{3var_banana} (\ie we did not know the transformations 
from the global coordinates and the relation $r_h=r(z,\vec{x})$),  
we could obtain the same result by looking at the vector orthogonal to the Killing flow in 
$(\vec{x},z)$ coordinates to construct $V^{\mu}$.\footnote{
In the cone coordinates, this is again very straightforward 
since $SO(d-1)$ symmetry reduces the problem to the $(z,R)$ plane  
with Killing vector $\nabla_{\{ \mu}\mathscr{K}_{\nu\}}=0$ given by
$\mathscr{K}=z\partial_z+R\partial_R$. In this case, we can also appreciate 
a nontrivial features of the cut-off surface $z=\epsilon$,
by noting from our expressions in eqs.~\eqref{cone_metric}-\eqref{3var_banana} 
that as we approach the horizon the other vector $N_z^{\mu}=(-1,N_z^i)$ 
becomes aligned with the Killing vector.}

To complete $I_{GHY}({\rm stretch})$, 
we need to compute the determinant of the induced metric, which we 
find takes the form
\beq\label{ind_metr_hor}
d^dx\,\sqrt{\sigma}=\frac{\rho^{d-2}\,d\Omega_{d-1}}{z^{d-2}}\,
\frac{\sqrt{f(r)}dx^1\,d\rho}{z^2 r_h}\times\bigg( 1+  \tilde{\cal D}(x_1,x_2,r_h;b)\bigg)\Bigg|_{z=z_\pm}
\eeq
where the difference function $\tilde{\cal D}$ vanishes exactly 
for $b=0$, and for $b\neq 0$ vanishes linearly when we expand around the insertion points. 
As we did before, we now specialize to the~$z=z_-$ component 
of the stretched horizon,  and extract the logarithmic form. 
This is again singled out by~$z_-^d$ in the denominator 
of eq.~\eqref{ind_metr_hor}, which similarly to eq.~\eqref{CFT_factor_2pt}
leads to~$r_h^d\, f_{2pt}$ in the domain $A$. The overall power of $r_h$ 
is then $r_h^{d-1}=4\GN\, S/\Omega_{d-1}$. Finally, 
by adding and subtracting $f_{2pt}$, we conclude that
\beq
I_{GHY}({\rm stretch})=S\,T_H\,\log\frac{|\vec{x}_{\bullet 1}-\vec{x}_{\bullet 2}|^2}{a^2}  + {\cal N}_\mt{stretch}\,.
\label{barn5}
\eeq
With the same argument as for ${\cal N}_{bulk}$. we 
can show that ${\cal N}_\mt{stretch}$ is a constant that 
does not dependent on the separation between insertion points.

%================================================================
\subsubsection{CFT from gravity}
%================================================================

Combining eqs.~\eqref{barn4} and \eqref{barn5} for the total 
onshell action, our final result reads
\beq\label{eval_totonshel}
I_{\begin{tikzpicture} \draw (0,0) rectangle (.5,.3); 
\draw[fill=black] (.14,.15) circle (1pt); 
\draw[fill=black] (.35,.15) circle (1pt); \end{tikzpicture} }= 
M \log\frac{|\vec{x}_{\bullet 1}-\vec{x}_{\bullet 2}|^2}{a^2} + {\cal N}
\eeq
where $M=\Delta$ and ${\cal N}={\cal N}_{bulk}+{\cal N}_\mt{stretch}$. It follows that 
\beq
\langle O(\vec{x}_{\bullet1})O(\vec{x}_{\bullet2})\rangle= 
e^{-I_{\begin{tikzpicture} \draw (0,0) rectangle (.5,.3); 
\draw[fill=black] (.14,.15) circle (1pt); 
\draw[fill=black] (.35,.15) circle (1pt); \end{tikzpicture}}} 
\simeq \frac{1}{|\vec{x}_{\bullet1}-\vec{x}_{\bullet2}|^{2\Delta}} \,,
\eeq 
as promised. 
In sum, taking into account  the backreaction from the insertion 
of two huge operators, the gravitational action, including a 
GHY term at the horizon, yields to the desired CFT two-point function \eqref{rewriting_goals}. 

A posteriori, we find that the onshell action in the two-point function geometry 
gives a result which is consistent with the following sequence of operations, 
\beq\label{calcs}
Z={\rm tr}( e^{-\beta \hat{H}})=e^{- \beta F_\mt{Gibbs}}\  
\rightarrow e^{-F_\mt{Gibbs} \delta\tau} 
\rightarrow e^{-E \delta\tau} 
\rightarrow e^{-\Delta \delta \tau}\,.
\eeq
Starting from the thermal partition function, which computes 
$F_\mt{Gibbs} = E - S\, T_H$,  we open up the thermal circle 
$\beta$ to an interval of length $\delta \tau$. This gives 
$e^{-F_\mt{Gibbs} \delta \tau}$ where now the endpoints  
play the role of the operator insertions, and induce a 
conical singularity in the bulk. To account for the latter,
we include the GHY term at the stretched horizon, 
and this changes $F_\mt{Gibbs}$ into $E$. Up to this point, 
the CFT has $S^{d-1}$ as its spacial slice 
and $E$ includes also Casimir energies in even $d$. 
By performing a Weyl transformation, we put the CFT on $\mathbb{R}^d$. 
This changes the AdS cut-off and yields the final ``dilatation amplitude'', 
$e^{-\Delta \delta\tau}$.\footnote{Time-like amplitudes of CFT 
states in holography can also be investigate using the formalism 
of \cite{Christodoulou:2016nej}. It would 
be interesting to relate this with our approach.} 
The actual computation which we performed shows 
how the $\log|\vec{x}_{\bullet 1}-\vec{x}_{\bullet 2}|$ 
dependence comes about and in particular the importance of 
excising small disks around the operators. 

We also have the normalization $\mathcal{N}$. It can be written down by 
following carefully the various steps above. Still, on its own it 
can always be absorbed in the definition of the operator. 
Physical quantities involve ratios, such as three-point couplings 
divided by two-point function normalizations. Then it is crucial 
to make sure that once we attack higher-point correlation functions, 
we compute them with the same scheme we are using for the two-point function. 
This is one of the motivations for the explorations that follow.

%===================================================================================================
\section{Fefferman-Graham Coordinates}\label{FG_expansion}
%===================================================================================================

In this section, we explore further aspects of the banana geometry. 
In particular, we will introduce Fefferman-Graham coordinates,
verify that the holographic stress tensor satisfies the conformal Ward Identity,
and discuss a nonperturbative feature of the Fefferman-Graham patch, 
which we call \emph{the wall}. With these insights, we will then 
revisit the notion of the cut-off surface at the boundary of AdS with marked points, 
and the computation of the onshell action as a proper surface integral.

Let us begin by recalling that Fefferman-Graham (FG) 
coordinates are defined by \cite{FG00,Fefferman:2007rka} 
\beq
ds^2_{FG}= \frac{d{\bf z}^2 }{{\bf z}^2} + {\bf h}_{ij}({\bf z},\vec{\bf x}) \,d\bx^i\,d\bx^j\,.
\label{choco2}
\eeq
The banana metric \eqref{3var_banana} is \textit{not} in this gauge, 
but can be brought to this form by a change of variables 
$z=z({\bf z},\vec{\bf x}),\, x^i=x^i({\bf z},\vec{\bf x})$, 
which is uniquely specified by requiring 
absence of mixed terms $d{\bz}\,d\vec{\bf x}$ and that $g_{\bz\bz}=1/\bz^2$ 
is fixed. In this gauge, ${\bf h}_{ij}({\bf z},\vec{\bf x})$ 
provides neatly the holographic evolution of the boundary metric, 
where the latter is defined  as the leading non-normalisable 
mode in the small ${\bf z}$ expansion.
In our case, since we are studying a vacuum solution with 
a flat Poincar\'e boundary, we will have 
\beq
{\bf h}_{ij}= \frac{1}{\bz^2} 
\(\delta_{ij} + \frac{2}{d}\,{\bf t}_{ij}\, {\bf z}^{d} +\cdots\)
\eeq
with ${\bf t}_{ij}$ traceless. Higher order terms are fixed 
once ${\bf t}_{ij}$ is given. Standard holographic 
renormalization methods, \eg \cite{Skenderis:1999nb,Skenderis:2002wp} 
show that ${\bf t}_{ij}$ is the (expectation value of the) stress tensor in the boundary CFT.

In order to generate FG coordinates, we can start 
from the small ${\bf z}$ expansion. For the banana on 
the line \eqref{3var_banana}, the FG gauge takes the form 
\beq
ds^2_{FG}= \frac{d{\bf z}^2 }{{\bf z}^2} + 
\sum_{a,b=\bx^1,\,\hrho} {\bf h}_{ab} d{\by}^a d{\by}^b+ {\bf h}_{\Omega\Omega}  d\Omega^2_{d-2}
\label{choco3}
\eeq
with metric components depending on the three variables 
${\bf z},\,{\bf x}^1$ and $\hrho=(\sum_{i=2}^d ({\bx}^i)^2)^{1/2}$, 
in addition to $M$ and $b$. Resumming in three-variables is a hard problem, in general.
An exception is $AdS_3$ where the series truncates, 
and we rediscover a result from Ba\~nados \cite{Banados:1998gg}
\beqa
ds^2 &=& {d{\rm AdS}_3 }^2
+ \bigg[ {\bf t}\, (d\by^1+i d\by^2)^2 + {\rm c.c.} \bigg] + \bz^2\, |{\bf t}|^2 ((d\by^1)^2+ (d\by^2)^2)
\label{barn6}\\
&&{\rm where}\qquad {\bf t}=-\frac{M}{4({\by}^1+i{\by}^2)(1+ b({\by}^1+i{\by}^2))}\,.
\nonumber
\eeqa
In the special case of the cone, $b=0$, the coordinates 
$\bx^1$ and $\hrho$ combine together to restore~$SO(d-1)$ symmetry,
and this allows us to find
\beq\label{resummed_metric}
ds^2_{FG}=\frac{1}{{\bf z}^2} \left[ d{\bf z}^2 + \frac{ (1-\frac{M{\bf z}^d}{4{\bf R}^d})^2 }{ (1+\frac{M{\bf z}^d}{4{\bf R}^d})^{  \frac{2(d-2)}{d}  } }d{\bf R}^2 +   (1+\tfrac{M{\bf z}^d}{4{\bf R}^d})^{ \frac{4}{d} }{\bf R}^2 d\Omega^2_{d-1}\right]
\eeq
with the change of variables from GtP cone to FG cone given by
\beq\label{FG_Roverz}
\frac{R}{z}=\frac{\bf R}{\bf z}\left(1+\frac{M}{4} \frac{{\bf z}^d}{{\bf R}^d} \right)^{\!\!\!\frac{2}{d} }\qquad;
\qquad 
R^2+z^2={\bf R}^2 \exp\Bigg[\int_0^{\frac{\bf z}{\bf R}} \frac{ 2xdx}{ k(x)} \Bigg]
\eeq
where $k(x)=x^2+{ (1-\frac{M}{4} x^d)^2 }\big/{(1+\frac{M}{4} x^d)^{\frac{2(d-2)}{d}}}$.\footnote{
See also \cite{Janik:2005zt,Janik:2006gp} where similar metrics appear.}

For the FG cone, $SO(d-1)$ symmetry implies that the image of a 
cylinder of radius~$r=R/z$ in global, is again a cone. 
However, the FG banana looks different. For example, in 
AdS$_3$ and AdS$_5$, we have\footnote{Note that when $M=0$, 
we recover the SCT map in the first term. On the other hand, 
with the limit $b\to0$, we recover our previous formula \eqref{FG_Roverz}. }
\begin{align}
\label{FG_bananas}
r^2_{AdS_3}\!=&\frac{  \tilde{\bf R}^2 }{{\bf z}^2} +\frac{M}{2} 
+\frac{M}{4}\bigg[ \frac{ 2 b({\bx}^1(1+b {\bf x}^1)-b \hrho^2 ) {\bf z}^2}{{\bf R}^2 {\bf R}_b^2 }+\frac{M{\bf z}^2}{4 {\bf R}^2 {\bf R}_b^2}\bigg]\\
r^2_{AdS_5}\!=&
\frac{  \tilde{\bf R}^2 }{{\bf z}^2} 
+ \frac{M {\bf z}^2 }{4 {\bf R}^2 {\bf R}_b^2} 
-\frac{M}{2}\bigg[
1
-\frac{ b\, {\bf  z}^2 (\tilde{\bf x}^1-b\, {\bf z}^2) }{{\bf R}^2 {\bf R}_b^2} 
- \frac{  ( \tilde{\bf R}^2 -2b\, \tilde{\bf x}^1\, {\bf z}^2 ) }{ b\, \hrho\, {\bf z}^2}
{\tan}^{-1}\bigg[  \frac{ b\, \hrho\, {\bf z}^2 }{  \tilde{\bf R}^2 -b\, \tilde{\bf x}^1\, {\bf z}^2  } \bigg] \bigg]+O(M^2) \notag
\end{align}
where $\tilde{\bf R}^2= (\tilde{\bf x}^1)^2+\hrho^2$ with $\tilde{\bf x}^1={\bf x}^1+ b\, ( {\bf R}^2 +{\bf z}^2)$, ${\bf R}^2=({\bf x}^1)^2 + \hrho^2$ and ${\bf R}_b^2=(1+b{\bf x}^1)^2 +b^2\,\hrho^2$.
  
The functional differences between a FG banana and 
a GtP banana \eqref{banana_eq} have a simple explanation: 
The GtP transformation is designed to foliate empty AdS, 
and knows nothing about the actual black hole metric. For 
example, it is $M$ independent. On the other hand, FG coordinates 
by construction are sensitive to the black hole geometry,
thus foliate the space accordingly. There is however a limitation:
the FG patch is not expected to cover the entire black hole geometry.
This statement is  simple to understand: since the small ${\bf z}$ 
expansion is fully determined by the knowledge of the boundary 
metric and  boundary stress tensor ${\bf t}_{ij}$, there is 
no freedom left to impose other boundary conditions, 
such as regularity conditions at the horizon.
This means that the FG patch must breakdown at some surface 
where the Jacobian of the coordinate transformation from global to FG vanishes. 
We will refer to this surface as \emph{the wall}. In practise, the wall 
encloses the horizon at a finite distance, and there is no way 
to access the horizon from the FG patch (on a real 
slice).\footnote{See \cite{Serantes:2022mgl} for a discussion 
about the radius of convergence in asymptotically AdS 
black holes in global coordinates.} 

In ${\bf x},\bz$ coordinates, the wall coincides with the 
surface where $\det {\bf h}_{ab}=0$. This can be seen by
examining the measure for the FG metric \eqref{choco3},
\beq
\frac{ \sqrt{ {\bf h} }}{\bf\ z} = \sqrt{g({\vec\bx,\bz})\,{\rm Jac}^2 (\vec\bx,\bz)}
\eeq
and further we have $g({\vec\bx,\bz})=\sqrt{g(r)}=r^{d-1}>0$ 
for the black hole metric \eqref{globalE}. 
In the FG cone coordinates, the wall is itself a cone given by
\beq
%\text{wall in the FG cone:}\ \ \ \ 
{\bf R}=\left(\frac{M}{4}\right)^{\!\!\frac{1}{d}}\, {\bf z}
\eeq
corresponding to $r_{wall}=M^{\frac{1}{d}}$. As expected, $r_{wall}>r_h$.
Note however that in general, the wall does not have to be a FG banana, nor 
the minimum in ${\bf z}$ of $r(\vec{{\bf x}},\bz)$ for fixed $\vec{\bf x}$. For example
in AdS$_3$, the wall is given by 
\beq
\(({\bf x}^1)^2+\hrho^2\)\( (1+b {\bf x}^1)^2+\hrho^2\)= 
\frac{M}{4}\, {\bf z}^2\,,
\eeq
which is not comparable with eq.~\eqref{FG_bananas}.

Even so, the wall and GtP bananas both originate from the insertion 
points at the AdS boundary, where ${\bf z}=z=0$ and $\vec{\bf x}=\vec{x}$. 
The novelty is what happens with respect to the cut-off, 
${\bf z}={\pmb \epsilon}$
 versus $z=\epsilon$. In fact, when we look at
${\bf z}={\pmb \epsilon}$ in the two-point function geometry 
with coordinates $z,\vec{x}$, this surface
cannot get arbitrarily close to the horizon, 
but intersects and stops at the preimage of the wall. 
This intersection is well defined and 
brings us to an important consideration:
Imagine we want to construct a black hole numerically,
in global coordinates. We are used to specify a boundary condition 
at infinity and a boundary condition in the interior. 
However, in the two-point function geometry, all of the boundary conditions 
are specified on $z=\epsilon$, and of course, they are distinct depending on whether 
we stay far or close to the operators. This distinction 
is quantified by the wall between the FG patch and the rest of the geometry.
In this sense, the usual notion of bulk evolution is modified in an interesting way.

%============================================================================================
\subsection{Holographic stress tensor }\label{holo_stress_t_sec}
%============================================================================================

To compute expectation values in the holographic CFT, 
the small ${\bf z}$ expansion of the FG metric is all we need, 
and this is straightforward to obtain in even the banana geometry.
The computation of the holographic stress tensors that follows is an example.
Nicely enough, for the cone coordinates,  
we can check all the various steps directly with pencil and paper.

Using the definition of the holographic stress tensor 
given in \cite{Skenderis:1999nb,Skenderis:2002wp}, we have
\beq\label{holo_stress}
\langle T_{ij}(\vec{\bf x}) \rangle_{   
\begin{tikzpicture} \draw (0,0) rectangle (.5,.3); 
\draw[fill=black] (.14,.15) circle (1pt); 
\draw[fill=black] (.35,.15) circle (1pt); \end{tikzpicture}  }= 
\frac{1}{8\pi \GN}\, \lim_{{\pmb \epsilon}\rightarrow 0}\, \frac{1}{{\pmb \epsilon}^{d-2}} 
\bigg[  {\cal K}_{ij}-{\cal K} {\bf h}_{ij}+
\frac{2}{\sqrt{\bf h}}\frac{\delta I_{ct}}{\delta {\bf h}^{ij}}
\bigg]_{{\bf z}={\pmb \epsilon}}\qquad{\rm with}\quad {\cal K}_{ij}
= \tfrac{1}{2} {\bf z}\,\partial_{\bf z} {\bf h}_{ij}
\eeq
and where, as in \eqref{counter_term_action}, the counterterm action is
\beq\label{counter_term_action2}
I_{ct}= \int_{{\bf z}={\pmb \epsilon}}\!\!d^d{\bf x} \,\sqrt{\bf h} \, 
\Bigg( (d-1)+\frac{1}{2(d-2)}\,{\cal R}[{\bf h}] +\cdots \Bigg)\,.
\eeq
Individual contributions in $\langle T_{ij} 
\rangle_{\begin{tikzpicture} \draw (0,0) rectangle (.5,.3); 
\draw[fill=black] (.14,.15) circle (1pt); 
\draw[fill=black] (.35,.15) circle (1pt); \end{tikzpicture}  }$ are divergent, 
but the combined sum in eq.~\eqref{holo_stress} is finite. 
For a flat boundary,  only the first term in eq.~\eqref{counter_term_action} 
is needed to resolve the divergence.  Note that had we done the computation 
in global coordinates, using FG coordinates with $\mathbb{R}\times S^{d-1}$ asymptotics, 
all counterterms would have contributed to the final result, \eg see \cite{Emparan:1999pm,Liu:2004it}.
So even though we started from the very same AdS-Schwarzschild black hole, 
the details of the $\langle T_{ij} \rangle_{   \begin{tikzpicture} 
\draw (0,0) rectangle (.5,.3); 
\draw[fill=black] (.14,.15) circle (1pt); 
\draw[fill=black] (.35,.15) circle (1pt); \end{tikzpicture}  }$ 
computation are quite different. For example, we will not 
be sensitive to Casimir energies of the $S^{d-1}$, 
and moreover, the spacetime dependence of
$\langle T_{ij}\rangle_{\begin{tikzpicture} 
\draw (0,0) rectangle (.5,.3); 
\draw[fill=black] (.14,.15) circle (1pt); 
\draw[fill=black] (.35,.15) circle (1pt); \end{tikzpicture}  }$ 
is quite more interesting.

With the insertion points $\vec{\bf x}_{\bullet1}$ 
and $\vec{\bf x}_{\bullet2}$, the final result is
\beq 
\langle T_{ij}(\vec{\bf x}) \rangle_{   \begin{tikzpicture} \draw (0,0) rectangle (.5,.3); \draw[fill=black] (.14,.15) circle (1pt); \draw[fill=black] (.35,.15) circle (1pt); \end{tikzpicture}  } = 
{\bf t}_{ij}(\vec{\bf x}) = \frac{d \, M}{(d-1) \Omega_{d-1} } 
\, \frac{ |\vec{\bf x}_{\bullet 1}-\vec{\bf x}_{\bullet 2}|^{d} }{ |\vec{\bf x}-\vec{\bf x}_{\bullet1}|^{d} |\vec{\bf x}-\vec{\bf x}_{\bullet2}|^{d} }\, \lambda_{ij}(\vec{\bf x};\vec{\bf x}_{\bullet1},\vec{\bf x}_{\bullet2})  
\label{holo_stress_tensor}
\eeq
where again $\Omega_{d-1}=  2\pi^\frac{d}{2} /\Gamma[\frac{d}{2} ]$, 
and the traceless tensor $ \lambda_{ij}$ is given by
\begin{align}
 \lambda_{ij}(\vec{\bf x};\vec{\bf x}_{\bullet1},\vec{\bf x}_{\bullet2})&=  -u_i u_j +\frac{\delta_{ij}}{d}\qquad{\rm with}\ \  u_i= \frac{ |\vec{\bf x}-\vec{\bf x}_{\bullet1}|\, |\vec{\bf x}-\vec{\bf x}_{\bullet2}| }{ |\vec{\bf x}_{\bullet 1}-\vec{\bf x}_{\bullet 2}|}
 \Bigg[\frac{ (\vec{\bf x}-\vec{\bf x}_{\bullet1})_{i} }{|\vec{\bf x}-\vec{\bf x}_{\bullet1}|^2  } -\frac{ (\vec{\bf x}-\vec{\bf x}_{\bullet2})_{i} }{| \vec{\bf x}-\vec{\bf x}_{\bullet2} |^2  } \Bigg]\,.
\end{align}
% 
%
%
%\rcmFA{Please check sign. I compared to eq.~(23) in \cite{Petkou:1994ad}.}{\color{red} FA: A cleaner paper of Petkou, \url{https://arxiv.org/pdf/hep-th/9410093.pdf} has the $+\delta_{ij}$ see there formula (2.14) and (2.21), so it coincides with what I am saying below. Also \url{https://arxiv.org/pdf/1608.04948.pdf} formulae (24) and (26), where in the second one he has a minus sign. It looks that [25] had a typo} \rcm{My only suggestion then is to remove [25] from our references below -- and perhaps add the other two papers that you found. FA OK, fixed}

%{\color{blue}
%\begin{mdframed}
%FA: I think the conventions for $T_{ij}$ are $T_{ij}=E u_i u_j + P (\delta_{ij}+u_i u_j)$ with $u^2=-1$ so to have a projector on the pressure. In the cone this is $u^i=+i x^i/R$. Now, when we do the holographic computation we get
%\beq
%T_{ij}dx^i dx^j= -(d-1) M dR^2  + M R^2 d\Omega^2= -(d-1) M \frac{x_i x_j}{R^2} + M(\delta_{ij}-\frac{x_i x_j}{R^2})
%\eeq
%\end{mdframed}}

Now interpreting
eq.~\eqref{holo_stress_tensor} as
\beq
\langle T_{ij}(\vec{\bf x}) \rangle_{\begin{tikzpicture} 
\draw (0,0) rectangle (.5,.3); 
\draw[fill=black] (.14,.15) circle (1pt); 
\draw[fill=black] (.35,.15) circle (1pt); \end{tikzpicture}  }=
\frac{\langle T_{ij}(\vec{\bf x})\,O(\vec{x}_{\bullet1})\,O(\vec{x}_{\bullet2})\rangle}{\langle O(\vec{x}_{\bullet1})\,O(\vec{x}_{\bullet2})\rangle}\,,
\label{barn7}
\eeq
we see that this expression precisely reproduces the conformally 
invariant three-point function of the energy momentum tensor and
two scalar operators \cite{Cardy:1987dg,Petkou:1994ad,Liu:1998bu,Penedones:2016voo} with
\beq
\Delta= M \,.%L_{AdS} 
\eeq 
This is a simple yet very important consistency check for 
our two-point function geometries! Moreover, it gives us an 
independent derivation of the relation between the conformal 
dimension and the mass of the black hole, which we used 
in the onshell action computation.

%===================================================================================
\subsection{The bulk, the total derivative and the cut-off}\label{action_derv_cof_sec}
%===================================================================================

We now want to comment on the onshell action.
The idea is to divide the geometry 
into a FG patch where the cut-off surface lives, 
and a global patch close to the horizon where the stretched horizon lives.
This organization is conceptually helpful because,
as we saw in section \ref{oneshell_sec}, both 
contributions are crucial in order to obtain the expected CFT
dependence on the dimension $\Delta=M$. The approach described 
in the following sections is not restricted to the Schwarzschild black hole but 
also generalizes to charged black holes in AdS \cite{USthree}.

Our starting point is the observation that when the bulk action 
in global coordinates can be written onshell as a total derivative, 
we can apply the GtP transformation directly on the integrand.
So let us  assume that the onshell action in global reads
\beq\label{Sabra_action}
I_{ \begin{tikzpicture} \draw (0,0) ellipse (.2cm and .1cm); 
\draw[rounded corners] (-.2,0) -- (.0,-.5) -- (.2,0); \end{tikzpicture}}=\int\!d\Omega_{d-1}
\int\!d\tau_{} \int\!dr \times\frac{d}{dr} \mathcal{A}(r)
\eeq
for some function ${\cal A}(r)$. For the AdS-Schwarzschild geometry, 
this is \cite{Liu:2004it}\footnote{In fact, for all $SO(d-1)$ symmetric 
and static backgrounds, which solve for Einstein gravity coupled 
to a general (two-derivative) action of scalar and vector fields, 
one can cast ${\cal A}(r)$ as a particular functional of the 
(time and radius) components of the metric \cite{Liu:2004it}.} 
\beq\label{exprecalA}
{\cal A}(r)=-2 r^{d-2}+2r^{d-2} f(r)= 2 r^{d}-2M 
\eeq
Going through the GtP map \eqref{poincareCoords} and the 
SCT map \eqref{SCT} to reach  the banana geometry \eqref{3var_banana} 
with the insertion points on a line, we find:
%
%
%\begin{align}
%I_{ \begin{tikzpicture} \draw (0,0) rectangle (.5,.3); \draw[fill=black] (.14,.15) circle (1pt); \draw[fill=black] (.35,.15) circle (1pt); \end{tikzpicture} }\ %=\ -iI_{bulk}
%&=\,\, 
%\int\!d\Omega_{d-1}\int\!dR dz \times \left[ -\frac{R}{R^2+z^2} \partial_z +\frac{z}{R^2+z^2}\partial_R \right]{\cal A}\big(r=\tfrac{R}{z}\big) \\
%&=\,\, 
%\int\!d\Omega_{d-2}\int\!\frac{x_2^{d-2} dx_2 dx_1dz}{ \tilde{R}^{d}} \times \!\!\sum_{i=z,1,2} \frac{ p_i(z,x_1,x_2)\, \partial_i{\cal A}\big( r=\frac{\tilde R}{z}\big)}{(R^2+z^2)( (1+b x_1)^2+b^2 x_2^2+b^2 z^2)} 
%r={\scriptstyle \sqrt{ \tilde \Omega^2 (\frac{ \tilde R^2}{\tilde z^2} -1)+1}}\big) 
%
%\label{fa_nnunit_th}
%\end{align}
%
%
\begin{align}
\label{fa_nnunit_th}
I_{ \begin{tikzpicture} \draw (0,0) rectangle (.5,.3); 
\draw[fill=black] (.14,.15) circle (1pt); 
\draw[fill=black] (.35,.15) circle (1pt); \end{tikzpicture} }\ %=\ -iI_{bulk}
&=\,\, 
\int\!d\Omega_{d-1}\int\!dR \,d\tz \times 
\left[ -\frac{R}{R^2+\tz^2} \partial_{\tz} +\frac{\tz}{R^2+\tz^2}\partial_R \right]{\cal A}\big(r=\tfrac{R}{\tz}\big) \\
&=\,\, 
\int\!d\Omega_{d-2}\int\!d\rho\, dx^1\,dz \times \!\!\sum_{i=z,x^1,\rho} u_i\, \partial_i{\cal A}\big( r(z,\vec{x})\big)\quad{\rm with}\ \ u_i=-\frac{ 1}{d-2}\,\frac{\rho^{d-2}}{z^{d+1}}
\frac{   \partial_i r^{2-d}  }{  
(\frac{x^2}{z^2}+1)\frac{\OOmega^2}{z^2}}  \notag
\label{fa_nnunit_th}
\end{align}
%
%with $\tilde R^2=x_2^2+\tilde{x}_1^2$, with $\tilde{x}_1=x_1+b(x_1^2+x_2^2+z^2)$.
%
where $r(z,\vec{x})=\sqrt{\Theta^2 \left( \frac{x^2}{z^2}+ 1\right) - 1}$ 
and $\OOmega^2=1+ 2\, b \cdot {x}  + b^2(  x^2 +  z^2 )$ 
with $x^2=(x^1)^2+\rho^2$, as given in eq.~\eqref{banana_eq}. 
Note that the dot product here is taken using $\delta_{ij}$.
Of course, the second expression reduces to the first one when $b=0$.
Then, it is simple to check, and perhaps expected, that the 
vector $u^i$ is divergence free.\footnote{Note 
also that $\sum_i\partial_{i} r\partial_{i} r=(\frac{R^2}{z^2}+1)\frac{\OOmega^2}{z^2}$.} Thus 
we can evaluate the onshell action by using 
the divergence theorem, reducing it to a surface integral, namely
\beq\label{total_der_action}
I_{\begin{tikzpicture} \draw (0,0) rectangle (.5,.3); 
\draw[fill=black] (.14,.15) circle (1pt); 
\draw[fill=black] (.35,.15) circle (1pt); \end{tikzpicture}}=
\int_{}d\sigma \,\frac{\vec{u}\cdot\vec{n}}{|\vec{n}|}\,\mathcal{A}\big(r(z,\vec{x})\big)\ 
\eeq
where $\vec{n}$ is an outward-pointing normal 
to the integration surface $\sigma$.
For the cone, this is just a line integral in the $(\tz,R)$ plane.

The choice of integration surface is crucial. For example,    
in the cone we could pick the union of the two cones corresponding 
to $\frac{R}{z}=r_\infty$ with a large $r_{\infty}\rightarrow\infty$ for the asymptotic AdS cut-off, 
and $\frac{R}{z}=r_h(1+\epsilon')$, for the 
stretched horizon. By computing the onshell action between 
these surfaces, we would actually be
repeating the same  computation as in global coordinates.\footnote{The 
way terms combine together is slightly different,
but we checked that we do get the same result as in global 
coordinates. In particular, we find a Casimir energy contribution for even $d$.}
This is \emph{not} the computation which we want for the two-point 
function. Rather, we want to compute  the onshell action 
with a cut-off at ${\bf z}=\pmb{\epsilon}$ in the FG patch.

In the following, we describe the computation of $I_{\begin{tikzpicture} 
\draw (0,0) rectangle (.5,.3); 
\draw[fill=black] (.14,.15) circle (1pt); 
\draw[fill=black] (.35,.15) circle (1pt); \end{tikzpicture}}$ 
with the appropriate boundary surface $\sigma$. We will 
distinguish the three boundary components as: 
the asymptotic cut-off  surface described by coordinates
$\vec{c}=(x^1(\bz=\pmb{\epsilon},\vec{\bf x}),\rho(\bz=\pmb{\epsilon},\vec{\bf x}),z(\bz=\pmb{\epsilon},\vec{\bf x}))$;
the banana corresponding to the stretched horizon; 
and two pieces of connecting tissue between the previous components, 
one for each of the insertion points.

%==========================================================================
\subsubsection{The asymptotic cut-off}
%==========================================================================

\begin{figure}[t]
\begin{center}
{\includegraphics[scale=0.4]{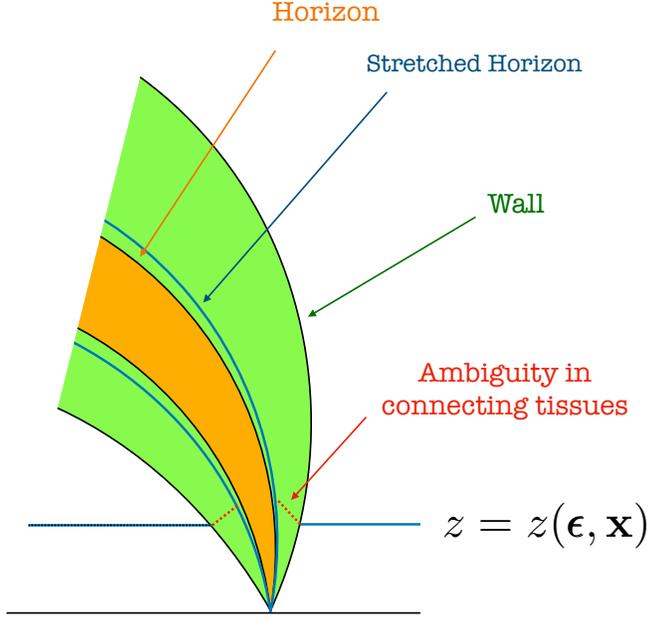}}
\end{center}
\caption{A section of the banana geometry to illustrate the integration surface $\sigma$. 
The cut-off surface $z=({\bf z},{\bf x}),x=({\bf z},{\bf x})$ can be extended up to the preimage of the wall, 
the choice of connecting tissues from there to the stretched horizon results in a choice of normalisation.
\label{bananaSection}}
\end{figure}

The asymptotic cut-off surface $\bz=\pmb{\epsilon}$ is described by 
$\vec{c}=(x^1(\pmb{\epsilon},\vec{\bf x}),\rho(\pmb{\epsilon},\vec{\bf x}),
z(\pmb{\epsilon},\vec{\bf x}))$, in terms of FG coordinates. 
From our discussion about the FG patch in the previous section, 
we know that the integration over $\vec{\bf x}$ 
can be  extended at most up to the intersection of 
${\bf z}=\pmb{\epsilon}$ and the wall -- see figure \ref{bananaSection}.
This observation explains more precisely what 
happened in section \ref{oneshell_sec} when we introduced 
a large enough radius $a$ for the disks encircling the 
insertion points in figure \ref{BananaRegulators}.
We understand now that the minimum value that $a$ 
can take is the preimage of $\vec{\bf x}$ 
on the wall at ${\bf z}=\pmb{\epsilon}$.

%-----------------------------------------

%-----------------------------------------

In the cone,  the surface integral on 
$\vec{c}=(R(\pmb{\epsilon},{\bf R}),
z(\pmb{\epsilon},{\bf R}))$ is very explicit, since we know 
the change of variables \eqref{FG_Roverz}.
We find 
\beq
{\cal A}_{cone}(\pmb{\epsilon},{\bf R})= 
\frac{ 2{\bf R}^d }{ \pmb{\epsilon}^d }-M +\frac{M^2}{4^2} \frac{\pmb{\epsilon}^{d}}{ {\bf R}^d }
\eeq
Then, the normal vector $\vec{n}$ is a 
rotation of $\partial_{{\bf R}}\vec{c}$, 
and the very definition of the 
surface integral cancels $|\vec{n}|$, 
so the measure of the integral becomes
\beq\label{cutoff_done}
\int_{ {\bf z}=\pmb{\epsilon}}\!\! d\sigma\,\frac{\vec{u}\cdot\vec{n}}{|\vec{n}|} = 
\int\frac{d{\bf R}}{\bf R} \left(  
1-\frac{ {\pmb{\epsilon} }^2 }{ {\bf R}^2 } \frac{1}{k(\frac{\pmb{\epsilon}}{\bf R})} \right)=
\int\frac{d{\bf R}}{\bf R} \left( \frac{1}{(1+\frac{ \pmb{\epsilon} ^2 }{ {\bf R}^2})} - \frac{d-1}{d} M \frac{ \pmb{\epsilon}^{d+2} }{  {\bf R}^{d+2}}+\ldots \right)\,.
\eeq
The whole $k(x)$ is given in eq.~\eqref{FG_Roverz}.
When taking the $\pmb{\epsilon}\rightarrow 0$ limit, 
we have to be careful in isolating the AdS 
contribution in $\vec{u}\cdot\vec{n}$ from  the rest 
because ${\cal A}_{cone}$ itself comes with a 
Laurent series in $\pmb{\epsilon}$.
What  we want to ensure is that 
the corrections to AdS proportional to  
$M$ do not interfere with the structure of ${\cal A}_{cone}$ 
as a series in $\pmb{\epsilon}$. This implies 
that we have to resum the AdS contribution in 
$\pmb{\epsilon}$ before taking the limit to zero.
This is what we made explicit on the right-hand 
side of eq.~\eqref{cutoff_done}.

The GHY and counterterm contributions at the 
asymptotic boundary can by definition be evaluated 
in the FG patch directly. Putting all of these together
cancels the usual UV divergence from AdS and a finite 
result remains,%\footnote{The GHY and counterterm contributions 
%in FG coordinates are not simply constants.}
\beq
\lim_{\pmb{\epsilon}\rightarrow 0}\frac{1}{16\pi \GN}
\int_{{\bf z}=\pmb{\epsilon}} d\sigma \,\frac{\vec{u}\cdot\vec{n}}{|\vec{n}|}{\cal A}(r) 
+I_{GHY}(\partial_{AdS}) + I_{ct}= -\frac{M}{16\pi \GN}
\int_{{\bf R}_\star}^{1/{\bf R}_\star} \frac{d{\bf R}}{\bf R}\,.
\eeq
We can take any ${\bf R}_{\star}\ge\left[\frac{M}{4}\right]^{\frac{1}{d}}\pmb{\epsilon}$, 
where the limit is the value fixed by the wall. It is worth mentioning 
that in the cone coordinates, we can compute the onshell action directly 
by evaluating $\sqrt{\det g_{FG}}$, from ${\bf z}=\pmb{\epsilon}$ to the ${\bf z}=$ wall. 
Since we know the preimage, \ie $r_{wall}=M^{\frac{1}{d}}$, we can then check 
this result against the surface integral done with ${\cal A}$. 
We find perfect agreement, as it should. In passing, 
we also notice that ${\cal A}(r_{wall})=0$.

We now repeat the cut-off surface integration for the banana. 
This time we do not have the full change of variables, 
however, we can proceed by using the series expansion in ${\bf z}$. The measure factor 
can be found to generalize the right-hand side of eq.~\eqref{cutoff_done} 
into\footnote{To compute $\vec{u}\cdot\vec{n}$, 
we observe that $\vec{n}=-\partial_{\bf z} + M O({\bf z}^{d+1})$, thus we  
replace the coordinate dependence of $u^z$ with 
bold font variables, and check that 
the actual FG expansion only modifies this by terms $M O({\bf z}^{d+1})$.} 
\beq
\vec{u}\cdot\vec{n}\big|_{{\bf z}=\pmb{\epsilon}}=\frac{\hrho^{d-2}}{\tilde{\bf R}^{\frac{d}{2} } }\,
\frac{ \hrho^2 + \tilde{\bf x}^1( \tilde {\bf x}^1- 2 b \pmb{\epsilon}^2) }{ ({\bf R}^2 +\pmb{\epsilon}^2 )\,\Theta^2({\bf x}^1,\hrho,\pmb{\epsilon}) } + M O(\pmb{\epsilon}^{d+1})
\eeq
where ${\bf R}^2=\hrho^2 +(\bx^1)^2$, $\tilde {\bf R}^2=\hrho^2+(\tilde{\bf x}^1)^2$, 
and $\tilde{\bf x}^1={\bf x}^1+b\, ({\bf R}^2+{\bf z}^2)$.
Then ${\cal A}_{banana}=\frac{ 2\tilde{\bf R}^d }{ \pmb{\epsilon}^d }-M+O(Mb\pmb{\epsilon}^d,M^2\pmb{\epsilon}^d)$. 
It follows again that
\beqa
&&\lim_{\pmb{\epsilon}\rightarrow 0}\frac{1}{16\pi \GN}
\int_{{\bf z}=\pmb{\epsilon}} d\sigma \,\frac{\vec{u}\cdot\vec{n}}{|\vec{n}|}{\cal A}(r) 
+I_{GHY}(\partial_{AdS}) + I_{ct}
\label{barn9}\\
&&\qquad\qquad \qquad\qquad= 
-\frac{M}{16\pi \GN}\int_{\begin{tikzpicture} 
\draw (0,0) rectangle (.5,.3); 
\draw[fill=black] (.14,.15) circle (1pt); 
\draw[fill=black] (.35,.15) circle (1pt); \end{tikzpicture} } 
\frac{ \hrho^{d-2}\,d\hrho\, d{\bf x}^1 d\Omega_{d-2}}{ 
(({\bf x}^1)^2+\hrho^2)^{d/2}\,( (1+b{\bf x}^1)^2+\hrho^2) ^{{d}/{2}}}\,,
\nonumber
\eeqa
where in the remaining integral, we recognise the 
expression for $f_{2pt}$ familiar from section 
\ref{oneshell_sec}, \eg compare with eq.~\eqref{CFT_factor_2pt}.

%==========================================================================
\subsubsection{From cut-off to stretched horizon} 
%==========================================================================

At this point, we must consider the two components 
of the integration surface $\sigma$ comprising 
connecting tissue extending between the asymptotic cut-off 
surface and the stretched horizon -- see figure \ref{bananaSection}. 
The simplest choice is to go straight into the bulk.
Other choices are possible, for example, 
we can use them to match the $z=\epsilon$ 
surface \emph{close} to the insertion points.   
Compared to computations usually done for probe objects,
the freedom in cutting the geometry passed the FG wall is something new.
To see why consider a  geodesic connecting the insertion points.
The length of such a geodesic, \ie $({\bf z}(s),{\bf x}(s))=\ell\,(\sin s,\cos s)$ with $s\in (0,\pi)$, 
computes the two-point function. The ambient space is just empty 
AdS and the cut-off is ${\bf z}=\pmb{\epsilon}$, to be 
understood physically as a UV cut-off or lattice spacing. 
The range of integration is read off from the equation ${\bf z}(s)=\pmb{\epsilon}$. 
Equally, we can parametrise the integration with respect to the $\bf{x}$ coordinate.
Note at this point, the obvious fact that empty AdS is already in the FG gauge. 
In particular, there is no extra space which corresponds to integrating 
over the two connecting tissues between the cut-off and the horizon.
It is also clear now that the role of the connecting tissues is 
simply to change the normalization of the operators, since
different tissues yield different normalization constants to the final action.

%==========================================================================
\subsubsection{Stretched horizon}
%==========================================================================

Finally, the stretched horizon contribution is the surface integral on
$\vec{c}=(x^1,\rho,z=z(x^1,\rho))$, where $z=z(x^1,\rho)$ solves 
the equation $r(z,\vec{x})=r_h(1+\epsilon')$ as we did in \eqref{zmp_def}.
The normal vector $\vec{n}$ is the cross 
product $\partial_{x^1}\vec{c}\times \partial_{\rho}\vec{c}$, and
of course, it becomes proportional to $\partial_i r(z,\vec{x})$.
Then, similarly to what we did in section \ref{oneshell_sec}, we look at
$z=z_-(x^1,\rho)$, and we extract the logarithmic form close to the insertion points.
The relevant integral is
\beq\label{intermediate_banana_integral}
I_{hor}=-\frac{1}{16\pi \GN}\int_{A}\frac{\rho^{d-2} d\rho\, dx^1\, d\Omega_{d-2}}{ \tilde{R}_{-}^{d-1}} \,\frac{ b^2 (1+r_h^2) {\cal A}(r_h) }{ 
(z_+^2-z_-^2) (R^2+z_-^2)( (1+b x^1)^2+b^2 \rho^2+b^2 z_-^2) }
\eeq
where $\tilde{R}_-=\tilde{R}(x^1,\rho,z=z_-(x^1,\rho))$. Close 
to the insertion points, the logarithmic form in 
eq.~\eqref{intermediate_banana_integral} is 
\beq
I_{hor}=
-\frac{{\cal A}(r_h)}{16\pi\GN}\int_{A}\frac{\rho^{d-2} 
d\rho\, dx^1\,d\Omega_{d-2}}{((x^1)^2+\rho^2)^{d/2}( (1+b x^1)^2+ b^2 \rho^2)^{{d}/{2}} } 
+{\color{red} \cdots}\,,
\label{barn8} \eeq
where again we recognize appearance of $f_{2pt}$. The rest of the terms denoted by the ellipsis are again free of divergences and will 
only contribute to the normalization.

%==========================================================================
\subsubsection{The onshell action revisited}
%==========================================================================

At this point, we can assemble again the total onshell action,
\beq
I_{ \begin{tikzpicture} \draw (0,0) rectangle (.5,.3); \draw[fill=black] (.14,.15) circle (1pt); \draw[fill=black] (.35,.15) circle (1pt); \end{tikzpicture} }= 
 I_{ bulk }\ + I_{GHY}(\partial_{AdS})+ I_{GHY}({\rm stretch}) +\ I_{ct}
 \eeq
where the contribution in $I_{bulk}$ coming from the asymptotic boundary
(\ie $\bz=\pmb{\epsilon}$) is given in eq.~\eqref{barn9},
and the contribution from the stretched horizon is given 
in eqs.~\eqref{intermediate_banana_integral} and \eqref{barn8}. 
Moreover, we have argued that the boundary components 
connecting the cut-off surface at the asymptotic boundary 
and at the stretched horizon only contribute to the 
overall normalization of the two-point function. 
To finalize, we need to include the GHY term at the 
stretched horizon, but this is the same as \eqref{barn5}.  
Therefore, by mechanically substituting in all of the contributions, 
we establish the same result which we found in section \ref{oneshell_sec} 
for the CFT two-point function, 
\beq
I_{\begin{tikzpicture} \draw (0,0) rectangle (.5,.3); 
\draw[fill=black] (.14,.15) circle (1pt); 
\draw[fill=black] (.35,.15) circle (1pt); \end{tikzpicture} }=   
\Delta \int_{\begin{tikzpicture} \draw (0,0) rectangle (.5,.3); \draw[fill=black] (.14,.15) circle (1pt); \draw[fill=black] (.35,.15) circle (1pt); \end{tikzpicture} }\!\!d^dx\,f_{2pt} \ + {\cal N}.
\eeq
where $\Delta=M$ for the operators, and as we understood the normalization 
depends on the choice of connecting tissues. 

A bonus of our discussion here, compared to that of section \ref{oneshell_sec}, is 
that by rewriting the bulk contribution to the onshell action as a manifest surface 
integral (\ie the integrand became a total derivative), we could make precise the FG renormalization scheme for the cut-off.
Another great advantage of working with a total derivative 
will be the following observation, or ``how to compute the 
total action without really trying".

Given an ansatz for a gravitational background in global 
coordinates, we can recast Einstein's equation as the 
equation of motion from an effective Lagrangian. With the 
latter, we look for a conserved (and finite) Noether charge  
${\cal Q}(r)$ corresponding to a scaling symmetry
\cite{Iyer:1994ys,Gubser:2009cg} in this effective Lagrangian. 
Since this Noether charge is conserved, it must be a constant 
satisfying $\partial_r{\cal Q}(r)=0$. For the AdS-Schwarzschild 
in \eqref{globalE}, this procedure yields
\beq
{\cal Q}(r)=-\tfrac{1}{2}r^{d-1}f'(r)+r^{d-2}f(r)-r^{d-2}=-\tfrac{d}{2}M\,.
\label{choco4}
\eeq
We note that the middle expression above applies for a general 
ansatz of the form given in eq.~\eqref{globalE} (\ie without specifying 
the specific solution for $f(r)$), while the final constant proportional 
to $M$ comes from substituting in eq.~\eqref{blackening_fact}.

The utility of the Noether charge is to relate the boundary contributions 
to the onshell action and the stretched horizon contribution. To see this, we
first replace  $r^{d-1}f'(r)$ term (\ie the first contribution in the 
middle expression above) with the GHY term on a surface of constant 
$r$ in global coordinates, namely
\beq
{\cal G} \equiv 2\sqrt{f(r)} r^{d-2} K = r^{d-1}f'(r)+2(d-1) r^{d-2} f(r)\,. 
\eeq
where the extrinsic curvature is $K=r^{-d+2}\partial_{r}(r^{d-2}\sqrt{f(r)})$. %\rcmFA{What is this really? $K$?}
Similarly, we replace the $r^{d-2}$ term (\ie the third contribution in 
the middle expression in eq.~\eqref{choco4}) using ${\cal A}(r)$ in 
eq.~\eqref{exprecalA}. Thus we arrive at the following expression for 
the Noether charge
\beq\label{Noether_charge_rel}
{\cal Q}(r)= -\tfrac{1}{2}{\cal G}(r)+\tfrac{1}{2}{\cal A}(r)+(d-1)r^{d-2}f(r)\,.
\eeq
As our notation indicates, we may evaluate the Noether charge at any radius and so
it is interesting to evaluate the above at the horizon $r=r_h$. There 
the last term vanishes since $f(r_h)=0$, and we are left with the first two. 
In this form, we don't have to know the exact gravity solution, but only 
read off how the constant Noether charge is related to quantities which 
we need to evaluate at the horizon. In particular, we have the surface 
term coming from the radial integral in bulk action 
and the GHY term on the stretched horizon. 
That is, we may use eq.~\eqref{Noether_charge_rel} to re-express the 
surface term in eq.~\eqref{barn8} along with the GHY term on the stretched 
horizon in terms of the Noether charge, as desired. Hence combining 
these contributions with the expression in eq.~\eqref{barn9}, the onshell action becomes
\begin{align}
I_{ \begin{tikzpicture} \draw (0,0) rectangle (.5,.3); 
\draw[fill=black] (.14,.15) circle (1pt); 
\draw[fill=black] (.35,.15) circle (1pt); \end{tikzpicture} }&= 
 I_{ bulk }\ + I_{GHY}(\partial_{AdS})+ I_{GHY}({\rm stretch}) +\ I_{ct} \\
 &=
 \frac{1}{16\pi \GN}\bigg[-M -2\lim_{r\rightarrow\infty}{\cal Q}(r) \bigg] \int_{\begin{tikzpicture} \draw (0,0) rectangle (.5,.3); \draw[fill=black] (.14,.15) circle (1pt); \draw[fill=black] (.35,.15) circle (1pt); \end{tikzpicture} }\!\!d^dx\,f_{2pt} \ +\ {\cal N}\,.
 \label{2pt_theorem}
 \end{align}
 Then the integral can be evaluated as in eq.~\eqref{barn10}, 
 and after restoring the mass normalization $\alpha$ given by eq.~\eqref{value_alpha}, we recover
\beq
I_{\begin{tikzpicture} \draw (0,0) rectangle (.5,.3); \draw[fill=black] (.14,.15) circle (1pt); \draw[fill=black] (.35,.15) circle (1pt); \end{tikzpicture} }=   \Delta\, \log\frac{|\vec{x}_{\bullet 1}-\vec{x}_{\bullet 2}|^2}{a^2} + {\cal N}
\eeq
using $\Delta=M$.
Hence we have again demonstrated our claim of deriving the 
two-point function correlator for huge operators from gravity.

%===================================================================
%\subsubsection{Analytic continuation to negative mass}
%====================================================================

%We conclude this section with a nice observation about positive vs negative mass. 
%The onshell action of the two-point function geometry always gives
%\beq\label{negative_mass}
%S_{\begin{tikzpicture} \draw (0,0) rectangle (.5,.3); \draw[fill=black] (.14,.15) circle (1pt); \draw[fill=black] %(.35,.15) circle (1pt); \end{tikzpicture} }=   |M| \log\frac{|\vec{x}_{\bullet 1}-\vec{x}_{\bullet 2}|^2}{a^2} + {\cal %N}
%\eeq
%thus works for both $\pm M$, even though the physics is quite different. 
%Let us show how. If the mass is negative 
%there is no horizon, and more precisely, the blackening factor diverges as $+\tilde{M}/r^{d-2}$ at $r=0$, 
%where $\tilde M\ge 0$ and we have set $M=-\tilde M$ in \eqref{blackening_fact} to avoid confusion.
%In order to account for the boundary operators we introduce again a membrane and a GHY term in the action.
%That this should be the case is also clear from considering 
%the limit $r_h\rightarrow 0$, which is a reasonable limit. 
%Note in fact that the GHY term ${\cal G}$ is still finite. At this point we look at \eqref{2pt_theorem}
%and sum all three terms. The energy $E=-(d-1)\tilde{M}$ is now negative. However, 
%the last contribution $2(d-1)r_h^{d-2} f(r_h)$, that was vanishing for an horizon, 
%now equals to $+2(d-1)\tilde{M}$, and leads to our claim.

%===========================================================================================
\section{Geodesics in the black hole two-point function geometry}\label{geodesic_sec}
%============================================================================================

As we have been discussing, the insertion of huge operators 
results in backreaction on the AdS geometry, and we found 
that our two-point function geometry simply corresponds to 
a Euclidean black hole presented in a somewhat unusual way. 
In this section, we study both qualitatively and quantitatively how 
the motion of geodesics that shoot in from the asymptotic boundary 
of AdS is affected by the presence on the
Euclidean black hole. These geodesics would correspond to the 
insertion of additional light operators and so this is a first 
step towards investigating higher point functions with our geometric approach.

\begin{figure}[t]
\begin{center}
{\includegraphics[scale=0.58]{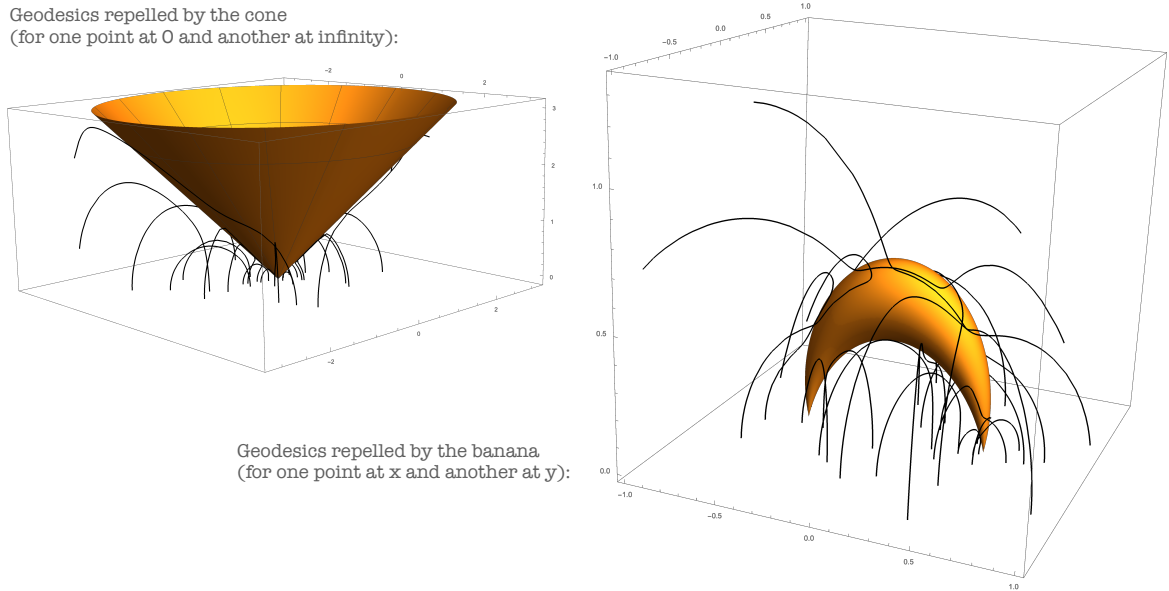}}
\end{center}
\vspace{-5cm}
\caption{Geodesics moving in the black hole background when 
the stretched horizon is a cone ({\bf left}) and when it is a banana ({\bf right}).
Unless finely tuned, geodesics do not collide with the horizon. 
\label{geodesicsFig}}
\end{figure}

Figure \ref{geodesicsFig} illustrates a variety of 
geodesics originating from points on the asymptotic  boundary, 
with a given velocity. A prominent feature of the plot 
is that the geodesics shown there do not reach
the horizon. In fact, the only geodesics that collide with the 
horizon are finely tuned. 
The latter is clearly understood by considering the metric 
\eqref{globalE} in global coordinates. In this setting, it is 
straightforward to show that the only geodesics reaching $r=r_h$ 
must have a vanishing velocity in the $\tau$ direction. Further, 
the angular momentum can not be very large (\ie $|\ell|\le r_h$).

\begin{figure}[t]
\begin{center}
{\includegraphics[scale=0.5]{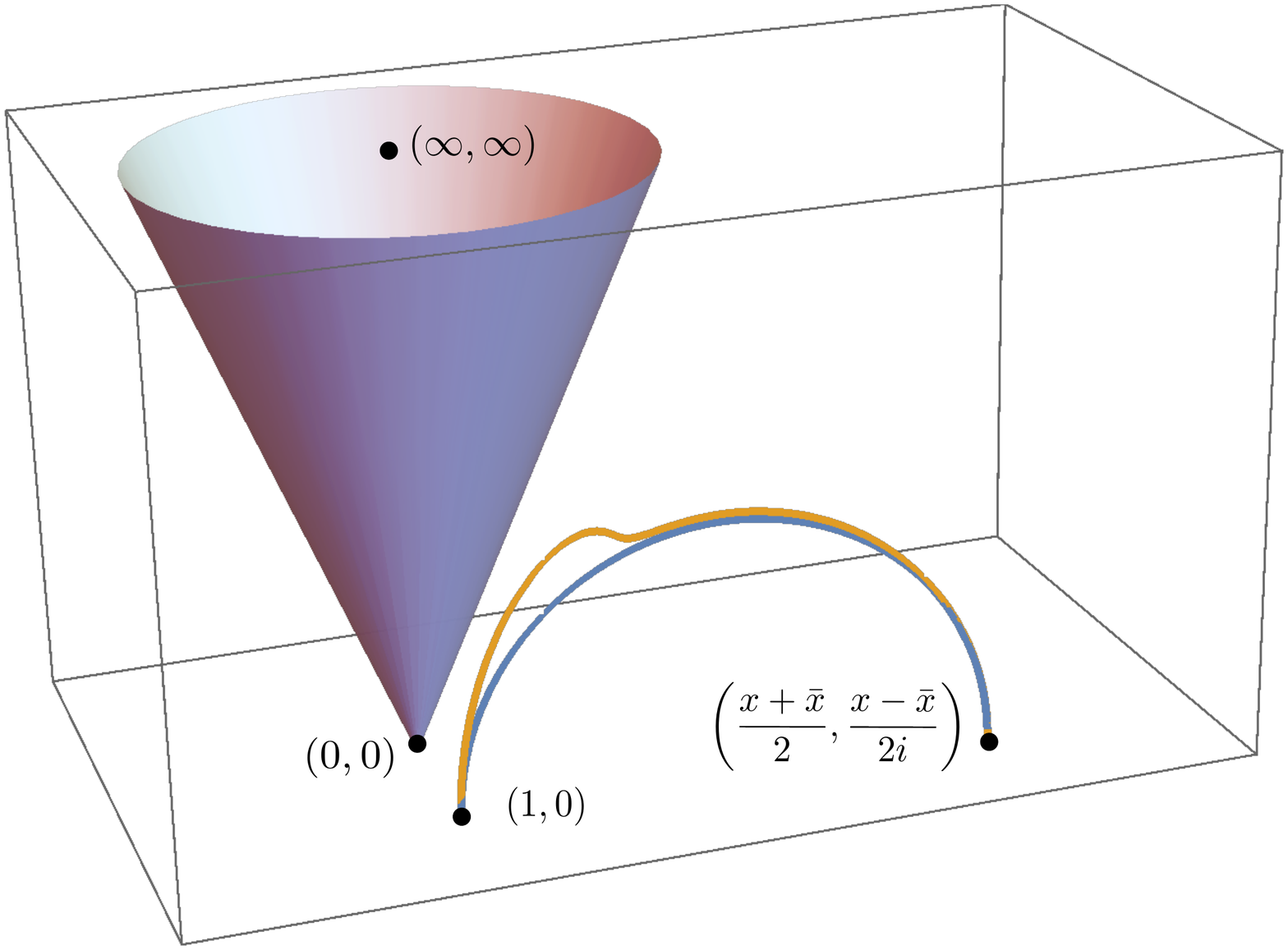}}
\end{center}
\vspace{-2.2cm}
\caption{Geodesic slightly deformed by a small black hole cone. 
The reduced four point function is the difference between the length 
of the geodesic without the cone (in blue) and the geodesic deformed 
by the cone (in orange). Here for concreteness we plotted a case 
where $(x,\bar x)=(7/6 + 7i/2,7/6 - 7i/2)$ and $M=1/5$. 
\label{deformedGeodesic}}
\end{figure}

We can also use geodesics to provide probes for a 
four-point function of two light particles of dimension $m$ in the presence 
of two maximally heavy operators of dimension $M$, \ie the black hole banana. 
This correlator is obtained by evaluating 
\beq
S(M)=m \int\!ds\, \sqrt{g_{\mu\nu}[M]\, \dot x^\mu(s)\, \dot x^{\nu}(s)} 
\eeq
where now the geodesic $x^\mu(s)$, differently from what we did above, 
has boundary conditions such that it approaches the two insertions 
points at the regulated boundary, namely 
$z=\epsilon$ for $\vec{x}_{\bullet1}$ and $\vec{x}_{\bullet2}$.
By conformal invariance we can arrange the external points as 
in figure \ref{deformedGeodesic}. 
Thus, the correlator is more precisely given by
\beq
\langle \phi_H(0) \phi_L(1) \phi_L(x) \phi_H(\infty) \rangle = e^{-(S(M)-S(0))}
\eeq
where the right-hand side is finite in $\epsilon$ after subtracting the AdS value.

The computation of $x^{\mu}(s)$ and the action $S$ 
are described in appendix \ref{geodesicAp}. 
In the $M\to 0 $ limit, the result simplifies significantly. 
For example, in AdS$_5$/CFT$_4$ we find
   \beq
S(M)-S(0) = 
\tfrac{1}{4} m M \times  (x-1) (\bar{x}-1) \times  \frac{h(x)-h(\bar x)}{x-\bar x} +O(M^2)\,,  \label{geodesic4pt} 
   \eeq
   where
\beq
 h(x)= \frac{6}{{x}-1}-\frac{\left( {x}^2+4  {x}+1\right) \log
   \left( {x}\right)}{\left( {x}-1\right)^2}
   \eeq
Quite nicely, the right hand side of \eqref{geodesic4pt}  
coincides \textit{precisely} with the t-channel conformal block 
for dimension $\Delta=4$ and spin $J=2$ in four spacetime 
dimensions,\footnote{The four-dimensional blocks read
\beq
\mathcal{F}(\Delta,J,x,\bar x)= \frac{x \bar x}{x-\bar x} ( 
h_{\frac{\Delta+l}{2}}(x) 
h_{\frac{\Delta-l-2}{2}}(\bar x) - ( x\leftrightarrow \bar x) ) \,, \qquad h_\lambda(x)=x^\lambda 
{}_2F_1(\lambda,\lambda,2\lambda,x)\,. 
\eeq}  
\beq
\log \frac{\langle \phi_H(0) \phi_L(1) \phi_L(x) \phi_H(\infty) \rangle}{\langle \phi_H(\infty) \phi_H(0)\rangle \langle \phi_L(x) \phi_L(1) \rangle}= \frac{m M}{120} \mathcal{F}(4,2,1-x,1-\bar x)+ O(M^2)
\label{block}
\eeq 
This is the expected result for a graviton exchange 
dual to the stress-tensor! 

In the AdS$_3$, the action can be computed without 
expanding at small $M$ and the result reads
\beq
S(M) - S(0) = -m \log\left( (1-M) \frac{z (1-w)^{\frac{1}{2} - \frac{1}{2 \sqrt{1-M}}} \bar{z} (1-\bar{w})^{\frac{1}{2} - \frac{1}{2 \sqrt{1-M}}}}{w \bar{w}} \right)
\eeq
where $1-w = (1-z)^{\sqrt{1-M}}$.
This is nothing but the (logarithm of) the semiclassical 
heavy-light limit of the Virasoro block, as discussed 
in \cite{Fitzpatrick:2015zha}. More details are given  in appendix~\ref{geodesicAp}.

Similar conclusions can be obtained in other dimensions. 
Our computation here fits very well with  
a more general discussion about Witten diagrams and 
geodesics given in \cite{Hijano:2015zsa}.

%====================================================================================
\section{Discussion}\label{discuss_sect}
%====================================================================================

Euclidean correlation functions of very heavy operators have so 
far been largely unexplored in AdS/CFT.
These are dual to full-fledged backreacted geometries with 
disturbances of the metric running all the way to marked points 
on the Poincar\'e boundary of AdS. Each one of such points 
corresponds to the insertion of a huge operator which spreads out in the bulk and
changes the AdS geometry in some way.

\begin{figure}[t]
\begin{center}
{\includegraphics[scale=0.58]{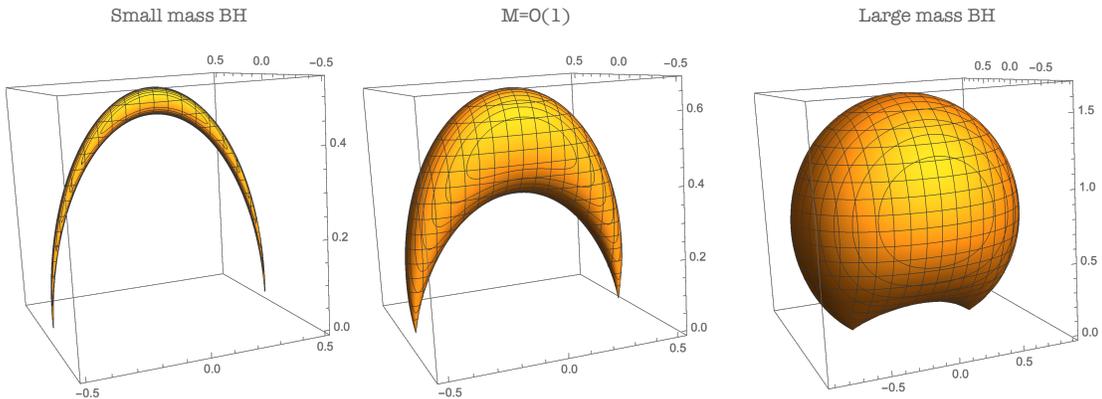}}
\end{center}
\vspace{-7cm}
\caption{Thin bananas are effectively geodesics. Fat bananas occupy a lot of space.}
\label{StepsBH}
\end{figure}

In this paper, we studied two-point function geometries, and showed 
that the geometry is nicely understood in terms 
of a banana foliation, with the tips of the bananas anchored at the insertion points.
For smooth geometries, we can shrink the bananas to the geodesic 
connecting the insertion points. For black holes instead, we can 
follow the bananas but only up to the location of the stretched horizon.
When the black hole is super light, this stretched horizon 
is approximately occupying the space of a geodesic, on the other hand, 
when the black hole is very massive it occupies a lot of the AdS space. 
In all cases, we demonstrated that the onshell action of these two-point black holes
reproduces the CFT result for two-point functions \eqref{start}. 
In particular, the renormalized onshell action has nontrivial 
spacetime dependence, coming from a logarithmic form at the marked points, 
and matches the dependence on the dimension $\Delta=M$ as a consequence 
of a crucial interplay between the boundary contribution at the AdS 
cut-off, and the boundary term  at the stretched horizon.

For holographic two-point functions in AdS, we believe we have unveiled a pretty complete picture. 
The bananas are special foliation of global AdS, which we obtained from the GtP map, and
therefore this map provides a solution generating technique
from global AdS geometries to two-point function geometries. 
It applies to all consistent truncations ansatz in string theory \cite{USthree}, 
thus AdS bubbles \cite{Chong:2004ce,Liu:2007xj}, charged clouds \cite{Bobev:2010de},
and more generally, electric solutions of ${\cal N}=2$ gauged SUGRA with spherically symmetric matter distributions. 
But there are also gravity solutions which are not consistent 
truncation ansatz, with interesting topology changes. A well known example 
are the Lin-Lunin-Maldacena geometries \cite{Lin:2004nb} describing all
maximally heavy half-BPS operators in $\mathcal{N}=4$ super Yang-Mills. 
%See also \cite{Skenderis:2007yb}.
% 
It would be interesting to generalize our results to this case too, 
especially in the light of the 
crucial interplay between bulk and boundary 
contributions that yield in the end the correct two-point function.

Our main motivation for this work, though, was to set up the formalism 
in general, and apply it to higher point correlation functions. 
Something preliminary we can say about these is the following. We 
can always solve  the FG form of a given multipoint geometry in a 
series expansion, once we know the expectation value of the stress-tensor. 
For example, for three-operators we would find,  
\begin{figure}[t]
\begin{center}
{\includegraphics[scale=0.5]{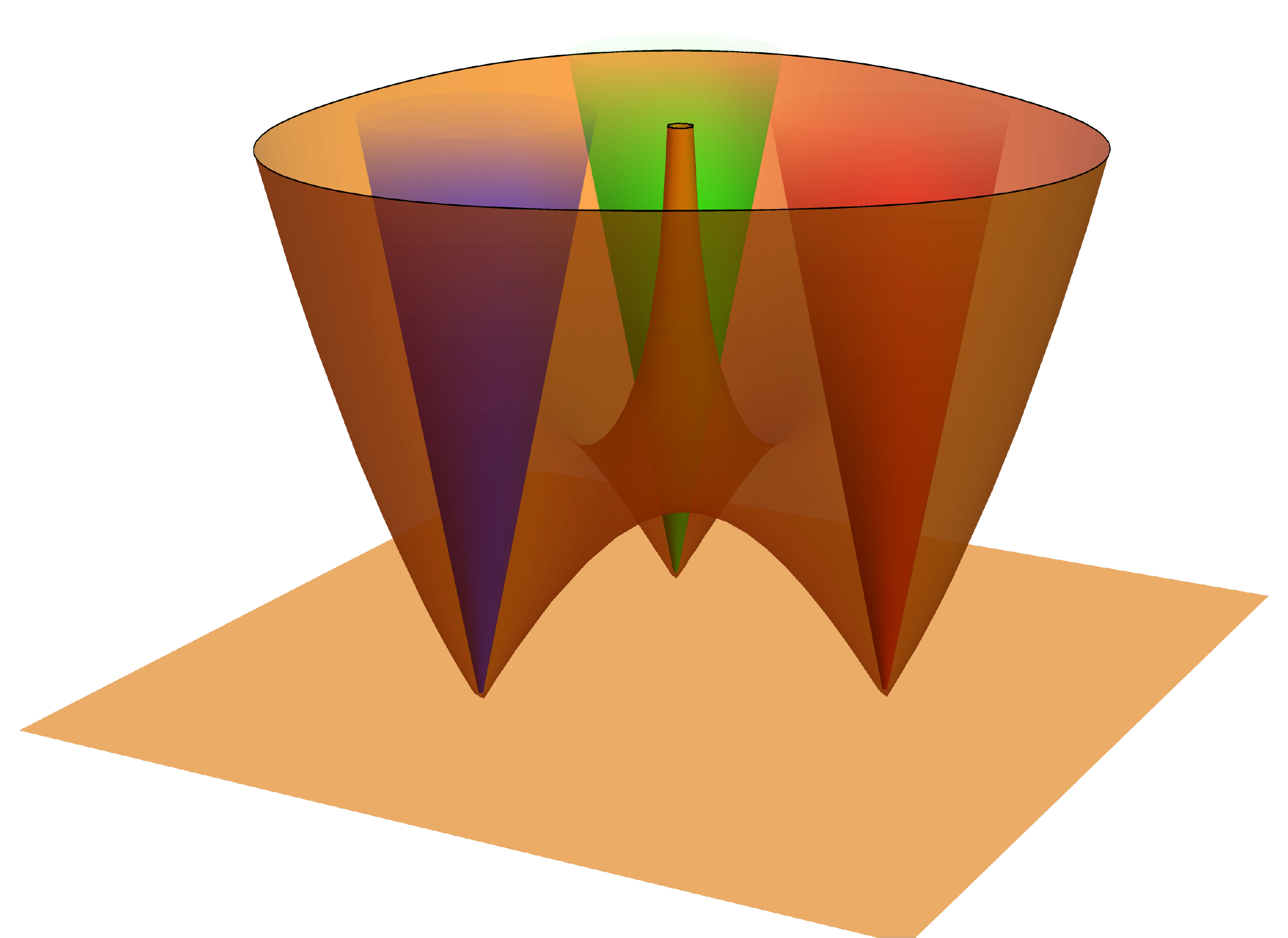}}
\end{center}
\vspace{-0.6cm}
\caption{The $\det(g_\texttt{Banados}) = 0$ surface in the three 
point function case, with $d=2$. The horizons (depicted in RGB), 
begin as cones. What fate awaits them far beyond the wall?
\label{walls}}
\end{figure}
\beq
ds^2\Big|_{FG}=\frac{d{\bf z}^2 +d\vec{\bf x}^2}{{\bf z}^2} + \ldots + {\bf z}^d \frac{\langle T_{ij}({\bf x})O_{\Delta}({\bf x}_1) O_{\Delta}({\bf x}_2)O_{\Delta}({\bf x}_3)\rangle}{\langle O_{\Delta}({\bf x}_1) O_{\Delta}({\bf x}_2)O_{\Delta}({\bf x}_3)\rangle} + \ldots
\eeq
with the higher order contributions (in ${\bf z}$) determined by Einstein's equations.
However, as we showed in this paper, already for the two points  
the black hole geometry is a completion of the FG patch, 
and in particular the horizon lies beyond the wall where $\det g_{FG}=0$.
Once again, the information from the stretched horizon was crucial in order 
to recover the CFT result as function of $\Delta=M$. So, here 
comes the question. For three- and higher point functions,  
what lies behind this wall?

Do Euclidean three- and higher point functions of huge operators 
have anything to do with physics of black holes merging? 
How sensitive are these geometries  
to the microscopic details of the operators? How would firewalls or fuzzballs manifest themselves?  Do averages play any role in 
these correlators in lower dimensions? What about in higher dimensions? 

This would be a good point to stop, but we cannot resist ourselves from adding two more small comments.

%======================================================================
\subsection{Comment 1: Three Dimensions}
%======================================================================

Three dimensions is a great laboratory for developing intuition.
In fact, in three dimensions, the exact FG metric was 
found by Ba\~nados \cite{Banados:1998gg} and simply reads
\beq
ds^2_{} = %\frac{dz^2+dx d\bar{x}}{z^2} 
dAdS_3+ T({\bf x}) d{\bf x}^2 + \bar{T}(\bar{\bf x}) d\bar{\bf x}^2 
+ {\bf z}^2 T({\bf x}) \bar{T}(\bar{\bf x}) d{\bf x} d\bar{\bf x} \label{Banados}
\eeq
where $T$ ($\bar T$) are the (anti)holomorphic boundary stress tensor. 
For example, for a scalar three-point function, we have 
\beq
T({\bf x}) = -\frac{1}{({\bf x}-x_1)({\bf x}-x_2)({\bf x}-x_3)} \sum_{i\neq j \neq k}\frac{M_i x_{ij} x_{ik}}{4({\bf x}-x_i)}\qquad ;\qquad \bar T(\bar {\bf x})=T(\bar {\bf x})
\eeq
With this assignment, $ds^2_{}$ in eq.~\eqref{Banados}
describes the three-point function geometry for three black holes 
in AdS$_3$/CFT$_2$ of masses $M_1,M_2,M_3$ inserted at locations $x_1,x_2,x_3$. 
As in the two-point function case, this metric has a wall beyond 
which we need better coordinates to figure out what the geometry does. 
This wall is much richer than in the two-point function case -- see figure \ref{walls}.
What lies behind this wall? 
What is the full three-point function geometry and what is the 
corresponding structure constant in the dual field theory? 
This challenge is the subject of our companion paper \cite{USthree}.

%==============================================================================
\subsection{Comment 2: LLM Three-Point Functions}
%=============================================================================

What about higher dimensions? What about three-point 
function geometries in ${\cal N}=4$ SYM? -- or other maximally symmetric theories in various dimensions? 

In global AdS$_5\times$S$^5$ the most 
general half-BPS geometries are LLM geometries \cite{Lin:2004nb} (see also 
\cite{Skenderis:2007yb} for a nice review and for the holographic renormalization 
aspects of these geometries).
In ${\cal N}=4$ SYM, they correspond to operators in the Schur basis \cite{Corley:2001zk} of the form
\beq\label{Schur_operators}
\mathcal{O}\propto \chi_{\underline{\lambda}}(y \cdot \phi)
\eeq
with Young diagram $\underline{\lambda}$ of $O(N^2)$ boxes.  
The vector $y$ is a six dimensional null vector picking up a particular combination 
of the six scalars $\phi^I$ of the gauge theory. 
While it would be fascinating to construct 
general three-point functions of the operators \eqref{Schur_operators} from group theory alone, 
one might start wondering if are there any hints or expectations that would help our intuition?
Well, as we discuss below, there are some.

A three-point function, 
\begin{figure}[t]
\begin{center}
{\includegraphics[scale=0.45]{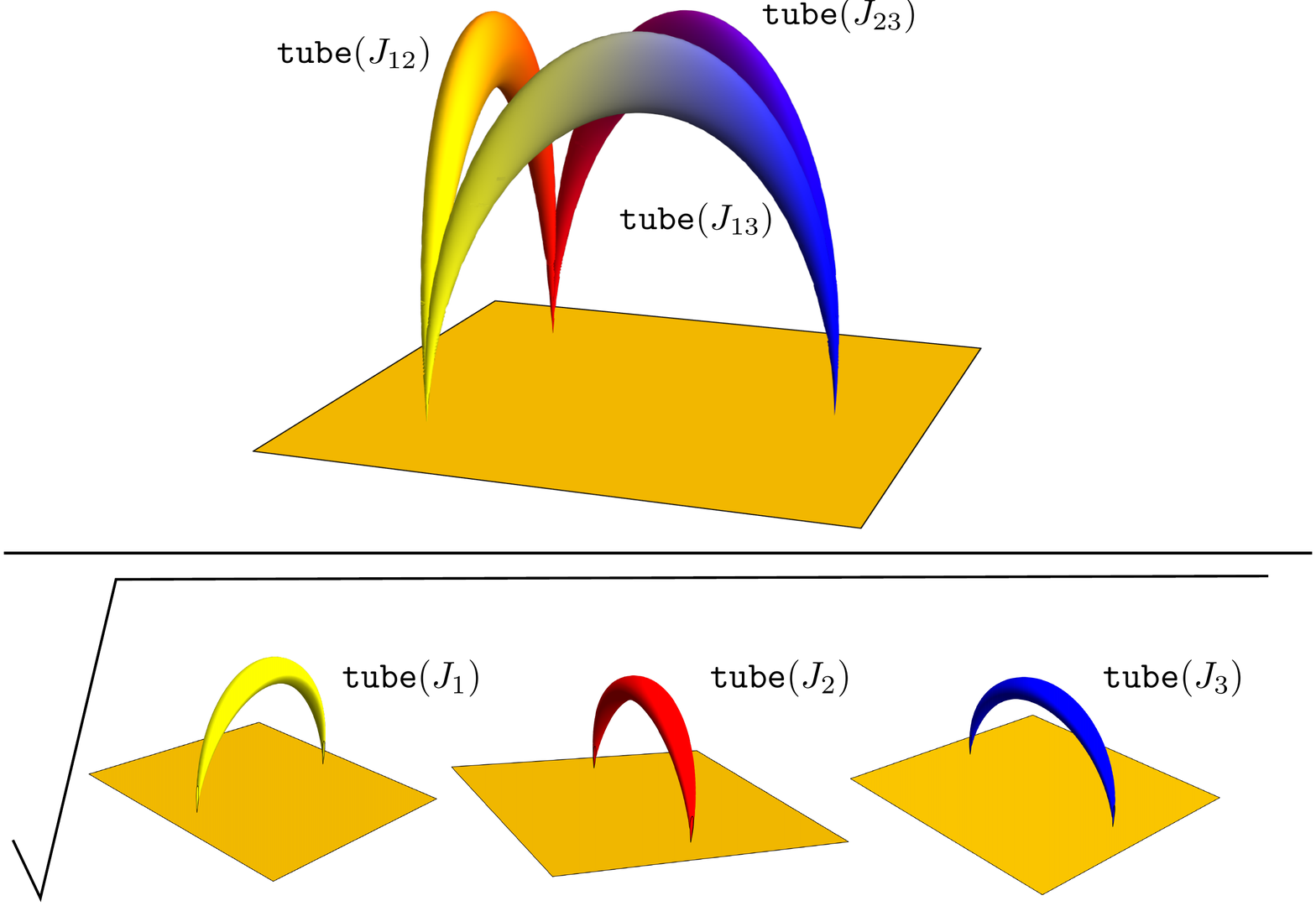}}
\end{center}
\vspace{-1cm}
\caption{A suggestive explanation for a result like \eqref{Ctubes}, 
where the geometry is foliated by conifolds shooting out of 
the insertion points like pairwise two-point function geometries.} 
\label{CtubesFig}
\end{figure}
\beq
\< \mathcal{O}_1 \mathcal{O}_2 \mathcal{O}_3 \> = \prod_{i<j} \Big(\frac{y_i\cdot y_j}{(x_i-x_j)^2}\Big)^{J_{ij}} C_{\underline{\lambda}_1,\underline{\lambda}_2,\underline{\lambda}_3} \,,
\eeq
depends on a fixed kinematical factor, built out of the $J_{ij}$, 
\ie  the number of propagators between operators $i$ and $j$,
\beq
%J_{31} =\frac{J_1+J_3-J_2}{2}\,,\qquad  J_{12}= \frac{J_1+J_2-J_3}{2}\, , \qquad  J_{23}= \frac{J_3+J_2-J_1}{2}   \,.
J_{12}= \frac{J_1+J_2-J_3}{2}\, ,\qquad J_{13} =\frac{J_1+J_3-J_2}{2}\,,\qquad    J_{23}= \frac{J_3+J_2-J_1}{2}   \,,
\eeq 
and by the structure constant $C$, which here is coupling independent, 
and therefore it can be computed in free $\mathcal{N}=4$ SYM by Wick contractions. 
Computing $C$ is thus a well-posed but not trivial combinatorial problem 
for any triplet $\underline{\lambda}_{i=1,2,3}$.
It  would indeed be great to solve this problem completely, or at least for 
operators with very large Young diagrams.\footnote{Everything we discuss 
in this section should have a nice embedding in twisted holography 
\cite{twistedHolography} which would be fascinating to work out.}--
In a simple instance, \ie when the operators are fully symmetric labelled 
by a Young diagram with a single row of $J$ boxes, we did manage to find the solution (see appendix \ref{combinatorics} for details)
\beq
\!\!\!\!\tilde{C}^\texttt{\,symmetric}_{{J_1} {J_2}{J_3} } \equiv \frac{
 _3F_2(-J_{12},-J_{13},-J_{23};1,1{- N}-\frac{J_1+J_2+J_3}{2} ;1)({ N})_{\frac{J_1+J_2+J_3}{2}} 
}{  (-)^{\frac{J_1+J_2+J_3}{2} } \sqrt{(- N-J_1+1)_{J_1} (- N-J_2+1)_{J_2} (- N-J_3+1)_{J_3}}}  \la{symC} \,.
\eeq
The $\tilde{C}$ is normalised by two-point functions, and the hypergeomtric function $\,_3F_2$  
accounts for genuine interactions among the three operators; 
when the $J_{i=1,2,3}\in\mathbb{N}$ it evaluates to a 
non trivial polynomial of $N$. 
In the limit when $J_i=N^2 j_i$
the result simplifies to:
\beq
%\log C^\text{symmetric}_{\chi_{J_1} \chi_{J_2} \chi_{J_3} } 
% \simeq -N^2  \texttt{Action} \qquad;\qquad  
 %\texttt{Action} =
\log \tilde{C}^\texttt{symmetric}_{{J_1}{J_2}{J_3} } \simeq 
-N^2 \Big[\sum_{ij} \texttt{tube}(j_{ij})-\tfrac{1}{2}\sum_{i} \texttt{tube}(j_{i})   \Big] \qquad{\rm with}\quad \texttt{tube}(j)=j \log j\,.
 \label{Ctubes}
\eeq
Even though $\tilde{C}^\texttt{\,symmetric}_{{J_1}{J_2}{J_3} }$ 
is telling us something only about the limit of a regular geometry, 
where each individual operator is dual to a thin ring that is 
taken off to infinity with respect to the AdS$_5\times$S$^5$ 
droplet \cite{Skenderis:2007yb}, the expression we got is quite suggestive.

In fact, we can rewrite the asymptotics as
\beq
\lim_{N\gg 1}\underbrace{ 
            \frac{ J_1! J_2! J_3!}{ (\frac{ J_1+J_2-J_3}{2} )! (\frac{ J_1+J_3-J_2}{2} )! (\frac{ J_2+J_3-J_1}{2} )! }  
\times   \frac{1}{\sqrt{  J_1! J_2! J_3! }  }} _{\ \ \ \ \#{\tt\, Wick\ contractions\ at\ 3pt }\ \ \ \times\ \ \ {\tt  2pt\ normalisation} }\ \simeq\ e^{N^2  \left[ \frac{1}{2}\sum_{i}  j_i \log j_i - \sum_{ij} j_{ij} \log j_{ij}) \right] } 
\eeq
showing that it coincides with the leading asymptotics of 
the number of Wick contractions for three-points,
when $J_i=N^2 j_i$, divided by two-point normalizations.
In other words, we get the a purely combinatorial result.
Note now that a LLM geometry generically involves multi-particle states 
and that the ${\tt tube}$ function is the building block for the two-point function normalization. So what if when $J\gg N$ the operator ${\cal O}$ breaks apart and the bulk 
geometry is foliated as in figure \ref{CtubesFig}? with 
conifold geometries emerging from the insertion points?

\begin{figure}[t]
\begin{center}
{\includegraphics[scale=0.4]{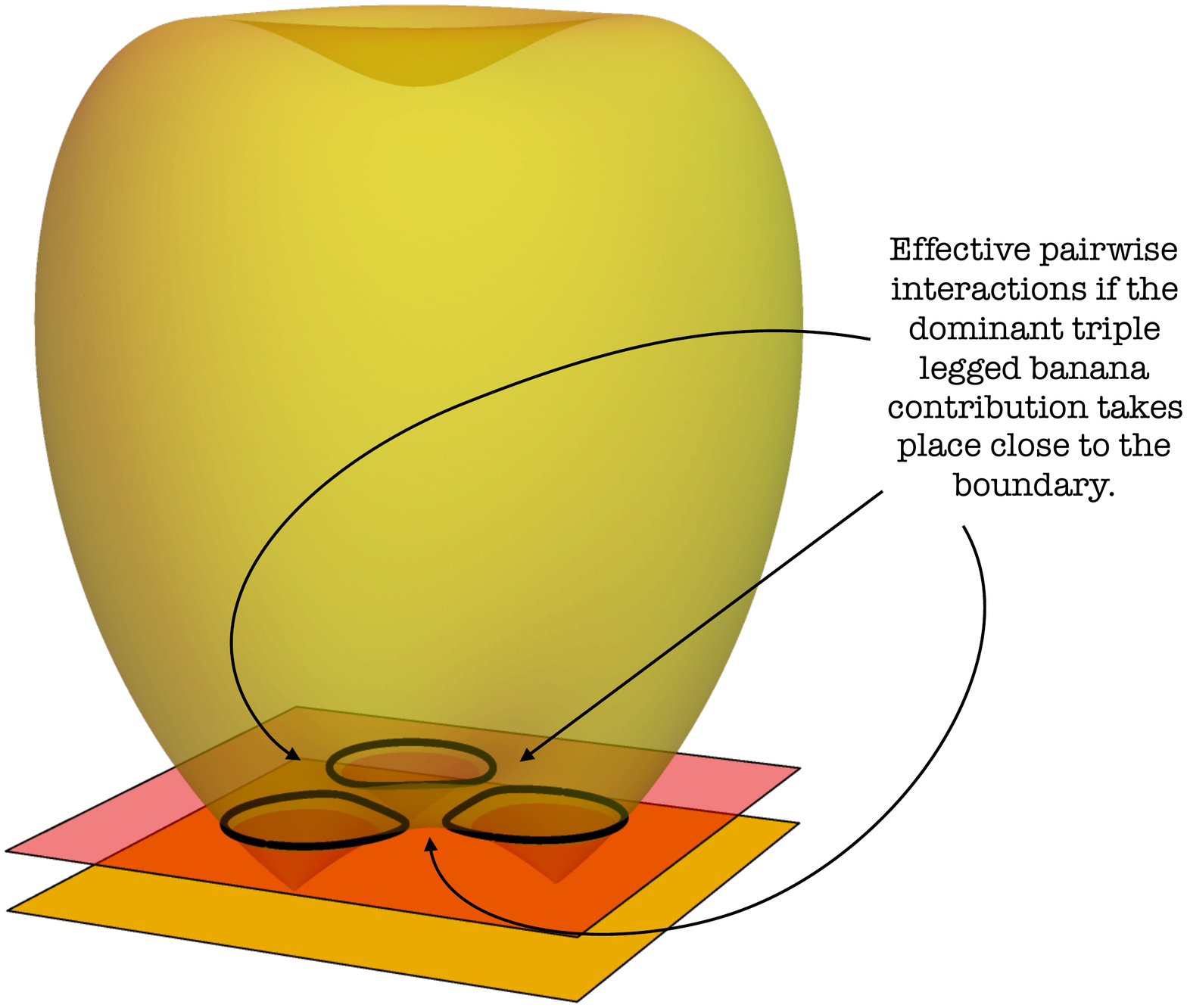}}
\end{center}
\vspace{-0.7cm}
\caption{Another scenario where operators in the bulk merge, 
but the dominant contribution to the action is effectively pairwise because the geometry change is huge.}  
\label{scenario}
\end{figure}

Another scenario, which was inspired to us by a beautiful talk of 
David Simmons-Duffin at KITP this January, is represented 
in figure \ref{scenario} -- see \cite{Benjamin:2023qsc}. 
There we imagine that each operator shoots from the boundary 
a huge geometry change, like the fat bananas in the figure,
so that interactions among the three operators are 
effectively pairwise. This scenario could also reproduce 
something like eq.~\eqref{Ctubes}.

We look forward to continuing to explore huge correlators, from the boundary, as well as from the bulk.

\section*{Acknowledgments} 
We thank Stefano Baiguera, Scott Collier, Sergei Dubovsky, 
Tom Hartman, Davide Gaiotto, Luis Lehner, Juan Maldacena, 
Don Marolf, Leopoldo Pando-Zayas, Joao Penedones, Eric Perlmutter for useful 
comments and discussions. FA also thanks Kostas Skenderis, 
Marika Taylor and David Turton for discussions.
Research at Perimeter Institute is supported in part by 
the Government of Canada through the Department of Innovation, Science, and Economic
Development Canada and by the Province of Ontario through the Ministry of Colleges
and Universities. 
RCM  and PV are supported in part by Discovery Grants from 
the Natural Sciences and Engineering Research Council of Canada, 
and by the Simons Foundation through the ``It from Qubit'' 
and the ``Nonperturbative Bootstrap'' collaborations, respectively  (PV: \#488661).
This work was additionally supported by   
FAPESP Foundation through the grants 2016/01343-7, 2017/03303-1, 2020/16337-8.
JA, FA and PV thank the organizers of the 
conference ``Gravity from Algebra: Modern Field Theory 
Methods for Holography", and acknowledge  KITP for hospitality. 
Research at KITP is supported 
in part by the National Science Foundation under Grant No. NSF PHY-1748958.
FA is supported by the Ramon y Cajal program through the fellowship RYC2021-031627-I funded 
by MCIN/AEI/10.13039/501100011033 and by the European Union NextGenerationEU/PRTR.

%Simons Foundation (PV: \#488661) \rcm{Perhaps drop the reference to Simons given the previous sentence?}

\appendix

%==============================================================================
\section{Euclidean Rotation}
\label{AdSunitary_app}
%==============================================================================

In this section we describe the GtP map with marked points $0$ and $\infty$ as a 
certain \emph{rotation} in embedding coordinates. This rotation is precisely the one 
used in \cite{Janik:2010gc} to map spinning strings in global AdS
to Euclidean strings in Poincar\'e AdS with two-point function boundary conditions. 
Here we will promote it to a local change of coordinates.

In embedding coordinates,\footnote{We 
use $-X_{-1}^2-X_0^2 +\sum_i X_i^2=-1$ for Lorentzian and 
$-X_{-1}+X_0^2+\sum_i X_i^2=-1$ for Euclidean.}  AdS admits the following standard parametrisations
\beq
\label{coord_ads_emb}
\begin{array}{l}
  \rule{.5cm}{0pt}  {\tt global\ Lorentzian}  \rule{2cm}{0pt}  {\tt Euclidean\ Poincare} \\[.2cm]
  \hline
\begin{array}{rl c|c cl}
  & & & & & \\
    X_{i}&=r \ \Omega_i\,,\ \ i\ge1& \quad & \quad  &  X_{i+1}&=x^i/\tz\,,\ \ {i\ge 1} \\[.1cm]
                       &    \,  \vdots           & \quad & \quad  & \ \ X_1&=\frac{1}{2\tz}(1 - R^2 - \tz^2)\\[.1cm]
X_0&=\sqrt{1+r^2}\,  \sin t & \quad & \quad  &  \ \ X_0&=x^0/\tz \\[.1cm]
X_{-1}&=\sqrt{1+r^2}\, \cos t & \quad & \quad  & \ \ X_{-1}&=\frac{1}{2\tz}(1 + R^2 + \tz^2)
\end{array}
\end{array} 
%\\
%\end{array}
%\end{array}    & \qquad;\qquad &
%\begin{array}{c}
%{\tt poincare}\\[.2cm]
%\begin{array}{cl}
%X_{i+1}&=\frac{1}{z} x_{i\ge 1} \\[.1cm]
%X_1&=\frac{z}{2}(-1 +\frac{1}{z^2} - \frac{R^2}{z^2}) \\[.1cm]
%X_0&=\frac{i}{z} x_0\\[.1cm]
%X_{-1}&=\frac{z}{2}(+1 +\frac{1}{z^2} + \frac{R^2}{z^2}) 
%\end{array}
%\end{array}
%\end{array}
\eeq
where $\Omega_i$ are spherical coordinates, and $R^2= \sum_{i=0}^{d-1}(x^i)^2$.
Black holes will live in $D=d+1$ dimensions, with $d$ being the spacetime dimension of the AdS boundary.

\begin{figure}[t]
\begin{center}
{\includegraphics[scale=0.58]{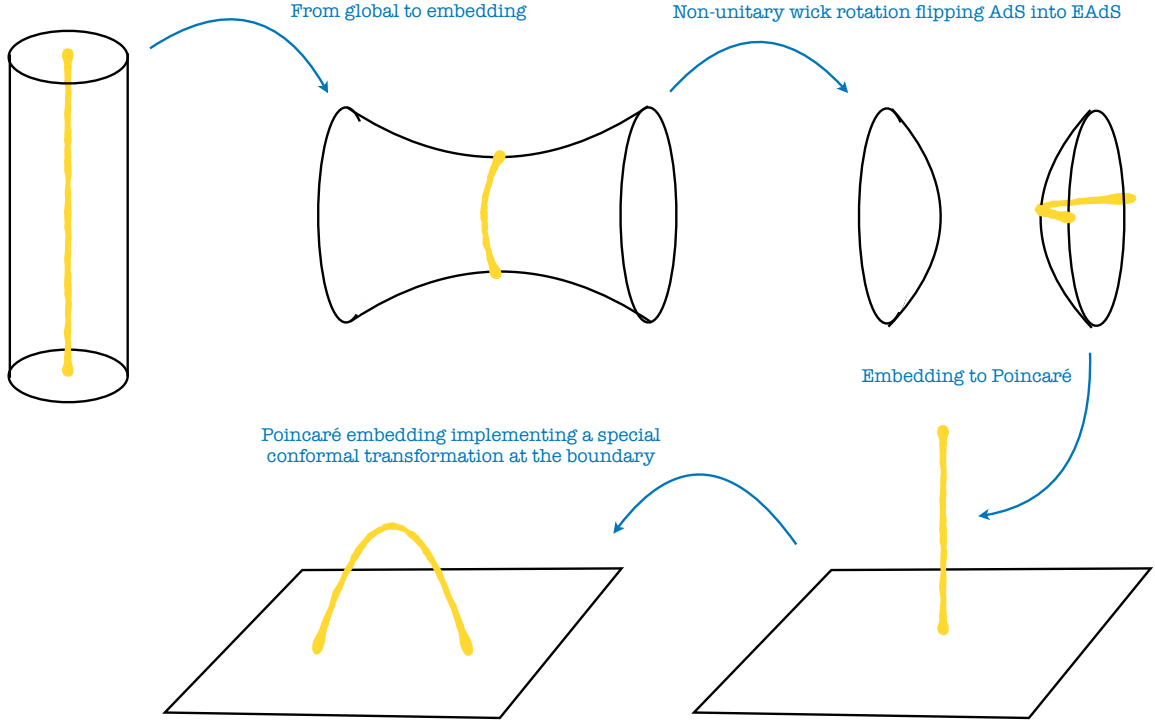}}
\end{center}
\vspace{-2.5cm}
\caption{Schematic for the sequence of transformations from {\tt global} AdS with Lorentzian signature to the euclidean Poincar\'e patch. Starting with a tubular object in the middle of~AdS, 
we end up with a banana shape with two insertion points at the boundary of AdS. The direct transformation from the top left to the bottom right pictures is eq.~(\ref{poincareCoords}). 
%Inside global AdS the geometry is deformed close to the horizon of the black hole, i.e.~the yellow thick line.  
%The black hole horizon is localized inside this tube; this is what we call the \texttt{tube}.  
\label{stepsBH} } %Away from it we see pure AdS.
\end{figure}
Recall the Schwarzschild black hole solution in global with Lorentzian signature is,
\beq\label{global_start}
{\tt global}\quad ;\quad 
%ds_D^2= \frac{ dr^2 }{ \Big(1+r^2-F[r]\Big)} -\underbrace{ \Big(1+r^2-F[r]\Big)}_{\equiv\, g_{\tau\tau} } dt^2 + r^2 d\Theta^2_{D-2} \quad;\quad F[r]=\frac{M}{r^{(D-3)}}%\qquad\qquad\qquad   \texttt{global} \
ds_D^2= \frac{ dr^2 }{ g_{\tau\tau}(r) } - g_{\tau\tau}(r) dt^2 + r^2 d\Omega^2_{D-2} 
\eeq
We will change signature by considering $\tau=-i t$.
Our starting point is then the following relations
\beq\label{differentialsglobal}
\begin{array}{l} 
%\displaystyle 
dt= \frac{X_{-1}\, dX_0 -X_0\, dX_{-1} }{1+r^2} \\[.3cm]
%\displaystyle
dr = \frac{ X_0\, dX_0 +X_{-1}\, dX_{-1} }{r } \\[.3cm]
%\displaystyle 
d\Omega_{D-2}^2=-\frac{dr^2}{r^2}+ \tfrac{1}{r^2} \sum_{i\ge 1} dX^2_i
\end{array} 
\qquad;\qquad  r=\sqrt{ X_{-1}^2+X_0^2-1} 
\eeq
which we can use to rewrite eq.~\eqref{global_start}. 
Then, the rotation of \cite{Janik:2010gc} into Euclidean 
AdS is $X_0\rightarrow i X_1$ and $X_1\rightarrow X_0$. 
Once we perform this rotation on eq.~\eqref{differentialsglobal}, we will 
rewrite the result by the using Euclidean Poincar\'e patch, \ie the second column of table (\ref{coord_ads_emb}). 
This gives the change of coordinates in the differential form, 
\beq
\begin{array}{rl}
\left[\begin{array}{c} d\tau \\[.2cm] dr \end{array}\right] =& 
\frac{1}{\tz} \left[ \begin{array}{cc}  \frac{R\,\tz}{\tz^2+R^2} &  \frac{  \tz^2}{\tz^2+R^2} \\[.2cm] 1 &- \frac{R}{\tz} \end{array} \right] 
\left[ \begin{array}{c} dR \\[.2cm] d\tz \end{array}\right]
 \end{array}
\eeq
where $R,\Omega_i$ are polar coordinates in Poincar\'e AdS. As it turns our the $d\Omega$ are invariant, since 
in fact we are preserving rotational symmetry. Upon integration, we find
\beq%\label{poincareCoords}
\tau = \frac{1}{2} \log(\tz^2+R^2)\qquad{\rm and}\qquad
r = \frac{R}{\tz}\,,
\eeq
which is precisely the AdS-unitary transformation in eq.~\eqref{poincareCoords}.

%========================================================================

\section{Geodesics in the banana background} \label{geodesicAp}

%========================================================================

In this appendix, we show how to solve the problem of a geodesic the background of a two-point function black hole, and for concreteness we will
%\ref{chargesSec} and \ref{expansionSec} and \ref{radialGeo} 
specialize to AdS$_5$ and AdS$_3$.
%for which the dual is four dimensional. In section \ref{ads3Geo} we focus on three dimensional gravity with a two dimensional CFT dual. 
Other odd bulk dimensions (\ie even boundary dimensions) can be analyzed 
with the same tools that we provide, but expressions become more complicated. We are working with the cone metric eq.~\eqref{cone_metric}.
%This is not surprising: we know that conformal blocks can only be writen in terms of simple elementary functions in even dimensional conformal field theories.  

\subsection{AdS$_5$ geodesics}
%\subsection{Two charges solve the problem}  
\la{chargesSec}

We can always put the four external points on a two-dimensional plane, 
through a conformal transformation, therefore the geodesic can be restricted 
to explore a three-dimensional space, the plane plus the holographic direction. 
We have $\tz(s)$, $R(s)$ and $\phi(s)$ or equivalently $\eta \equiv \tz/R$, $\rho\equiv\log(R)$ and $\phi$.
The latter will simplify the problem since they are essentially the same as in global coordinates \eqref{globalE}.
%{\color{red} It makes sense to use $\eta^2$.}

Now, the cone geometry \eqref{cone_metric} is invariant under dilatations $(R,\tz)\to \lambda (R,\tz)$ and $SO(d-1)$ rotations. 
For the three-dimensional problem, we have two conserved charges for geodesics which we denote as $\mathcal{E}$ and $\mathcal{J}$. By using these charges, we obtain first order equations
\beqa
&&\!\!\!\!\!\!\!\!\!\!\!\!\!\!\!\!\!\! \phi' = { \pm}\frac{\mathcal{J} \eta \eta'}{\sqrt{\mathcal{J}^2 M \eta^6-\left(\mathcal{J}^2+M\right) \eta^4-\eta^2
   \left(\mathcal{J}^2+\mathcal{E}^2-1\right)+1}} \la{phiDot}\\
&&\!\!\!\!\!\!\!\!\!\!\!\!\!\!\!\!\!\!\rho'=   \pm \frac{\mathcal{E}\eta\eta'}{\left(1+\eta^2-M \eta^4\right) \sqrt{\mathcal{J}^2 M \eta^6-\left(\mathcal{J}^2+M\right) \eta^4-\eta^2 \left(\mathcal{J}^2+\mathcal{E}^2-1\right)+1}}-\frac{\eta \eta'}{\eta^2+1} \la{rDot}
\eeqa
In total the solution is parametrized by four integration constants, ${\cal E},{\cal J}$ and $C_\rho,C_\phi$ from these first order equations.

It is straightforward to solve eq.~\eqref{rDot} for empty AdS with ${\cal J}=0$. This gives the semicircle in the form, 
\beq
\big[ R(\eta)-R_0\big]^2+\big[ \tz=\eta R(\eta)\big]^2=L^2\qquad;\qquad 
\begin{array}{c} R=\frac{C}{{\cal E}\pm \sqrt{1-({\cal E}^2-1) \eta^2 }}\\[.5cm] {\cal E}=\frac{R_0}{L}\quad;\quad C=\frac{R_0^2-L^2}{L} \end{array}
\eeq
The signs $\pm$ provide a parametrization for the two branches of the semicircle, each one going from one insertion point at $z=0\rightarrow\eta=0$ 
up to the turning point $\eta_\star$, where the derivatives blow up. This is given by the solution of $1-({\cal E}^2-1) \eta_{\star}^2=0$.
Note that since $\eta\ge0$, in our conventions ${\cal E}\ge 1$.  The limit where the insertion points are close to each other is the limit ${\cal E}\rightarrow\infty$, since $\eta_{\star}$ and thus $\tz_{\star}$ goes to zero. Note also the connection with global coordinates, 
since we can rewrite
\beq
\rho'+\frac{\eta \eta'}{\eta^2+1}=\frac{d}{ds}\frac{\log\big(z(s)^2+R(s)^2\big)}{2}=\tau'(s)
\eeq
Of course, $\phi$ in the cone is the same as in global coordinates.

In the black hole case, the polynomial in the square root is a cubic in $\eta^2$ and has three zeros. 
One (and only one of them) of them is finite for $M\to 0$ and ${\cal J}\to 0$. This one 
defines the turning point $\eta_*>0$.
By construction this turning point has a good limit to empty AdS.

%\begin{figure}[t]
%\begin{center}
%\includegraphics[scale=1]{figurelogx.pdf}
%\end{center}
%\caption{We plot the right hand side of \eqref{Deltar} 
%on the line $x=\bar x$, ${\cal J}=0$. In this configuration the result can be written in terms of elementary functions, 
%$\log(x)= H_+ +H_- $ where 
%$ H_\pm =\mp \frac{\cal E}{\sqrt{1+4M}} \frac{ 1}{ \sqrt{1- \left(\mathcal{E}^2-1\right)\eta^2_\mp-M \eta_{\mp}^4}} \times \%(\coth^{-1}\left[\frac{ \sqrt{1- \left(\mathcal{E}^2-1\right)\eta_{\mp}^2-M \eta_{\mp}^4} }{ 2 - ({\cal E}^2-1)\eta_{\mp}^2} \right] \pm %\frac{i}{2} \) 
%$
%and $2M \eta^2_{\pm}= 1\pm \sqrt{1+ 4M}$. For empty AdS, $\log(x)$ blows up at ${\cal E}=1$, 
%which implies $1\leq x\leq \infty$, \ie there is a geodesic 
%for any two insertion points. 
%For the black hole with $M>0$, $\log(x)$ reaches a maximum.  
%% If we think of populating empty AdS with geodesics anchored at two insertion points, since the membrane effectively removes a portion of %empty AdS, some geodesics necessarily will cease to exists. 
%}
% \label{logxplot}
%\end{figure}

\begin{figure}[t]
\begin{center}
\includegraphics[scale=1]{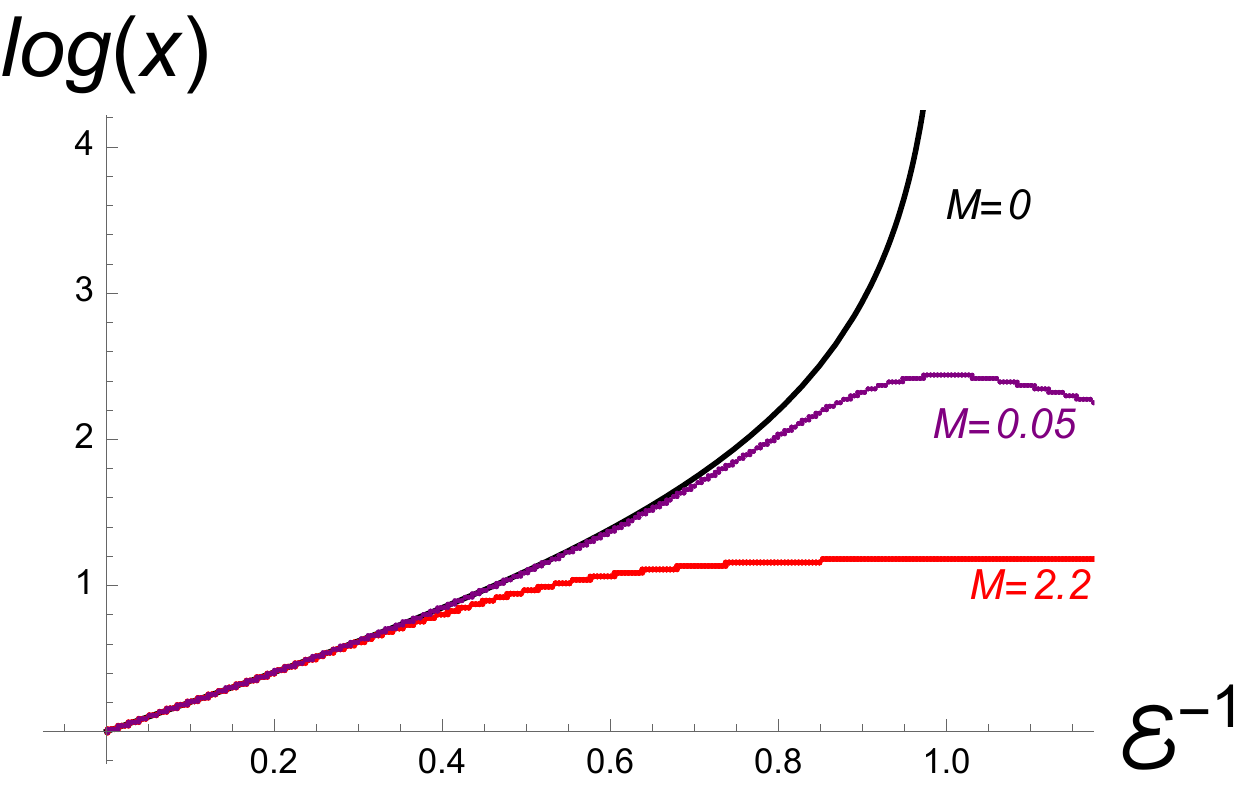}
\end{center}
\caption{We plot the right hand side of eq.~\eqref{Deltar} 
on the line $x=\bar x\ge 1$. 
The result can be written through elementary functions, and it reads \\[.2cm]
\begin{minipage}{\linewidth}
$ \log(x)= H_+ +H_-\quad;\quad
H_\pm =\mp \frac{\cal E}{\sqrt{1+4M}} \frac{ 1}{ \sqrt{1- \left(\mathcal{E}^2-1\right)\eta^2_\mp-M \eta_{\mp}^4}} \times \(\coth^{-1}\left[\frac{ \sqrt{1- \left(\mathcal{E}^2-1\right)\eta_{\mp}^2-M \eta_{\mp}^4} }{ 2 - ({\cal E}^2-1)\eta_{\mp}^2} \right] \pm \frac{i}{2} \) $
\end{minipage}\\[.2cm]
with $\eta^2_{\pm}= (1\pm \sqrt{1+ 4M})/(2M)$. For empty AdS, $\log(x)$ blows up at ${\cal E}=1$, 
which implies that there is a geodesic for any two insertion points, with no limits on their separation. 
For the black hole, $M>0$, the range of $\log(x)$ is bounded, thus real 
geodesics connect insertion points only up to some distance.  
% If we think of populating empty AdS with geodesics anchored at two insertion points, since the membrane effectively removes a portion of empty AdS, some geodesics necessarily will cease to exists. 
}
 \label{logxplot}
\end{figure}

%Derivatives can be rewritten as functions of ${\cal Z}$ rather than $s$. Then, by 
Eqs.~\eqref{phiDot} and \eqref{rDot} can be integrated to obtain the shifts. For the right-hand side, we sum the integrals from $0\leq \eta\leq \eta_\star$ for the first branch and the $\eta_\star\leq \eta\leq 0$ for the second branch. For the left-hand side, we then consider the external points as described in figure \ref{deformedGeodesic}. In sum we find
%
%\underbrace{\ \ \color{red}\log\sqrt{{x}/{\bar x} }\ \  }_{\Delta \phi}
%
\begin{align}
\!\!\!\!\!\!\!\underbrace{-i\log\sqrt{{x}/{\bar x} }}_{\Delta \phi} &= 2 \int\limits_0^{\eta_*}\!\! d\eta \frac{\mathcal{J} \eta }{\sqrt{\mathcal{J}^2 M \eta^6-\left(\mathcal{J}^2+M\right) \eta^4-\eta^2
   \left(\mathcal{J}^2+\mathcal{E}^2-1\right)+1}}    \label{Deltaphi} \\
\!\!\!\!\!\!\!\underbrace{\log \sqrt{ x \bar{x}}}_{\Delta \rho} &= 2 \int\limits_0^{\eta_*}\!\! d\eta  \frac{\mathcal{E} \eta }{\left(1+\eta^2-M \eta^4\right) \sqrt{\mathcal{J}^2 M \eta^6-\left(\mathcal{J}^2+M\right) \eta^4-\eta^2 \left(\mathcal{J}^2+\mathcal{E}^2-1\right)+1}}  \label{Deltar}
\end{align}
%
%\beqa
%&&\!\!\!\!\!\!\!\!\!\!\!\!\!\!\!\!\!\! \underbrace{\text{arccos} \frac{x+\bar{x}}{2 \sqrt{x \bar{x}}}}_{\Delta \phi} = 2 \int\limits_0^{\eta_*}\!\! d\eta \frac{\mathcal{J} \eta }{\sqrt{\mathcal{J}^2 M \eta^6-\left(\mathcal{J}^2+M\right) \eta^4-\eta^2
%   \left(\mathcal{J}^2+\mathcal{E}^2-1\right)+1}} \label{Deltaphi} \\
%   &&\!\!\!\!\!\!\!\!\!\!\!\!\!\!\!\!\!\! \underbrace{\log \sqrt{ x \bar{x}}}_{\Delta \rho} = 2 \int\limits_0^{\eta_*}\!\! d\eta  \frac{\mathcal{E} \eta }{\left(1+\eta^2-M \eta^4\right) \sqrt{\mathcal{J}^2 M \eta^6-\left(\mathcal{J}^2+M\right) \eta^4-\eta^2 \left(\mathcal{J}^2+\mathcal{E}^2-1\right)+1}}  \label{Deltar}
%\eeqa
These integrals express the physical insertion points $x,\bar{x}$ in terms of ${\cal E}$ and ${\cal J}$.
They can be computed but the result is written 
terms of elliptic functions so not very illuminating, and so we omit it.

A somewhat subtle point in the computation is that for some $x,\bar x$, a real solution to these equations will not exist, see figure \ref{logxplot}. Implicitly, we invert the relation~(\ref{Deltar}) in a finite domain where this inverse exists (with a real solution) and analytically continue from there to access the result for any $x,\bar x$.

%{\color{red} Where is the info that one boundary point is $1$, on the rhs?} This $\Delta r=r(Z=Z_{\star})-r(Z=0)$.
%Now $r(Z=0)$ is my insertion point, it coud be $x,\bar{x}$ or $1$. how do we distinguish?\\

%It is useful to consider that for $M={\cal J}=0$, the turning point is
%$\eta^2_\star=\frac{1}{{\cal E}^2-1}$, thus we should require ${\cal E}\ge 1$ in our conventions. The limit where the insertion points are close to each other is the limit ${\cal E}\rightarrow\infty$, since $\eta_{\star}$ and thus $z_{\star}$ goes to zero.

%The full geodesic is composed of two branches: It goes up from the boundary at $\mathcal{Z}=0$ to $\mathcal{Z}=\mathcal{Z}_*$ and then down to $\mathcal{Z}=0$. (Well, not exactly  $\mathcal{Z}=0$ since we should regulate the computation and start and end the geodesic at a distance $\epsilon$ from the boundary; we will put those epsilons in below when needed.) On the other hand, 
%These integrals can be computed but the result is in terms of elliptic functions so not very illuminating. 
%What is important is that these equations completely fix the two charges $\mathcal{J}$ and $\mathcal{E}$ in terms of the two cross-ratios $z$ and $\bar z$. To 
%

% Note however that 
% we do not need the solution to compute the action.
To obtain the action we can directly substitute eqs.~(\ref{phiDot}) and (\ref{rDot}) into the action to find 
\beq
S(M)= m\left( \int_{\epsilon}^{ \eta_*} +\int_{\frac{\epsilon}{\sqrt{x \bar x}}}^{ \eta_*}  \right) \frac{d\eta}{\eta} \frac{1}{\sqrt{\mathcal{J}^2 M \eta^6-\eta^4 \left(\mathcal{J}^2+M\right)-\eta^2 \left(\mathcal{J}^2+\mathcal{E}^2-1\right)+1}}
\eeq
Note that when computing $S(M)-S(0)$ the $\epsilon$ divergent terms coming from the lower integration limit 
drop out nicely and we can simply take $\epsilon \to 0$ there. This combination is what gives us the 
% \beq
% S(M)-S(0)= 2m  \int_{0}^{ \eta_*}   \frac{d\eta}{\eta} \left( \frac{1}{\sqrt{\mathcal{J}^2 M \eta^6-\eta^4 \left(\mathcal{J}^2+M\right)-\eta^2 \left(\mathcal{J}^2+\mathcal{E}^2-1\right)+1}}- (M\to 0)\right)
% \eeq
% for the 
relevant four point function with stripped off two-point function prefactors. 

We will now make progress analytically in the result by taking some limits. 

\subsubsection{Small black holes $M \to 0$} \la{expansionSec}
If we expand the equations above at small $M$ all integrals become trivial to compute in terms of functions like log and arctan. We then easily find 
\beqa
&&\!\!\!\!\!\!\!\!\!\! \!\!\!\!\!\!\!\!\!\!  S(M)-S(0)=M {\color{blue} \left(-\frac{\left(z^2+4 z+1\right) \left(\bar{z}-1\right) \log (z)}{4 (z-1) \left(z-\bar{z}\right)}+\frac{(z-1) \left(\bar{z}^2+4 \bar{z}+1\right) \log \left(\bar{z}\right)}{4 \left(z-\bar{z}\right)
   \left(\bar{z}-1\right)}-\frac{3}{2}
   \right)} \label{bigExpansion}\\\nn
 && \!\!\!\!\!\!\!\!\!\! \!\!\!\!\!\!\!\!\!\!  +
   M^2 { \color{magenta} \Big(
\tfrac{8 z^2 \bar{z}^2+z^2 \bar{z}+z \bar{z}^2-106 z \bar{z}+43 \bar{z}^2+\bar{z}+43 z^2+z+8}{16 \left(z-\bar{z}\right)^2}} \\\nn
   && { \color{magenta} \!\!\!\!\!\!\!\!\!\! \!\!\!\!\!\!\!\!\!\! -\tfrac{(z-1) \left(z^2 \bar{z}^4-2 z^2 \bar{z}^3+6 z^2 \bar{z}^2+24 z^2 \bar{z}+5 \bar{z}^4-10 z \bar{z}^3+24 \bar{z}^3-48 z
   \bar{z}^2+6 \bar{z}^2-10 z \bar{z}-2 \bar{z}+5 z^2+1\right) \log \bar{z}}{16 \left(z-\bar{z}\right)^3 \left(\bar{z}-1\right)} }\\\nn
   && { \color{magenta} \!\!\!\!\!\!\!\!\!\! \!\!\!\!\!\!\!\!\!\! +\tfrac{\left(\bar{z}-1\right) \left(z^4 \bar{z}^2-2
   z^3 \bar{z}^2-10 z^3 \bar{z}+6 z^2 \bar{z}^2-48 z^2 \bar{z}+24 z \bar{z}^2-10 z \bar{z}+5 \bar{z}^2+5 z^4+24 z^3+6 z^2-2 z+1\right)\log z}{16 (z-1) \left(z-\bar{z}\right)^3} }\\\nn
   && { \color{magenta} \!\!\!\!\!\!\!\!\!\! \!\!\!\!\!\!\!\!\!\! + \tfrac{\left(z^4 \bar{z}^3-2 z^4
   \bar{z}^2-5 z^4 \bar{z}+z^3 \bar{z}^4+16 z^3 \bar{z}^3+8 z^3 \bar{z}^2-32 z^3 \bar{z}-2 z^2 \bar{z}^4+8 z^2 \bar{z}^3+24 z^2 \bar{z}^2+8 z^2 \bar{z}-5 z \bar{z}^4-32 z \bar{z}^3+8 z \bar{z}^2+16 z
   \bar{z}-5 \bar{z}^3-2 \bar{z}^2+\bar{z}-5 z^3-2 z^2+z\right) \log z\log \bar{z}}{16 \left(z-\bar{z}\right)^4} }\\\nn
   && { \color{magenta} \!\!\!\!\!\!\!\!\!\! \!\!\!\!\!\!\!\!\!\! -\tfrac{\left(\bar{z}-1\right)^2 \left(z^6 \bar{z}+z^5 \bar{z}^2+16 z^5 \bar{z}-4
   z^4 \bar{z}^2-13 z^4 \bar{z}-6 z^3 \bar{z}^2-80 z^3 \bar{z}+20 z^2 \bar{z}^2-13 z^2 \bar{z}+25 z \bar{z}^2+16 z \bar{z}+\bar{z}+25 z^5+20 z^4-6 z^3-4 z^2+z\right) \log ^2z}{32 (z-1)^2
   \left(z-\bar{z}\right)^4} } \\  \nn
   && { \color{magenta} \!\!\!\!\!\!\!\!\!\! \!\!\!\!\!\!\!\!\!\! -\tfrac{(z-1)^2 \left(z^2 \bar{z}^5-4 z^2 \bar{z}^4-6 z^2 \bar{z}^3+20 z^2
   \bar{z}^2+25 z^2 \bar{z}+z \bar{z}^6+16 z \bar{z}^5+25 \bar{z}^5-13 z \bar{z}^4+20 \bar{z}^4-80 z \bar{z}^3-6 \bar{z}^3-13 z \bar{z}^2-4 \bar{z}^2+16 z \bar{z}+\bar{z}+z\right) \log
   ^2\bar{z}}{32 \left(z-\bar{z}\right)^4 \left(\bar{z}-1\right)^2} 
 \Big) } + O(M^3)
\eeqa
The first line is just the conformal block for the graviton exchange as we saw in the main text. The second line should correspond to an exchange of two gravitons. Indeed we find an infinite sum precisely compatible with that interpretation (it is infinite since the two massless gravitons can easily form states of any spin and twist)
\beqa
{\color{blue} \texttt{blue}} &=& c^{(1)}_{4,2} \times \mathcal{F}_{4,2}(w,\bar w) \qquad,  \qquad (w,\bar w)=(1-z, 1-\bar z) \\  
{\color{magenta} \texttt{magenta}} &=& \sum_{J=0}^\infty  \sum_{\Delta=\text{max}(8,4+J)}^\infty c^{(2)}_{\Delta,J}  \mathcal{F}_{\Delta,J}(w,\bar w)  
\eeqa
where $c^{(1)}=-1/120$. We could find several trajectories for $c^{(2)}$ but not a closed expression. For example
\beqa
&&c^{(2)}_{4+J,J} = -\frac{\sqrt{\pi } 2^{-2 J-5} (J-2) J \left(J^4+6 J^3+35 J^2+78 J-72\right) \Gamma (J-3)}{(J+2) (J+4) (J+6) \Gamma \left(J+\frac{3}{2}\right)}  \,, \qquad J=4,6,8,\dots \nn \\
&&c^{(2)}_{\Delta,0}=-\frac{\pi  4^{-\Delta -4} \left(32 \Delta ^6-429 \Delta ^5+1410 \Delta ^4+1660 \Delta ^3-10472 \Delta ^2+2144 \Delta +7680\right) \Gamma \left(\frac{\Delta }{2}-1\right)^2}{5 (\Delta -5) \Gamma
   \left(\frac{\Delta -1}{2}\right) \Gamma \left(\frac{\Delta +3}{2}\right)} \,,\nn \\   &&\qquad\qquad\qquad\qquad\qquad\qquad\qquad\qquad\qquad\qquad\qquad\qquad\qquad\qquad\qquad\qquad \Delta=8,10,12,\dots \nn
\eeqa
And so on. 
%These expressions have some nice reciprocity relations that Francesco pointed out. It is trivial to get many other such expressions. {\textbf{ To do: Explore analogy of the $M$ expansion to the expansion of the Virasoro block in terms of global blocks. }}

% {\color{blue} P: Rob, where is the BH metric with higher curvatures? We culd recompute the c's with that...}

It would be interesting to see what changes in higher curvature gravity. In particular, it would be interesting if some structure constants would become negative for swampland values of these curvature coefficients. 

%\subsubsection{NEW Radial geodesics}

\subsubsection{Radial Geodesics and OPE} 
%\la{radialGeo}
Another regime where we get remarkable simplifications is when we put the four points on a line. Then $x=\bar x$, and therefore ${\cal J}=0$. All formulae in section (\ref{chargesSec}) simplifies and 
for example the action simply reads
% (\ref{Deltar}) and (\ref{Deltaphi}) simply reduce to 
% \beq
% \!\! \!\! \tfrac{\sqrt{\sqrt{4 M+1}-1}}{\sqrt{8 M+2}}{
%  \tan ^{-1} \tfrac{2 \sqrt{2} \sqrt{M \left(\sqrt{4 M+1}+1\right)} \mathcal{E}}{4 M-\left(\sqrt{4 M+1}+1\right) \left(\mathcal{E}^2-1\right)} -\tfrac{\sqrt{\sqrt{4 M+1}+1}}{\sqrt{8M+2}} \tanh
%    ^{-1}\tfrac{2 \sqrt{2} \sqrt{M \sqrt{4 M+1}-M} \mathcal{E}}{\left(\sqrt{4 M+1}-1\right) \left(\mathcal{E}^2-1\right)+4 M}} = \log(z) \,. \la{lineEq}
% \eeq
% giving us $\mathcal{E}(z)$ and the action simply becomes 
\beq
S(M)-S(0)= -\frac{m}{2} \log(1 + 4 M - 2 \mathcal{E}^2 + \mathcal{E}^4 )
-(M \to 0) 
\eeq
which is thus parametrically a function of $x$. It is nice to consider the OPE expansion for $x\to 1$ so that the geodesic stays close to the boundary. This limit is the limit of large energy ${\cal E}\gg 1$,
and we can expand all expressions in that limit trivially to any desired OPE order and for any mass $M$. 
Using $w=1-x$ we find
\beqa
S(M)-S(0)&=&{\color{blue}-\frac{M }{40} }w^4{\color{blue}-\frac{M }{20} }w^5{\color{blue}-\frac{M }{14}} w^6{\color{blue}-\frac{5 M }{56}} w^7+\left({\color{blue}-\frac{5 M}{48}}{\color{magenta}-\frac{11 M^2}{14400}}\right) w^8\\
&+&
\left({\color{blue}-\frac{7 M}{60}} {\color{magenta}-\frac{11 M^2}{3600}}\right)
   w^9+\left({\color{blue}-\frac{7 M}{55}}{\color{magenta}-\frac{37 M^2}{4928}}\right) w^{10}+\left({\color{blue}-\frac{3 M}{22}}{\color{magenta}-\frac{1081 M^2}{73920}}\right) w^{11}\nn \\
   &+&\left({\color{blue}-\frac{15 M}{104}}{\color{magenta}-\frac{1011 M^2}{40768}}w-\frac{89 M^3}{1872000}\right) w^{12}+\left({\color{blue}-\frac{55 M}{364}}{\color{magenta}-\frac{39083 M^2}{1019200}}-\frac{89 M^3}{312000}\right)
   w^{13}\nn \\
   &+&\left({\color{blue}-\frac{11 M}{70}}{\color{magenta}-\frac{82079 M^2}{1478400}}-\frac{21991
   M^3}{22176000}\right) w^{14}+\left({\color{blue}-\frac{13 M}{80}}{\color{magenta}-\frac{4525 M^2}{59136}}-\frac{1657 M^3}{633600}\right) w^{15} \nn \\
   &+&O(w^{16}) \nn
\eeqa
% {\color{red} NOT SURE THIS IS RIGHT} For instance, as the two geodesic points colide we have $z\to 1$ or $w=1-z\to 0$. so that the right hand side of (\ref{lineEq}) vanishes that that equation will be solved for very small $\mathcal{E}$. (That makes sense; a very small geodesic has very little energy.) We can thus just expand the last two equations at small $\mathcal{E}$ to find that the subtracted action is %$S(M)-S(0)$ is given by 
% \beqa
% &&S(M)-S(0)=\\
% &&-\frac{1}{2} \log (4 M+1)+M w^2+M w^3\nn\\
% &&+\left(\frac{2 M^3}{3}+\frac{M^2}{6}+\frac{7 M}{8}\right) w^4+\left(\frac{4 M^3}{3}+\frac{M^2}{3}+\frac{3 M}{4}\right) w^5\nn \\
% &&+\left(\frac{16 M^5}{9}+\frac{56
%    M^4}{45}+\frac{11 M^3}{5}+\frac{31 M^2}{60}+\frac{13 M}{20}\right) w^6\nn\\
%    &&+\left(\frac{16 M^5}{3}+\frac{56 M^4}{15}+\frac{49 M^3}{15}+\frac{43 M^2}{60}+\frac{23 M}{40}\right) w^7\nn \\ 
%    &&+\left(\frac{64
%    M^7}{9}+\frac{2396 M^6}{315}+\frac{4234 M^5}{315}+\frac{1973 M^4}{252}+\frac{109 M^3}{24}+\frac{6233 M^2}{6720}+\frac{291 M}{560}\right) w^8\nn \\
%    &&+\left(\frac{256 M^7}{9}+\frac{9584 M^6}{315}+\frac{3032
%    M^5}{105}+\frac{1459 M^4}{105}+\frac{181 M^3}{30}+\frac{1921 M^2}{1680}+\frac{67 M}{140}\right) w^9+O\left(w^{10}\right) \nn
% \eeqa
Note that although this is not a small $M$ expansion, it is automatically organized as one as expected physically. Namely, the more subleading the OPE is, the more \textit{primary} gravitons are being exchanged and thus more powers of $M$ appear. The linear and quadractic terms in $M$ agree with general limit of the previous section when taken the line. 
 
% Another trivial limit where things simplify is $M\to 0$. In this limit we can reproduce the results of (\ref{expansionSec}) and easily go to higher orders in $M$. This is quite useful to debug the off-the line computations. 

\subsection{AdS$_3$ geodesics} \la{ads3Geo}

In this case the problem is simpler, since the blackening factor $f(r)=r^2+1-M$ is polynomial. 
We will introduce the quantity $r_h^2=M-1$. Note that there are two regimes: the black hole regime where $M>1$ 
and $r_h\ge 0$ is real, the defect regime where $0\leq M\leq 1$ and $r_h$ is imaginary. 
We will be careful in the following in finding expressions for the geodesic where the transitions is smooth. 

The AdS$_3$ equations of motions are
\begin{align}
\phi' &= {\pm}\frac{\mathcal{J} \eta \eta'}{\sqrt{r_h^2 \mathcal{J}^2 \eta^4 - (\mathcal{E}^2 + \mathcal{J}^2 + r_h^2)\eta^2 + 1}}\\
r' &= \pm \frac{\mathcal{E} \eta \eta'}{\left( 1-r_h^2 \eta^2 \right) \sqrt{r_h^2 \mathcal{J}^2 \eta^4 - (\mathcal{E}^2 + \mathcal{J}^2 + r_h^2)\eta^2 + 1}} - \frac{\eta \eta'}{\eta^2 +1}
\end{align}
The square root contains only a quadratic expression in $\eta^2$, so there are only two roots.  
\beq
\eta^2_{\pm}= \frac{ 
{\cal J}^2 +r_h^2 +{\cal E}^2\pm \sqrt{ ({\cal J}^2 +r_h^2 +{\cal E}^2)^2-4 {\cal J}^2 r_h^2} 
}{2{\cal J}^2 r_h^2}
\eeq
The turning point corresponds to $\eta_\star=\eta_-$, since this is the one that can go to the boundary $\eta=0=z$.
As in AdS$_5$ -- see eqs.~\eqref{Deltaphi} and \eqref{Deltar} -- 
we can integrate these expressions to find the shifts in $\phi$ and $r$. 
One might notice that the square root in $\eta_{\pm}$ is simply,
\beq
({\cal J}^2 +r_h^2 +{\cal E}^2)^2-4 {\cal J}^2 r_h^2\ \ = \prod_{\substack{ s_1=\pm\\ s_2=\pm}}({\cal J} + s_1 r_h+ i s_2 {\cal E})
\eeq
Then, the combinations that appear after doing the integrals are such that the result can be written as
\begin{align}
\underbrace{-i\log\sqrt{{x}/{\bar x} }}_{\Delta \phi} =
\frac{\log\frac{ 
({\cal J}+r_h+i{\cal E})({\cal J}+r_h-i{\cal E}) }{({\cal J}-r_h-i{\cal E})({\cal J}-r_h+i{\cal E})  }}{2r_h}\qquad;\qquad
\underbrace{\log \sqrt{ x \bar{x}}}_{\Delta\rho} = \frac{\log\frac{ ({\cal J}-r_h+i{\cal E})({\cal J}+r_h-i{\cal E}) }{({\cal J}-r_h-i{\cal E})({\cal J}+r_h+i{\cal E})  }}{2ir_h}
\end{align}
These function have a smooth transition from the defect to the black hole regime.
It is also interesting to see how both remain real. In the defect regime, $r_h$ is imaginary. By checking the argument of the log on the r.h.s.~of $\Delta\phi$, this is a phase, and so $\Delta\phi$ is real, similarly, by checking the argument of the log on the r.h.s.~of $\Delta\rho$, this is a radius, so $\Delta\rho$ is again real. In the black hole regime $r_h$ is real, and the opposite happens. 

%{\bf OLD}
%\begin{align}
%\underbrace{\text{arccos} \frac{x+\bar{x}}{2 \sqrt{x \bar{x}}}}_{\Delta \phi} &= \frac{1}{\sqrt{M-1}} \log \left(\frac{\sqrt{(\mathcal{E}^2 + \mathcal{J}^2 + M - 1) - 4 (M-1) \mathcal{J}^2}}{\mathcal{E}^2 + \mathcal{J}^2 + M-1 + 2 \sqrt{M-1} \mathcal{J}} \right)\\
%
%\underbrace{\log \sqrt{ x \bar{x}}}_{\Delta\rho} &= \frac{1}{\sqrt{M-1}} \left(\frac{\pi}{2} - \arctan\left(\frac{M-1-\mathcal{E}^2 - \mathcal{J}^2}{2\sqrt{M-1} \mathcal{E}}\right) \right)
%\end{align}
%{\color{red} I also think $\Delta r$ is not smooth.}

We can now compute the action for the geodesic, by substituting $\phi'$ and $r'$,
\begin{align}
S(M) &
= m \left(\int_{\frac{\epsilon}{1}}^{\eta_*} + \int_{\frac{\epsilon}{\sqrt{z \bar{z}}}}^{\eta_*} \right) \frac{d\eta}{\eta} \frac{1}{\sqrt{(M-1)\mathcal{J}^2 \eta^4 - (\mathcal{E}^2 + \mathcal{J}^2 + M - 1)\eta^2 + 1}}
\end{align}
This integral 
receives no contribution from $\eta_\star$, and it is log divergent, 
\begin{align}
S(M)&=-\frac{m}{2} \log\Bigg[\frac{\epsilon^4}{16 x \bar{x}}
\prod_{s_1=\pm,s_2=\pm}({\cal J} + s_1 r_h+ i s_2 {\cal E})\Bigg]
\end{align}
The next step is to invert ${\cal E},{\cal J}$ in terms of $\Delta\rho,\Delta\phi$, thus $x,\bar{x}$. 
For real values of ${\cal E},{\cal J}$ the range of $x,\bar{x}$ 
is restricted on a certain domain, which is less than the whole boundary of AdS.
With this understanding we find, 
\begin{align}
S(M)&=-\frac{m}{2}\log\Bigg[
\frac{\epsilon^4 r_h^4}{4 x\bar{x} }
\frac{1}{(\cosh(i r_h\Delta\rho)-\cosh(r_h\Delta\phi))^2}\Bigg]\\
&=
-\frac{m}{2} \log\Bigg[ \frac{\epsilon^4 r_h^4 x^{-ir_h-1}\bar{x}^{-ir_h-1 }  }{ 
(1-x^{ -ir_h } )^2(1-\bar{x}^{-ir_h})^2} \Bigg]
\end{align}
For $r_h=i$, \ie empty AdS, the result becomes $S(0)=-\frac{m}{2} \log[ \frac{\epsilon^4   }{ 
(1-x )^2(1-\bar{x})^2}]$.
The geodesic 4pt correlator in the BTZ background is
\begin{align}
\!\!\!\!\!\frac{\langle \phi_H(0) \phi_L(1) \phi_L(x) \phi_H(\infty) \rangle}{\langle \phi_H(\infty) \phi_H(0)\rangle \langle \phi_L(x) \phi_L(1) \rangle} &= e^{-(S(M)-S(0))}=\notag\\
&=\left[(1-M) \frac{(1-x) x^{\frac{\sqrt{1-M}-1}{2}} (1 - \bar{x}) \bar{x}^{\frac{\sqrt{1-M}-1}{2}}}{(x^{\sqrt{1-M}} - 1) (\bar{x}^{\sqrt{1-M}} - 1)} \right]^m
\end{align}
Note this is invariant under $x\rightarrow 1/x$ and $\bar{x}\rightarrow1/\bar{x}$, as it should. Quite nicely, the result is simply the Virasoro block in the t-channel orientation. In the defect regime, we can expand at small $M$ to get, 
\beq
\!\!\!\!\!\frac{\langle \phi_H(0) \phi_L(1) \phi_L(x) \phi_H(\infty) \rangle}{\langle \phi_H(\infty) \phi_H(0)\rangle \langle \phi_L(x) \phi_L(1) \rangle} =1 + \frac{m M}{24} \left( (1-x)^2 {}_2 F_1(2,2,4;1-x) + c.c. \right) + O(M^2)
\eeq
which at leading order is just the stress tensor global conformal block.

\section{Schur 3-pt functions: a new exact and non-extremal result} \label{combinatorics}
In this appendix we will derive a new exact formula for three-point function of half-BPS operators given by fully symmetric (or anti-symmetric) characters. As will any three-point function of half-BPS correlators, this correlation function can be computed at tree level, by working out the combinatoric of Wick contractions. 

There are plenty of exact three-point correlators of half-BPS operators computed in this way in the the literature but most of them -- if not all -- were computed for extremal correlators where the length of one operator is equal to the sum of the other two and thus there are no propagators between those two smaller operators. The main novelty of this appendix is that we will consider a maximally non-extremal correlator where all fields are connected to all fields.

The starting point reads
\beqa
&&
 \left\langle 
\prod_{i=1}^3 \text{det}[1 + t_i \,Y_i \cdot \Phi_i(x_i)]^{\pm 1} 
 \right\rangle = \label{hundreedtwo} \\ && \qquad = C \int \prod_{i \neq j} d\rho_{ij} e^{\,\frac{2 N}{g^2}\sum\limits_{i<j} \rho_{ij} \rho_{ji}} 
  (1 - \hat{\rho}_{12}\hat{\rho}_{12} -\hat{\rho}_{13}\hat{\rho}_{31} - \hat{\rho}_{23}\hat{\rho}_{32} + \hat{\rho}_{12}\hat{\rho}_{23}\hat{\rho}_{31} + \hat{\rho}_{13}\hat{\rho}_{32}\hat{\rho}_{21})^{\pm N} \nn
\eeqa This formula is derived following \cite{Jiang:2019xdz, Yang:2021kot}: One (1) introduces a set of auxiliary fields $\chi$ to cast the determinants as Gaussian integrals\footnote{Depending on what sign we want on the left hand side we use fermions or bosons.}; (2) integrates out the scalars $\Phi$ to obtain a quartic action in the auxiliary field; (3) introduce a second set of auxiliary fields $\rho$ to render that quartic interation quadractic using the usual Hubbard-Stratonovich trick; (4) integrate out the now quadractic fields $\chi$ to arrive at (\ref{hundreedtwo}). 

The dependence on the generating function variables $t_j$ and on the positions and polarizations on the right hand side of (\ref{hundreedtwo}) is hidden inside the hatted variables
\begin{equation}
\hat{\rho}_{ij} \equiv \rho_{ij} \times  \sqrt{4 t_i t_j \frac{Y_i\cdot Y_j}{x_{ij}^2}} \,. 
\end{equation}

Each determinant is a generating function of characters of fully anti-symmetric ($+$ sign) or symmetric ($-$ sign)  characters so to get the correlation function of the desired characters we simply need to pick up the corresponding powers of $t_j$ on the right hand side by using the usual Binomial expansion to expand out 
\beq
  (1 - \hat{\rho}_{12}\hat{\rho}_{12} -\hat{\rho}_{13}\hat{\rho}_{31} - \hat{\rho}_{23}\hat{\rho}_{32} + \hat{\rho}_{12}\hat{\rho}_{23}\hat{\rho}_{31} + \hat{\rho}_{13}\hat{\rho}_{32}\hat{\rho}_{21})^{\pm N}
\eeq
Because there are six terms in this expression we can cast it as a five-fold binomial sum where each of the six terms is raised to an integer power. The power of the last and next-to the last terms needs to be the same to get a non-trivial result and this kills one of the five sums. Since we are interested in fixed powers of $t_1$, $t_2$ and $t_3$ we set three extra constraints on the powers arising in these four sums killing three of them. So we are left with a single sum for the desired character three-point function (here specializing to the anti-symmetric case for simplicity):
\beqa
\langle \mathcal{O}_{\chi^{anti}_{j_1}} \mathcal{O}_{\chi^{anti}_{j_2}} \mathcal{O}_{\chi^{anti}_{j_3}} \rangle = C \left(\frac{Y_1\cdot Y_2}{x_{12}^2}\right)^{(j_1+j_2-j_3)/2} \left(\frac{Y_2\cdot Y_3}{x_{23}^2}\right)^{(j_2+j_3-j_1)/2} \left(\frac{Y_1\cdot Y_3}{x_{13}^2}\right)^{(j_1+j_3-j_2)/2} \times \\ \times  \sum_{n=0}^{N/2} \binom{N}{2n, \ \frac{j_1+j_2+j_3}{2}-3n} \binom{\frac{j_1+j_2+j_3}{2}-3n}{\frac{j_1+j_2-j_3}{2}-n, \ \frac{j_1+j_3+-j_2}{2}-n} \binom{2n}{n}
(-1)^{\frac{j_1+j_2+j_3}{2}}  \nn \\
\times \langle({\rho}_{12} {\rho}_{21})^{\frac{j_1+j_2-j_3}{2}} \rangle \langle({\rho}_{13} {\rho}_{31})^{\frac{j_1+j_3-j_2}{2}} \rangle \langle({\rho}_{23} {\rho}_{32})^{\frac{j_2+j_3-j_1}{2}} \rangle \nn 
\eeqa
where the averages in the last line are with respect to to the Gaussian measure in eq.~(\ref{hundreedtwo}) so they can be trivially evaluated. (For symmetric characters we get a similar expression with $N\to -N$ and the sum running up to infinity.) Once we evaluate the integrals we can simply preform the sum: We get an hypergeometric function. We should also normalize properly the two-point function to get a properly defined three-point function. Keeping track of all those simple normalization factors, we obtain eq.~(\ref{symC}).\footnote{Formula (\ref{symC}) is for symmetric characters. For anti-symmetric simply replace $N \to -N$ there.}

\end{document}